\documentclass[twocolumn, longauth]{aa}  

\usepackage{graphicx}
\usepackage{txfonts}
\usepackage{xcolor}
\usepackage{natbib}
\usepackage{subfig}
\usepackage{afterpage}
\usepackage{color}
\usepackage[citecolor=blue, colorlinks=True, urlcolor=blue]{hyperref}
\newcommand{\Reff}{$R_{\rm eff}$}
\newcommand{\seff}{$\sigma_{\rm eff}$}
\newcommand{\vlos}{$V_{\rm LOS}$}
\newcommand{\slos}{$\sigma_{\rm LOS}$}
\newcommand{\Vsys}{$V_{\rm sys}$}
\newcommand{\Mdyn}{$M_{\rm dyn}$}

\begin{document}

   \title{Looking into the faintEst WIth MUSE (LEWIS): Exploring the nature of ultra-diffuse galaxies in the Hydra-I cluster}
   \subtitle{II. Stellar kinematics and dynamical masses.}

   \author{Chiara Buttitta\inst{1}\fnmsep\thanks{\email{chiara.buttitta@inaf.it} }
          \and Enrichetta Iodice\inst{1}
          \and Goran Doll\inst{1,2}
          \and Johanna Hartke \inst{3,4}
          \and Michael Hilker\inst{5}
          \and Duncan A. Forbes \inst{6}
          \and Enrico M. Corsini \inst{7,8}
          \and Luca Rossi \inst{1,2}
          \and Magda Arnaboldi \inst{5}
          \and Michele Cantiello\inst{9}
          \and Giuseppe D'Ago \inst{10}
          \and Jesus Falc{\'o}n-Barroso \inst{11,12}
          \and Marco Gullieuszik \inst{7}
          \and Antonio La Marca \inst{13,14}
          \and Steffen Mieske \inst{15}  
          \and Marco Mirabile \inst{5,9,16}
          \and Maurizio Paolillo \inst{1,2}
          \and Marina Rejkuba \inst{5}
          \and Marilena Spavone \inst{1}
          \and Chiara Spiniello \inst{17,1}   
          \and Marc Sarzi \inst{18}
          }
        %--1
        \institute{INAF $-$ Astronomical Observatory of Capodimonte, Salita Moiariello 16, I-80131, Naples, Italy
        \and
        %%--2
        University of Naples ``Federico II'', C.U. Monte Sant'Angelo, Via Cinthia, 80126, Naples, Italy
        \and
        %%--3
        Finnish Centre for Astronomy with ESO (FINCA), FI-20014 University of Turku, Finland
        \and
        %%--4
        Tuorla Observatory, Department of Physics and Astronomy, FI-20014 University of Turku, Finland            
        \and
        %%--5
        European Southern Observatory, Karl$-$Schwarzschild-Strasse 2, 85748 Garching bei München, Germany
        \and
        %--6
        Centre for Astrophysics \& Supercomputing, Swinburne University of Technology, Hawthorn VIC 3122, Australia
        \and
        %%--7
        INAF $-$ Osservatorio Astronomico di Padova, Vicolo dell’Osservatorio 5, I-35122 Padova, Italy
        \and
        %%--8
        Dipartimento di Fisica e Astronomia ``G. Galilei'', Universit\`a di Padova, vicolo dell'Osservatorio 3, I-35122 Padova, Italy
        \and
        %%--9
        INAF $-$ Astronomical Observatory of Abruzzo, Via Maggini, 64100, Teramo, Italy
        \and
        %-10
        Institute of Astronomy, University of Cambridge, Madingley Road, Cambridge CB3 0HA, UK 
        \and
        %--11
        Instituto de Astrof\'isica de Canarias, Calle V\'ia La\'actea s/n, E-38205. La Laguna, Tenerife, Spain
        \and
        %--12
        Departamento de Astrof\'isica. Universidad de La Laguna, Av. del Astrof\'isico Francisco S\'anchez s/n, E-38206, La Laguna, Tenerife, Spain
        \and
        %--13
        SRON Netherlands Institute for Space Research, Landleven 12, 9747 AD Groningen, The Netherlands
        \and
        %--14
        Kapteyn Astronomical Institute, University of Groningen, Postbus 800, 9700 AV Groningen, The Netherlands
        \and
        %--15
        European Southern Observatory, Alonso de Cordova 3107, Vitacura, Santiago, Chile
        \and   
        %--16
        Gran Sasso Science Institute, viale Francesco Crispi 7, I-67100 L'Aquila, Italy
        \and
        %--17
        Sub-Dep. of Astrophysics, Dep. of Physics, University of Oxford, Denys Wilkinson Building, Keble Road, Oxford OX1 3RH, United Kingdom
        \and
        %--18
        Armagh Observatory and Planetarium, College Hill, Armagh, BT61 9DG UK
        }

\date{Received ....; accepted ...}

\abstract
% context heading (optional)
% {} leave it empty if necessary  
  {This paper focuses on a class of galaxies characterised by an extremely low surface brightness: the ultra-diffuse galaxies (UDGs). We used new integral-field spectroscopic data, obtained with the ESO Large Programme Looking into the faintEst WIth MUSE (LEWIS). It provides the first homogeneous integral-field spectroscopic survey performed by MUSE at the Very Large Telescope of a complete sample of UDGs and low-surface brightness galaxies within 0.4 virial radius in the Hydra I cluster, according to the UDG abundance-halo mass relation.}
% aims heading (mandatory)
  {Our main goals are addressing the possible formation channels for this class of objects and investigating possible correlations of their observational properties, including the stacked (1D) and spatially-resolved (2D) stellar kinematics. In particular, we derive the stellar velocity dispersion from the stacked spectrum integrated within the effective radius (\seff) and measure the velocity map of the galaxies in LEWIS. These quantities are used to estimate their dynamical mass (\Mdyn).} 
% methods heading (mandatory)
  {We extracted the 1D stacked spectrum inside the effective radius (\Reff), which guarantees a high S/N ratio, to obtain an unbiased measure of \seff. To derive the spatially-resolved stellar kinematics, we first applied the Voronoi tessellation algorithm to bin the spaxels in the datacube and then we derived the stellar kinematics in each bin, following the same prescription adopted for the 1D case. We extracted the velocity profiles along the galaxy major and minor axes and measured the semi-amplitude ($\Delta V$) of the velocity curve.}
% results heading (mandatory)
  {We found that 7 out of 18 UDGs in LEWIS show a mild rotation ($\Delta V \sim 25- 40$\,km s$^{-1}$), 5 do not have evidence of any rotation, and the remaining 6 UDGs are unconstrained cases. This is the first large census of velocity profiles for UDGs. On average, UDGs in LEWIS are characterised by low values of \seff\,($\leq 30$~km s$^{-1}$), comparable with available values from the literature. Two objects show larger values of \seff \,($\sim 30-40$~km s$^{-1}$). Such larger values might reasonably be due to the large rotation observed in these galaxies, which affects the values of \seff. In the Faber-Jackson relation plane, we found a group of UDGs consistent with the relation within the errorbars. Outliers of the Faber-Jackson relation are objects with non-negligible rotation component. On average, UDGs and LSB galaxies in the LEWIS 
  sample have larger dark matter content than dwarf galaxies (\Mdyn/$L_{V, {\rm eff}} \sim 10-100$~M$_\odot$/L$_\odot$) with similar total luminosity. We do not find clear correlations between the derived structural properties and the local environment.}
% conclusions heading (optional), leave it empty if necessary
  {By mapping the stellar kinematics for a homogenous sample of UDGs in a cluster environment, we have found a significant rotation for many galaxies. Therefore two classes of UDGs are found in the Hydra I cluster, based on the stellar kinematics, the rotating and non-rotating systems. This result, combined with the dark matter content and upcoming analysis of the star formation history and globular cluster population, can help to discriminate between the several formation scenarios proposed for UDGs.}


\titlerunning{II. Stellar kinematics and dynamical masses}
\authorrunning{C. Buttitta et al. }
\maketitle

%
%-------------------------------------------------------------------

\section{Introduction}
\label{sec:intro}

Several faint and large low surface-brightness (LSB) galaxies were first discovered in photographic plate surveys
in the 80-ies \citep[e.g. ][]{Sandage1984}. However, the interest in the science community was triggered nearly ten years ago, when a
large and intriguing population of galaxies, then dubbed ultra-diffuse galaxies (UDGs) was discovered in the Coma cluster 
\citep{vanDokkum2015, Koda2015, Yagi2016}. UDGs are empirically defined as galaxies with a central surface brightness fainter 
than $\mu_{0,g}\geq 24$~mag arcsec$^{-2}$ and an effective radius larger than \Reff\,$\geq$~1.5 kpc \citep{vanDokkum2015}.
Since then, many efforts have been devoted to derive the structural properties of these objects.
Several studies, based on new, deep imaging surveys, provided large samples of low-surface brightness (LSBs) galaxies, 
including UDGs \citep[see][and references therein]{Alabi2020, Forbes2020, Lim2020, Marleau2021,LaMarca2022a,Zaritsky2022, Buzzo2022, Buzzo2024}. 
With the increased statistics, UDGs are found to be $\sim 2.5 \sigma$ fainter and larger than the average 
distribution of the parent dwarf galaxy population. 
Therefore, UDGs are considered as the extreme LSB tail of the size–luminosity 
distribution of dwarf galaxies.

The considerable amount of imaging data collected so far showed that these galaxies span a wide range of 
structural and photometric properties. Observations strongly suggest that the UDGs might comprise
different types of galaxies, with different intrinsic properties, such as colours, globular cluster
(GC) content, age and metallicity, and dark matter (DM) amount \citep{Roman2017a,Leisman2017,
Ferre-Mateu2018,RuizLara2018,Prole2019b, Forbes2020, Gannon2021, Saifollahi2022, Buzzo2024}.

Because of their LSB nature, getting spectroscopic data for UDGs is a challenging task. 
To date, as opposed to the availability of deep images, we still lack a statistically significant 
sample of UDGs with spectroscopy, which strongly limits our knowledge of their stellar populations 
and DM content. Compared to the thousands of UDGs detected from imaging surveys, 
only $\leq100$ UDGs were analysed with spectroscopic data. 
However, these data revealed the existence of both metal-poor ($-0.5 \leq [M/H] \leq -1.5$~dex) 
and old systems \citep[$\sim9$~Gyr,][]{Pandya2018,Fensch2019,Ferre-Mateu2018,FerreMateu2023}, 
as well as younger star-forming UDGs \citep{Martin-Navarro2019}. 
Using the Dragonfly Ultrawide Survey, \citet{Shen2024} have recently provided spectroscopic confirmation 
for several UDGs, highlighting their quiescent nature and the presence of both old and intermediate-age stellar populations.

Compared to the stellar population analysis, the kinematic measurements are available for a few UDGs.
Recently, \cite{Gannon2024} have collected all the UDGs in the literature analysed with spectroscopic data. 
To date, a total of 18 UDGs have an estimate for the line-of-sight stellar velocity dispersion (\slos) measured 
directly with spectroscopic data or through the analysis of the velocity of the GC 
systems bound to the galaxy. On average, measurements suggest that UDGs have a low stellar velocity
dispersion within 1\Reff\,(\seff$\sim 5-50$~km s$^{-1}$). 

\cite{Chilingarian2019} studied the stellar kinematics of a sample of UDG-like galaxies. In this sample, 6 out of 9 
UDGs have effective radii smaller than 1.5 kpc, thus they do not fully satisfy the criteria proposed in \cite{vanDokkum2015}.
In addition, these UDG-like objects were observed through three slits. In these conditions, the spatial information for 
the stellar velocity field can be only partially derived.
Only DF44 and NGC\,1052-DF2, which are the most debated cases, especially in terms of DM content and GC 
populations, have resolved stellar kinematics \citep{Emsellem2019, vanDokkum2019b}. DF2 revealed a velocity 
profile close to the major photometric axis with a mild rotation of $\sim 6$\,km s$^{-1}$, whereas DF44 
shows no evidence of rotation.

Even considering the ever-growing statistics collected during the last 10 years,
the DM content of the UDGs remains one of the most debated topics and results point to a rather 
diverse population \citep{Kravtsov2024}. 
Most of the DM studies for UDGs (both from GCs analysis and spectroscopy) revealed that
UDGs have a larger DM content than dwarf galaxies of similar luminosity \citep{Toloba2018,vanDokkum2019,Forbes2021,Gannon2021},
but a few of them appear to be almost DM-free \citep{vanDokkum2018,Collins2021}.

%--- formation scenarios
Theoretical works show that the different types of observed UDGs require more than one formation channel
or, reasonably, a combination of internal or external physical processes, including
environmental effects. Gas-rich UDGs, with a dwarf-like DM halo, can originate from star formation
feedback or highly rotating DM halos
\citep{Amorisco2016,diCintio2017,Rong2017,Tremmel2020}. 
Gravitational interactions and merging between galaxies, as well as interactions with the environment, are `external' processes that might shape galaxies to become UDG-like systems, by removing the gas supply and/or puffing up their stellar component
\citep{Bennet2018,Muller2019,Tremmel2020,Carleton2021,vanDokkum2022}. 
In these scenarios, UDGs are expected to be red and quenched, gas-poor, and with different DM content and
metallicity. Blue, dusty, star-forming and DM-free UDGs, with moderate to low metallicity and UV emission, 
could originate from the collisional debris of merging galaxies
\citep{Lelli2015,Ploeckinger2018,Silk2019,Ivleva2024} or from ram-pressure-stripped gas clumps in the tails of the
so-called jellyfish galaxies \citep{Poggianti2019}.
All the above formation scenarios make specific predictions on the UDGs morphology,
colours, gas, DM content, and stellar population.

A few works in the literature have provided predictions of the kinematics of UDGs.
By analysing  IllustrisTNG simulations, \cite{Sales2020} found two
classes of UDGs in a cluster environment. The `born UDGs’ (B-UDGs),
formed from LSB galaxies which, after joining the cluster potential, lost their gas supply, and were quenched.
The `tidal-UDGs’ (T-UDGs) originated from tidal forces that acted on high-surface brightness galaxies in the cluster, 
removing their DM and puffing up their stellar component.
The two classes of UDGs are predicted to have different structural properties and location inside the cluster.
In detail, T-UDGs populate the centre of the clusters and, at a given stellar mass, have lower velocity dispersion, 
higher metallicity, and lower DM fraction with respect to the B-UDGs. 

Using the large volume of TNG50 simulations from field to galaxy clusters, \citet{Benavides2023}
found that UDGs have similar properties to the normal dwarf galaxies, i.e. comparable DM halos 
($M_{200}< 10^{11}$~M$_\odot$) and environmental trends, 
where field UDGs are star-forming and blue, whereas satellite UDGs are typically quiescent and red. In addition, 
they found that  massive UDGs ($M_\ast \gtrsim 10^{8.5}$~M$_\odot$) are rotation supported, whereas
dispersion-dominated systems populate the low mass regime.

Studying a sample of field UDGs in NIHAO 
simulations, \cite{Cardona-Barrero2020} analysed the kinematic support of UDGs by extracting 
the stellar velocity and velocity dispersion maps and computing the projected specific angular momentum 
\citep[i.e. $\lambda_{\rm R}$,][]{Emsellem2007}. They found that rotation-supported UDGs 
have a disk-like morphology, higher HI content, and larger radii. 

%--- Buzzo's paper
The existence of two classes of UDGs seems to be confirmed by observations too.
By combining structural properties from multi-band deep imaging data and 
stellar population properties derived from the spectral energy density fitting, 
for a sample $\sim$60 of UDGs, \citet{Buzzo2024} found a clear segregation 
in two classes of objects, named `class A' and `class B'. 
The UDGs of class A have lower stellar masses and 
prolonged star formation histories (SFHs), are more elongated, host fewer GCs, are younger and 
less massive, follow the mass-metallicity relation of classical dwarf galaxies, and live in less dense 
environments. Conversely, the UDGs of class B have higher stellar masses and rapid SFHs, they are 
rounder, host numerous GCs, are older and brighter, and follow the high-redshift mass-metallicity 
relation, suggesting an early quenching, and are found in dense environments. 
The observed properties of UDGs in the class A reconcile with a `puffed-up dwarf' formation scenario, 
i.e. a dwarf galaxy that experienced a physical process that provoked an expansion of its stellar distribution.
UDGs in Class B resemble the `failed galaxies’, i.e. a massive galaxy that lost its gas supply and
turned into a UDG-like system.

%--- key observables to discriminate between different formation scenarios
Based on the overview we provided above, it is clear that most of the key parameters necessary to 
discriminate between different classes of UDGs come from spectroscopy, to constrain 
the DM content from the stellar kinematics, and the age and metallicity from the 
stellar population analysis.

%--- LEWIS project can provide all of them
Using the data from the {\it Looking into the faintEst WIth MUSE}\footnote{\href{https://sites.google.com/inaf.it/lewis/home}
{https://sites.google.com/inaf.it/lewis/home}} (LEWIS, P.I. E. Iodice) project, we are making a decisive impact 
in this direction. LEWIS is an ESO Large Programme, approved in 2021, granted 133.5 hours with MUSE at the VLT.
LEWIS is the first homogeneous integral-field follow-up spectroscopic survey of 30 extreme LSB galaxies in the Hydra I cluster. The majority of LSB galaxies in the sample (22 in total) are UDGs. 
Thanks to the LEWIS project we are able to map, for the first time, {\it i)} the 2D stellar
kinematics, {\it ii)} the stellar population and {\it iii)} the GC content and their specific 
frequency, for a sample of UDGs in a galaxy cluster with integral-field (IF) 
spectroscopic data. The sample of UDGs and LSBs in LEWIS was first presented in \cite{Iodice2020, Iodice2021, LaMarca2022b}.
According to the UDG abundance-halo mass relation proposed in \cite{vanderBurg2017} and the virial mass of Hydra I cluster 
of galaxies \citep{LaMarca2022b}, we expect to observe $48\pm10$ UDGs within Hydra virial radius ($R_{200}\sim1.6$\,Mpc). 
Since the LEWIS sample extends out to $\sim0.4\,R_{200}$, we conclude that it is a nearly complete 
sample within this radius. The project description
and first preliminary results have been published in LEWIS Paper I \citep{Iodice2023}. 

%--- Stellar kinematics in LEWIS
This paper focuses on the stellar kinematics and aims to present the stacked and 
spatially resolved stellar kinematics and dynamical masses 
of UDGs and LSB galaxies in the LEWIS sample. 
The paper is organised as follows. 
In Section~\ref{sec:ob_data_reduction} we report the status of the observations and the confirmed galaxy membership. 
In Section~\ref{sec:struct_morph} we present the morphological classification of the LEWIS sample according to the structural parameters. 
In Section~\ref{sec:stellar_kin} we describe the recipe adopted to extract the stacked and spatially-resolved stellar kinematics. 
In Section~\ref{sec:results} we present the structural properties of the LEWIS sample in terms of stellar kinematics and DM content.
In Section~\ref{sec:discussion} we discuss correlations between the derived properties and cluster environment.
Finally, in Section~\ref{sec:conclusion} we report our conclusions and future perspectives.

%------------

%
\section{Galaxy sample, observations and data reduction}
\label{sec:ob_data_reduction}

The observations of LEWIS galaxies were carried out in service mode with the ESO integral-field spectrograph Multi Unit Spectroscopic Explorer \citep[MUSE, ][]{Bacon2010} at the ESO (Prog. Id. 108.222P, P.I. E. Iodice). 
MUSE was configured with the Wide Field Mode, covering a field-of-view (FOV) of $1' \times 1'$ and providing a spatial resolution of 0.2 arcsec pixel$^{-1}$. 
The nominal wavelength range of MUSE is 4800-9300\,\AA\ with a spectral sampling of 1.25 \AA\, pixel$^{-1}$ and an average nominal spectral resolution of FWHM = 2.51 \AA\, \citep{Bacon2017}.

The observing programme started in 2021 during the observational period P108 and is $\sim$92\% completed. 
The LEWIS project, galaxy sample, and observing strategy were presented in LEWIS Paper I \citep{Iodice2023}.
The original LEWIS sample is composed of 30 galaxies, 22 are classified as UDGs according to the definition by \cite{vanDokkum2015} and 8 are LSB galaxies \citep{Iodice2023}, assuming that they are at the distance of the Hydra I cluster. 
However, a few objects were discarded by the analysis due to different reasons: UDG2, UDG5 and LSB2 were excluded due to the presence of the Ferris wheel-like pattern, caused by the scattered light of a nearby bright star; the detection of LSB3 is difficult due to the proximity with the halo of NGC\,3311; the central Dominant (cD) galaxy of Hydra I \citep{Arnaboldi2012}; and the light distribution of UDG18 is heavily contaminated by the presence of a bright saturated star. 

Therefore, the final LEWIS sample is composed of 25 galaxies, 19 UDGs and 6 LSB galaxies, reported in Table~\ref{tab0_LEWIS}. 
To date, observations for 22 targets are completed while the remaining 3 objects (UDG16, UDG17, and UDG22) have partial observations. 
The data were reduced using the MUSE pipeline routine \citep{Weilbacher2020}, running in the {\sc ESOREFLEX} environment \citep{Freudling2013}. 
The steps of the standard data reduction included bias and overscan subtraction, lamp 
flat-fielding correction, wavelength calibration, determination of the line 
spread function (LSF), illumination correction, sky background subtraction, and 
flux calibration. For each object, the different exposures have been aligned and have been
combined to produce the final combined datacube. Since the resulting sky-subtracted datacube 
was characterised by the contamination of sky residuals, datacubes were cleaned by applying the Zurich Atmospheric Purge algorithm \citep[ZAP,][]{Soto2016}. 

As already described in \cite{Iodice2023}, we improved the standard data reduction by adding a few changes in a modified {\sc ESOREFLEX} workflow, as described below:

\begin{itemize}
    \item Custom mask: for each sample galaxy, we extracted the white light image by collapsing the datacube along the wavelength direction. We built a custom mask, detecting and masking all the possible light contamination from the foreground/background/spurious sources, including the contribution of the target. All sources are detected using the deep images available for this cluster \citep{Iodice2020b,LaMarca2022b}. 
    This mask improved the sky background estimate and it was directly injected in the {\sc ESOREFLEX} workflow, with additional parameters {\sc SkyFr\_1=0.75} and {\sc SkyFr\_2=0.1}\footnote{{\sc SkyFr\_1} and {\sc SkyFr\_2} are the fraction of spaxels in the sky image and scientific image used to evaluate the sky background, respectively.}.

    \item Normalisation of the exposures: we adapted a two-step approach to normalise the flux variations across the FOV and between exposures. To reduce slice-to-slice flux variations, for each exposure we used the {\sc autocalibration=deepfield} algorithm developed for the MUSE deep fields that calculate calibration factors that are applied to each pixelstable. When combining the exposures into the final datacube, we also accounted for flux variations of different exposures (e.g. due to different observing conditions and sky levels) with a multiplicative correction.
    
    \item ZAP with custom mask: we used the custom mask in the ZAP routine to improve the detection of the sky background filtering all the light contributions in the FOV. In addition, we realised that the automatic application of ZAP turned out to partially remove the flux of the galaxy target. Thus, we tested different combinations of parameters to minimise the subtraction of the signal from the target. 
    We used {\sc cfwidthSP}\footnote{{\sc cfwidthSP} is the window size for the continuum filter used to remove the continuum features for calculating the eigenvalues per spectrum.}
    between [30,50] and {\sc cfwidthSVD}\footnote{{\sc cfwidthSVD} is the window size for the continuum filter for the SVD computation.}
    between [10,30] \citep[see discussion in ][]{Soto2016}.
\end{itemize}

\begin{figure*}[!h]
    \centering
    \includegraphics[scale=0.45]{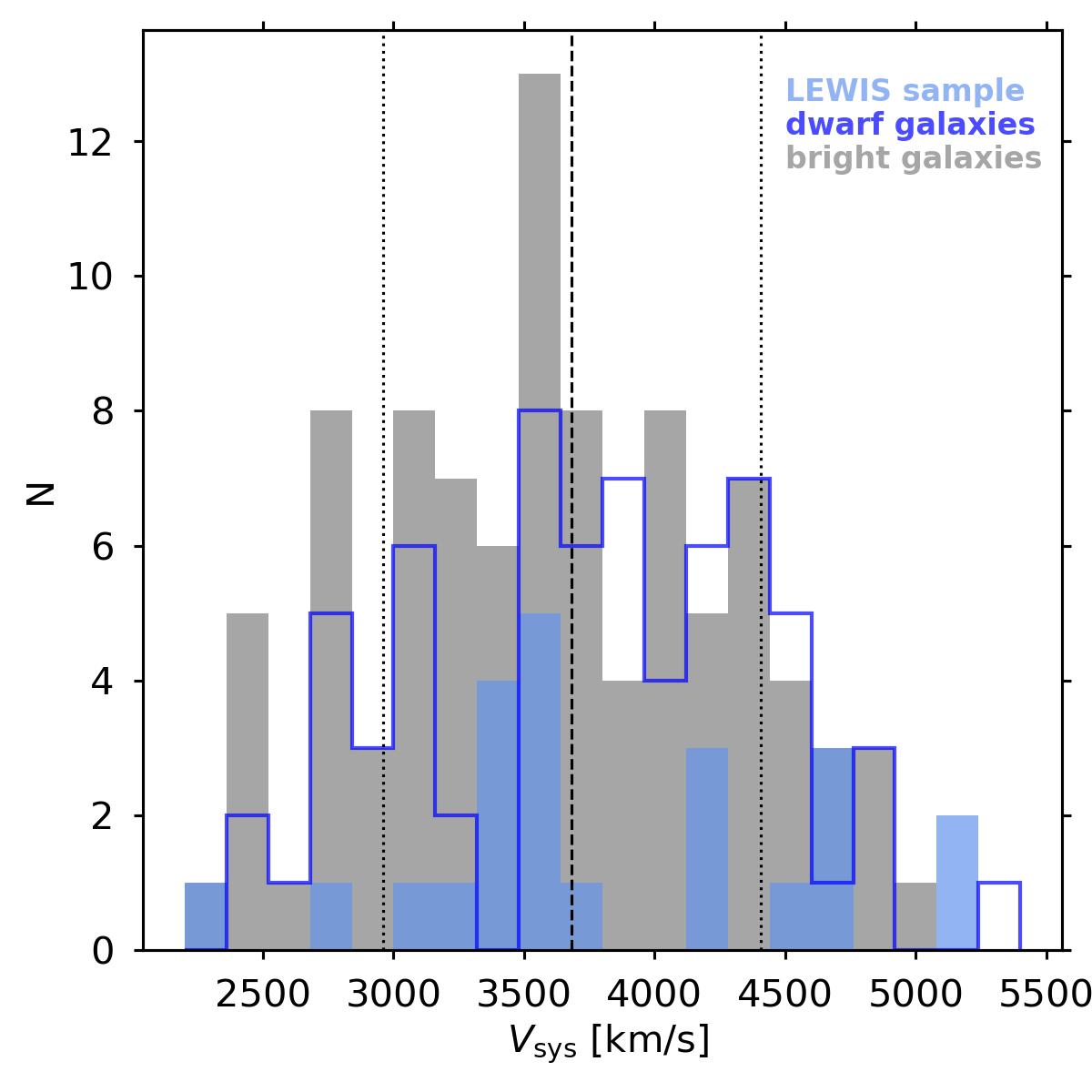}
    \includegraphics[scale=0.45]{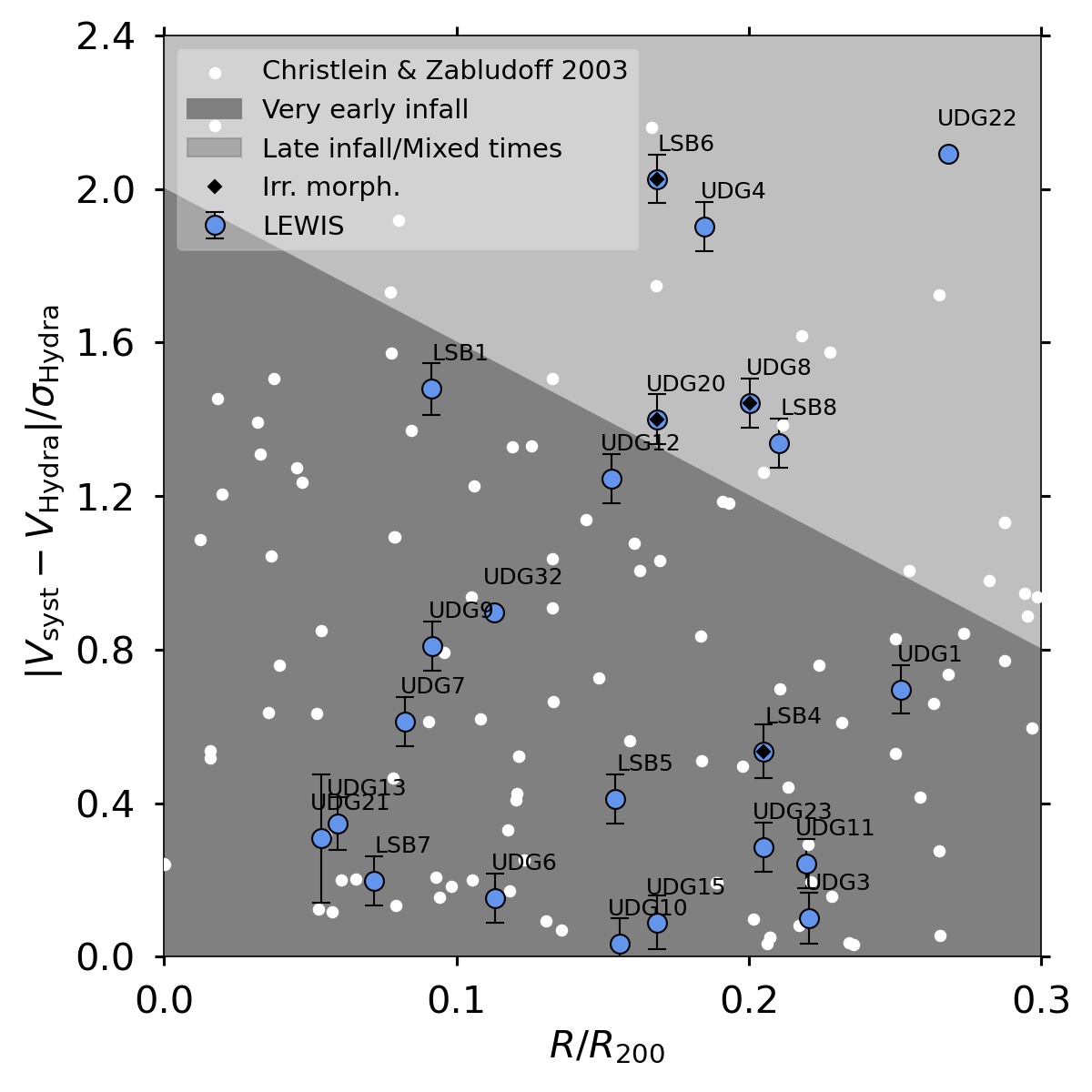}
    \caption{Left panel: Distributions of systemic velocity for galaxies in the LEWIS sample (light blue histogram), bright galaxies  ($m_B < 16$ mag, grey histogram) and dwarf galaxies (blue histogram) in Hydra I cluster. The vertical dashed line represents the average cluster velocity ($V_{\rm Hydra}$\,=\,$3683 \pm 46$ km s$^{-1}$, from \citealt{Christlein2003}), while the vertical dotted lines mark the cluster velocity dispersion ($\sigma_{\rm Hydra}=724\pm31$ km s$^{-1}$, from \citealt{Lima-dias2021}). 
    Right panel: Phase-space diagnostic diagram for galaxies in Hydra I. The systemic galaxy's velocity relative to the average velocity of the cluster normalised by the cluster velocity dispersion ($|V_{\rm syst}- V_{\rm Hydra}|/\sigma_{\rm Hydra}$) is shown as a function of the projected clustercentric distance normalised by the virial radius of the cluster ($R/R_{200}$). Galaxies from the LEWIS sample are marked with light blue circles, while galaxies from \cite{Christlein2003} are marked with white dots. The black diamonds mark the objects with an irregular morphology (UDG8, UDG20, LSB4, and LSB6) described in Section~\ref{sec:struct_morph}. The shaded regions represent the `very early' (dark grey) and `late infall/mixed times' (light grey) regions \citep{Rhee2017, Forbes2023}.}
    \label{fig:histogram_vsys_phase_space}
\end{figure*}

As already shown in \cite{Iodice2023}, the new data reduction reduces the sky background fluctuations and improves the quality of the data. We then proceeded with the validation of cluster membership of the targets by measuring the systemic velocity (\Vsys). To this aim, we masked all the possible light contamination in the FOV and derived the 1D stacked spectrum by co-adding all the spaxels within a circular aperture of radius \Reff. We fitted the spectrum using the Penalised Pixel-Fitting code algorithm \cite[{\sc pPXF}, ][]{Cappellari2004, Cappellari2017}. We chose the optical wavelength range between $4800-7000$~\AA , masking all the spectral regions contaminated by strong sky line residuals. At this stage, the choice of the wavelength range does not impact on the estimate of \Vsys\, and it is independent of the spectral resolution of MUSE. We adopted the E-MILES stellar library \citep{Vazdekis2012, Vazdekis2016}, which has a spectral resolution of FWHM~$= 2.51$ \AA \citep{FalconBarroso2011}, comparable with the average MUSE instrumental resolution. 

In the left panel of Figure~\ref{fig:histogram_vsys_phase_space} we show the systemic velocity distribution of the sample galaxies, compared with the systemic velocity distribution of bright ($m_B < 16$ mag) and dwarf galaxies in the Hydra I cluster \citep{Christlein2003}. To date, 15 galaxies of the LEWIS sample have \Vsys\ consistent with the average cluster velocity ($V_{\rm Hydra}$\,=\,$3683 \pm 46$ km s$^{-1}$, from \citealt{Christlein2003}), lying within 1$\sigma$  of the cluster velocity distribution ($\sigma_{\rm Hydra}=724\pm31$ km s$^{-1}$, from \citealt{Lima-dias2021}), whereas the remaining 8 objects have \Vsys\ within 2$\sigma_{\rm Hydra}$. 

In the right panel of Figure~\ref{fig:histogram_vsys_phase_space} we report an updated version of the phase-space diagnostic diagram presented in \cite{Forbes2023}. The systemic galaxy's velocity relative to the average velocity of the cluster normalised by the cluster velocity dispersion ($|V_{\rm syst}- V_{\rm Hydra}|/\sigma_{\rm Hydra}$) is shown as a function of the projected clustercentric distance normalised by the virial radius of the cluster ($R/R_{200}$, with $R_{200}\sim$~1.6\,Mpc, \citeauthor{Lima-dias2021} \citeyear{Lima-dias2021}). Based on location in the phase space, we classified galaxies in the LEWIS sample into two classes: `very early infall' (or ancient infallers, dark grey region) and `late infall/mixed times' (light grey region). 
According to \cite{Rhee2017}, the `very early infallers' are those galaxies which entered the cluster a long time ago, thus they are virialized with the cluster, have lower relative velocities, and are located at smaller virial distances. Galaxies classified as `late infallers/mixed times' have instead larger relative velocities, because they are approaching the cluster for the first time or have already passed the first pericenter. The majority of the galaxies in LEWIS are composed of `very early infall' (17 galaxies) that were in the cluster for at least $\sim$ 7 Gyr \citep{Rhee2017, Forbes2023}. The remaining galaxies, classified as `late infall/mixed times', might have already passed through the cluster and could be backsplash galaxies, or they have not yet entered the cluster (Table~\ref{tab0_LEWIS}).

%------------Tab1
\begin{table*}
\setlength{\tabcolsep}{10pt}
\renewcommand{\arraystretch}{1.23}
\caption{LEWIS sample: UDG and LSB galaxies in the Hydra I cluster.} 
\label{tab0_LEWIS}
\vspace{10pt}
\begin{center}
\begin{tabular}{lccccc}
\hline\hline
Object & $\mu_{0}$           &  \Reff  &    $\epsilon$   &    Type    &  Location in Hydra I   \\ 
       &   [mag arcsec$^{-2}$]  & [kpc]   &                 &            &                        \\
 (1)   &       (2)           & (3)     &   (4)           &     (5)    &       (6)              \\
\hline 
UDG1  &  24.2$\pm$0.1 & 1.75$\pm$0.12 &   0.28    & UDG                  &     Very early infall        \\
UDG3  &  25.2$\pm$0.2 & 1.88$\pm$0.12 &   0.21    & UDG                  &     Very early infall        \\
UDG4  &  24.9$\pm$0.1 & 2.64$\pm$0.12 &   0.30    & UDG                  &     Late infall/Mixed times  \\
UDG6  &  24.1$\pm$0.1 & 1.37$\pm$0.12 &   0.59    & UDG/cLSB             &     Very early infall        \\ 
UDG7  &  24.4$\pm$0.4 & 1.66$\pm$0.12 &   0.14    & UDG                  &     Very early infall        \\
UDG8  &  23.2$\pm$0.6 & 1.40$\pm$0.12 &   0.53    & Dwarf/Extended dwarf &     Late infall/Mixed times  \\
UDG9  &  24.2$\pm$0.2 & 3.46$\pm$0.12 &   0.41    & UDG                  &     Very early infall        \\
UDG10 &  24.3$\pm$0.3 & 2.29$\pm$0.10 &   0.14    & UDG                  &     Very early infall        \\
UDG11 &  24.4$\pm$0.1 & 1.66$\pm$0.12 &   0.30    & UDG                  &     Very early infall        \\
UDG12 &  25.1$\pm$0.2 & 1.64$\pm$0.12 &   0.29    & UDG                  &     Very early infall        \\
UDG13 &  24.2$\pm$0.2 & 1.60$\pm$0.20 &   0.15    & UDG/cLSB             &     Very early infall        \\
UDG15 &  25.0$\pm$0.3 & 1.51$\pm$0.15 &   0.27    & UDG/cLSB             &     Very early infall        \\
UDG16 &  25.9$\pm$0.2 & 1.75$\pm$0.12 &    -      & UDG                  &         -                    \\
UDG17 &  24.9$\pm$0.1 & 1.50$\pm$0.20 &    -      & UDG/cLSB             &         -                    \\
UDG20 &  26.0$\pm$0.3 & 1.97$\pm$0.12 &   0.45    & UDG                  &     Late infall/Mixed times  \\
UDG21 &  24.0$\pm$0.4 & 1.50$\pm$0.12 &   0.20    & Transition           &     Very early infall        \\
UDG22 &  25.3$\pm$0.2 & 3.60$\pm$0.12 &    -      & UDG                  &     Late infall/Mixed times  \\
UDG23 &  24.3$\pm$0.3 & 2.47$\pm$0.20 &   0.16    & UDG                  &     Very early infall        \\
UDG32 &  26.2$\pm$1.0 & 3.80$\pm$1.00 &    -      & UDG                  &     Very early infall        \\
\hline
\hline
LSB1  & 23.9$\pm$0.2 & 0.81$\pm$0.90  &   0.12    &  Transition            &   Very early infall         \\
LSB4  & 24.7$\pm$0.1 & 1.48$\pm$0.12  &   0.35    &  UDG/cLSB              &   Very early infall         \\
LSB5  & 23.9$\pm$0.1 & 1.42$\pm$0.12  &   0.49    & Dwarf/Extended dwarf   &   Very early infall         \\
LSB6  & 23.0$\pm$0.2 & 4.00$\pm$1.00  &   0.48    & Extended dwarf         &   Late infall/Mixed times   \\
LSB7  & 22.7$\pm$0.1 & 1.97$\pm$0.10  &   0.26    & Extended dwarf         &   Very early infall         \\
LSB8  & 23.2$\pm$0.2 & 1.51$\pm$0.20  &   0.17    & Dwarf/Extended dwarf   &   Late infall/Mixed times   \\
\hline
\end{tabular}
\end{center}
\tablefoot{ Column 1 reports the target name of the galaxy in the LEWIS sample. Columns 2 and 3 report the central surface brightness $\mu_0$ and effective radius \Reff in the $g$ band from \citet{Iodice2020b} and \citet{LaMarca2022a}, respectively. Column 4 reports the average ellipticity of the galaxy measured from the isophotal analysis performed on the MUSE reconstructed image. Column 5 reports the morphological classification of the galaxy according to the structural parameters and their errorbars. Column 6 reports the location of the galaxy within the Hydra I cluster, according to the infall diagnostic diagram \citep{Forbes2023}. Missing values correspond to the galaxies with on-going observations (UDG16, UDG17, and UDG22). UDG32 is a special case and will be presented in detail in a dedicated paper (Hartke et al. subm.).}
\end{table*}

\section{Structural and morphological classification}
\label{sec:struct_morph}

In this section, we describe the structural classification of the LEWIS galaxy sample and 
then provide a detailed morphological classification.   

We classified galaxies in the LEWIS sample into four classes according to the value of the central surface brightness $\mu_{0,g}$ and effective radius \Reff\ and their associated errors.
According to the \cite{vanDokkum2015} definition, UDGs have \Reff\,$\geq 1.5$~kpc and $\mu_{0,g} \geq 24$~mag arcsec$^{-2}$. Classical dwarfs are brighter and more compact (with \Reff\,$<1.5$ kpc and $\mu_{0,g} < 24$~mag arcsec$^{-2}$). 
Compact LSB (cLSB) galaxies are fainter than classical dwarfs but have smaller effective radii than UDGs (with \Reff\,$<1.5$ kpc and $\mu_{0,g} \geq 24$~mag arcsec$^{-2}$). Extended dwarf galaxies have comparable luminosity to dwarf galaxies, but they are more extended (with \Reff\,$\geq 1.5$ kpc and $\mu_{0,g} < 24$~mag arcsec$^{-2}$). 
We report a double classification for those galaxies that are located in between two classes taking into account the errorbars, and we flagged as `transition' those that are consistent within the uncertainties with all the classes in the \Reff\ - $\mu_0$ plane. In Table~\ref{tab0_LEWIS} we report the structural and morphological properties of the LEWIS sample derived by \cite{Iodice2020b} and \cite{LaMarca2022a}, and recently presented in the LEWIS Paper~I \citep{Iodice2023}. 

We performed an isophotal analysis on the reconstructed MUSE image of each target in the LEWIS sample. This allows us to recover the mean geometric parameters of the galaxy and investigate their morphology. 
To this aim, we masked all the foreground and background objects, spurious sources, GC candidates, and bad pixels in the reconstructed image. 
We fitted galaxy isophotes using the {\sc ellipse} task in {\sc photutils} Python software \citep{photutils}. First, we allowed the centre, ellipticity ($\epsilon$), and position angle (PA) of the fitting ellipses to vary. Then, we fitted again the galaxy isophotes fixing the centre coordinates with the median values of the $x$ and $y$ coordinates of the inner ellipses and we recovered the mean PA and $\epsilon$ of the galaxy. The values are consistent with the results derived by \cite{Iodice2020a} and \cite{LaMarca2022b} from VST data.

We found that the majority of the galaxies in the LEWIS sample have a regular morphology, characterised by a nearly circular or slightly elongated shape ($\epsilon \sim 0.1-0.3$) and a light distribution characterised by a constant core and a decrease at larger radii \citep[see also][]{Iodice2020b, LaMarca2022b, LaMarca2022a}. We report here the only few cases which present an irregular morphology and/or peculiar features:

UDG6 is located between the northern overdensity \citep{LaMarca2022a} and cluster centre and is characterised by an elongated shape ($\epsilon\sim0.6$). By inspecting the datacube across the reconstructed image, we found several clumpy regions characterised by strong emission lines (\ion{H}{$\beta$}, [\ion{O}{III}], \ion{H}{$\alpha$}, [\ion{S}{II}]). 
Among the various targets in LEWIS, it is the UDG with the bluest colour 
\citep[$g-r=0.32 \pm 0.2$ mag, see][]{Iodice2020b}. 
Due to this peculiarity, UDG6 will be studied in detail in a dedicated paper (Rossi et al., in prep.).

UDG8 is located on the northern side of the cluster and presents an off-centred elongated structure in the inner region (Figure~\ref{udg8_2D}). The isophotal analysis revealed a tilt of the isophotes in the inner region, characterised by a local maximum in the ellipticity profile ($\epsilon$\,>\,0.5) and a constant position angle ($\Delta$PA<10$^\circ$) in the same region, which can be interpreted as a bar signature.

UDG20 is located on the eastern side of the cluster and presents an irregular morphology (Fig~\ref{udg20_2D}). It is characterised by two luminosity peaks, one on the North-East side and the other on the South-West side, symmetric with respect to the galaxy's geometric centre. We inspected these features individually by extracting a stacked spectrum of the two regions and we found that the spectrum of the North-East region is characterised by a moderate \ion{H$\alpha$}{} absorption line, while the spectrum of the South-West region presents a weak \ion{H$\beta$}{} absorption line and a prominent \ion{H$\alpha$}{} emission line at the same redshift of the galaxy ($V\sim4690$\,km s$^{-1}$). These findings suggest that these overdensities are bound to the galaxy and should be considered as part of it.

LSB4 is located on the northern side of the cluster and presents a disturbed morphology (Fig~\ref{lsb4_2D}). It has an elongated arc-like structure in the northern outskirts of the main body of the galaxy. This suggests that this galaxy is suffering from an external interaction. According to its structural parameters, this galaxy is located between the LSB and UDG regimes in the \Reff$~-~\mu_0$ plane.

LSB6 is located on the western side of the cluster centre and presents an elongated shape ($\epsilon\sim0.5$, Fig~\ref{lsb6_2D}). The isophotes in the inner regions are characterised by a `boxy' shape, while in the outskirts, they are more `disky'. The isophotal analysis revealed a twist of the isophotes, characterised by a change in the orientation of $\Delta{\rm PA} \sim10^\circ$ from the inner to outer regions.

\begin{figure*}
    \centering
    \includegraphics[scale=0.28]{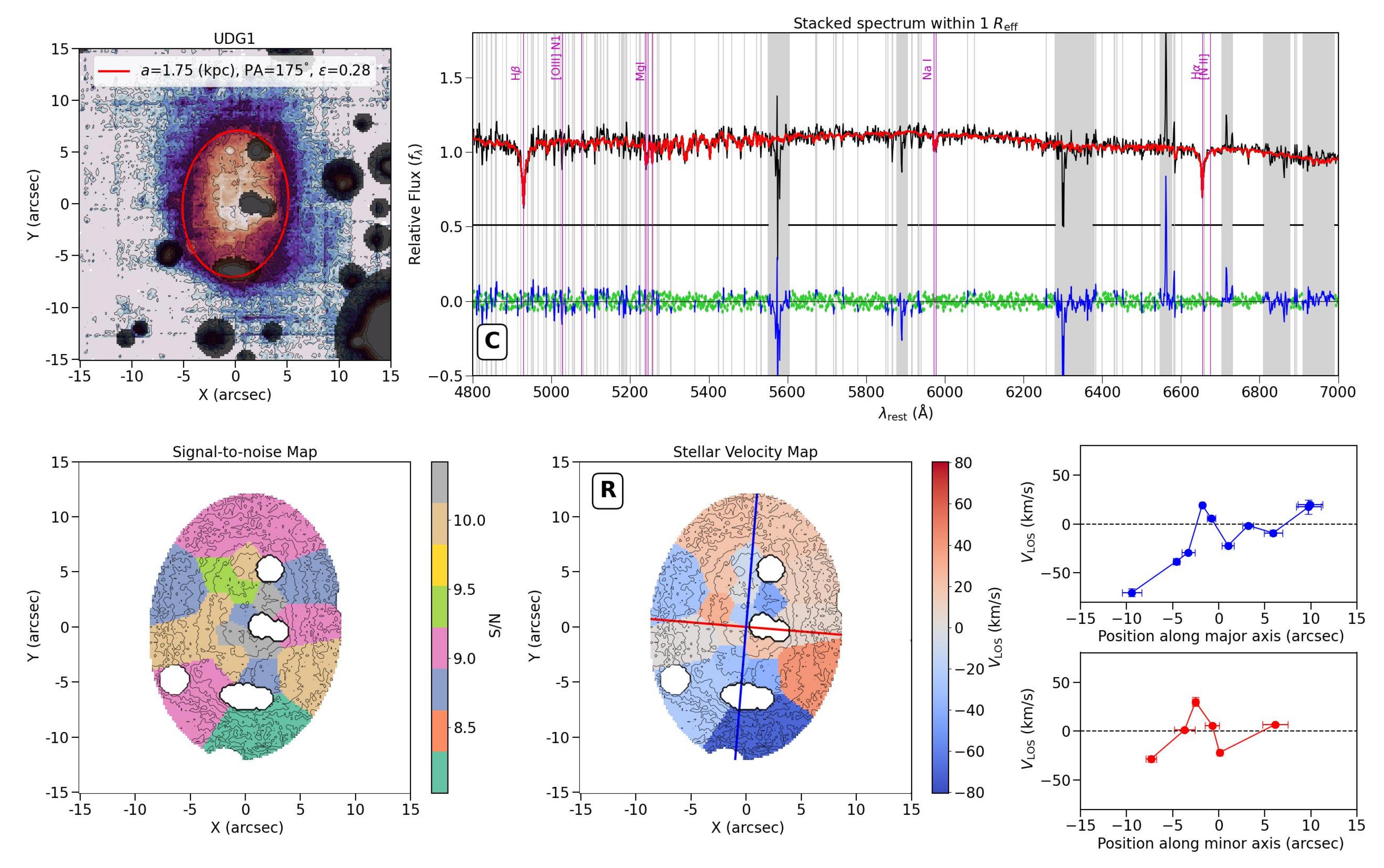}
    \caption{Stacked (1D) and spatially-resolved (2D) stellar kinematics of UDG1. Top left panel: MUSE reconstructed image of UDG1. The red ellipse represents the elliptical region used to extract the stacked spectrum within 1\Reff, while the grey circles are the masked image regions. The top label reports the values of the semi-major axis ($a$), position angle (PA) and ellipticity ($\epsilon$) of the ellipse. Top right panel: MUSE 1\Reff\ stacked spectrum (black solid line) for UDG1. The bottom label reports the fit type (Table~\ref{tab1_stellar_kinematics}). The main absorption features are marked with magenta lines and labels. The red solid line represents the best-fitting spectrum obtained with {\sc pPXF}. Green points are the residuals between the observed and its best-fitting spectrum. The grey areas are the masked spectral regions excluded from the fit. Residual points excluded from the fit are marked in blue. Bottom left panel: Voronoi binned map of the S/N. Bottom middle panel: Stellar velocity map subtracted from systemic velocity \Vsys . The top label reports the type of rotation (Table~\ref{tab1_stellar_kinematics}). The blue and red lines mark the photometric major and minor axes of the galaxy, respectively. Bottom right panel: Velocity profiles extracted along the major (top) and minor (bottom) axes and subtracted from \Vsys.}
    \label{fig:UDG1_stellar_kinematics}
\end{figure*}

UDG32 is located in the filaments of the spiral galaxy NGC~3314A. It is one of the most diffuse and faintest UDGs observed in the Hydra I cluster ($\mu_{0,g} = 26 \pm 1$ mag arcsec$^{-2}$). Its nature and location suggest that this galaxy might have originated from ram-pressure stripped (RPS) material from NGC\,3314A \citep{Iodice2021}. The MUSE data from LEWIS will allow us to understand whether this galaxy is at the same location in the cluster phase space as the stellar filaments. A detailed analysis of this interesting object will be presented in a forthcoming paper (Hartke et al. subm.).

 The majority of the galaxies with irregular morphology (UDG8, UDG20, and LSB6) are classified as `late infall/mixed times' (black diamonds, right panel of Figure~\ref{fig:histogram_vsys_phase_space}). The presence of an isophotal twist or disrupted structures is a hint of the possible interaction between the galaxy and cluster environment.

\section{Stellar kinematics}
\label{sec:stellar_kin}

One of the science goals of the LEWIS project is to constrain the stellar kinematics 
and dynamical structure of the UDGs and LSB galaxies in the Hydra I cluster. 
The effective velocity dispersion \seff\ requires a spectrum with a sufficiently high 
signal-to-noise (S/N) to be measured. Paper I showed that the most recent version of
{\sc pPXF} can recover a reliable measurement of the velocity dispersion for LSB 
galaxies, even if the value is below the actual spectral resolution of MUSE as long as the
S/N of the spectrum is high enough. In particular, using Monte Carlo simulations, they 
demonstrated that the true value of \seff \ could be overestimated if spectra have S/N~<~10, 
whereas for S/N~>~15, the fitted parameters are unbiased. For this reason, we estimated the 
effective velocity dispersion \seff\ from the 1D\, \Reff\ stacked spectrum, which allows us 
to obtain a high S/N. 
The quality of the spectra, i.e. the S/N, together with a precise
estimate of spatial and spectral variation of the instrumental resolution are essential to accurately measure the stellar
velocity dispersion \citep[see also][]{Chilingarian2007}. \cite{Cappellari2017} demonstrated that with high-quality 
spectra (S/N\,>\,3000 per spaxel), it is possible to recover an accurate measurement of the velocity dispersion even below the
instrumental resolution. Recent papers demonstrated that also in the low S/N regime, values for \slos\, can 
be accurately derived \citep{Eftekhari2022, Iodice2023}. 
Conversely, deriving the line-of-sight stellar velocity \vlos\ 
is less affected by the quality of the spectra, thus it is possible to relax the S/N threshold of the spectra for this measurement. 
\subsection{Line-of-sight velocity and effective velocity dispersion}
\label{sec:1D_stellar_kin}

\begin{figure*}
    \centering
    \includegraphics[scale=0.42]{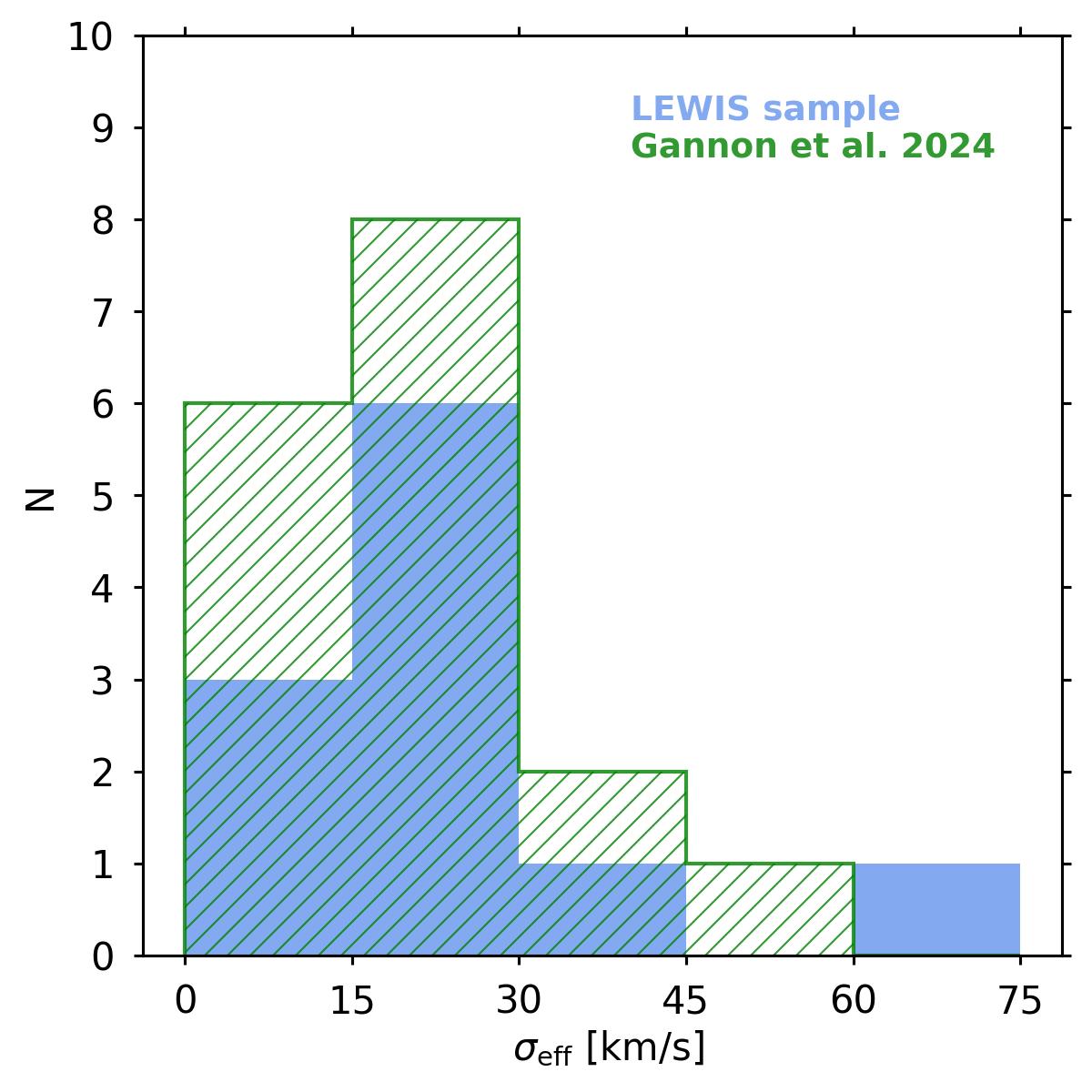}
    \includegraphics[scale=0.42]{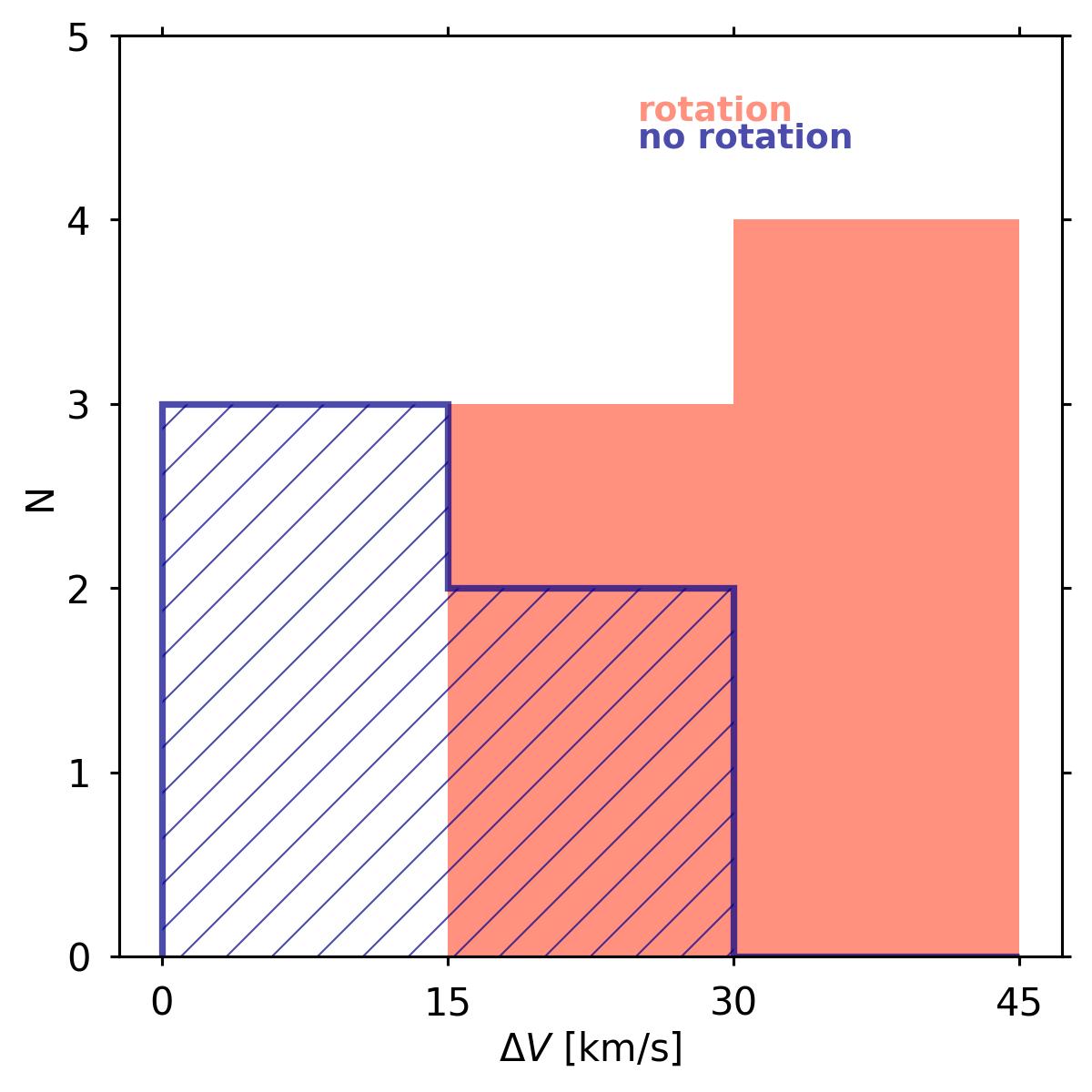}
    \caption{Left panel: Distributions of the velocity dispersion \seff\ of UDGs from literature (green histogram) and LEWIS sample. The light blue histogram corresponds to the distribution of galaxies with constrained fit (C fit type). Right panel: Distribution of the semi-amplitude of rotation curve ($\Delta V$) of UDGs in the LEWIS sample. The pink and blue histograms correspond to the UDGs with a mild rotation along any axis (R and IR type) and those with no rotation (NR type), respectively.}
    \label{fig:distribution}
\end{figure*}

We measured the stacked stellar kinematics of each galaxy in the LEWIS sample.
We masked all the possible light contamination in the FOV and defined an elliptical aperture with the centre coinciding with the galaxy geometric centre ($x_{\rm c}$,$y_{\rm c}$), PA and $\epsilon$ derived from the isophotal analysis and semi-major axis equal to \Reff . We obtained the 1D stacked spectrum by co-adding all the spaxels within this elliptical aperture to derive both \vlos\ and \seff\ (Figure~\ref{fig:UDG1_stellar_kinematics}, top left). 

We started masking out all the spectral regions affected by the contamination from residuals of sky line subtraction, and we estimated the S/N of the spectrum in the spectral range used to derive the stellar kinematics (Table~\ref{tab1_stellar_kinematics}). We computed the dimensionless average S/N in the fitting wavelength range after removing the pixels contaminated by residuals or noisy regions, and after additional filtering of the pixels through a sigma-clipping technique. The reported quantity is an average estimate, the S/N would be even higher when restricting the wavelength range around a prominent absorption line. We fitted the spectrum with the {\sc pPXF} algorithm, choosing the optical wavelength range between $4800-7000$~\AA\ (Figure~\ref{fig:UDG1_stellar_kinematics}, top right). In a few cases, when the Calcium absorption line triplet (\ion{CaT}{}) is clearly visible i.e. clean from contamination of sky line residuals, we extended the fit up to $8900$~\AA\,. 
Since this is the range where MUSE reaches the best spectral resolution (FWHM$\sim$2.6\,\AA, see Fig.~\ref{app:LSF}), the measured values of \seff\, could be more accurate. As done in \cite{Iodice2023} (see Table\,2), we have performed tests on different fitting spectral ranges and we found consistent values of \seff\, within errorbars.
We adopted the E-MILES stellar library, which has a spectral resolution similar to the average instrumental one and covers a large range in age (from 30 Myr to 14 Gyr) and total metallicity ($-2.27 \leq$ [$M$/H] $\leq 0.4$ dex). We adopted the MUSE LSF that we measured from the sky emission lines (see Appendix~\ref{app:LSF} for details). As addressed in Paper I, we used additive and multiplicative Legendre polynomials to correct discrepancies in the flux calibration and reduce imperfections in the spectral calibration affecting the continuum shape. We performed a preliminary fit fixing the grade of additive (ADEG) and multiplicative (MDEG) polynomials to 10.

We generated 10000 perturbations of the original spectrum by randomly removing and replacing the 30\% of masked spectral regions and by randomly adding 20\% of the Poissonian noise without significantly varying the S/N of the perturbed spectra (on average $\Delta$S/N$\sim$1.5). We fitted the perturbed spectra keeping fixed the \vlos\ to the best-fitting value, and adopting ADEG and MDEG Legendre polynomials as free parameters. We restricted their possible values of polynomial degrees in a range of $[0, 12]$ with a step of 2, also including the case of ADEG=-1 which corresponds to not using additive polynomials. This test was made to explore different combinations for the Legendre polynomials and to find the best combination (or a confident range of possible values) for the degrees of additive and multiplicative polynomials that ensure a reliable estimate for \vlos\ and \seff . On average, the adopted interval of degree for the additive and multiplicative polynomials ranges between 6 and 10, although it is wider for highest S/N sources.

We finally fitted the 1D stacked spectrum with {\sc pPXF} by fixing the grades of additive and multiplicative polynomials to the values we obtained from the previous analysis. We estimated the errors on the fitted parameters (\vlos , \seff) by generating 1000 perturbed spectra using the same approach explained before, and fitting both  \vlos\ and \seff\ . We estimated the associated errors by calculating the 16th and 84th percentiles of the distributions of the fitted parameters. Alternatively, uncertainties on the fitted kinematic parameters could be estimated by using the shape of the stellar template and the information of the S/N \citep[see][for details]{Chilingarian2020}.

In Table~\ref{tab1_stellar_kinematics} we reported for each sample galaxy the S/N of the spectrum, the wavelength range used to derive the best fit, \vlos\ and \seff\ values.

The reliability of the fitted parameters, especially of \seff, depends on the quality of the data. Therefore we decided to split the sample galaxies into three types, according to the value of the average S/N of the spectrum and accuracy of the final {\sc pPXF} fit: 1) galaxies with constrained fit (C), when S/N\,>\,15 and thus the estimate of \seff\ is reliable; 2) with intermediate-quality fit (I), when 10\,$\leq$\, S/N\,$\leq$\,15 and value of \seff\ could be slightly overestimated; 3) with unconstrained fit (U) when the S/N\,<\,10 and estimate of \seff\ is heavily overestimated. The constrained-fit galaxies are bright objects: the S/N of all perturbed spectra are above the trustful threshold and a broad range of ADEG and MDEG guarantees a reliable estimate of \seff . The intermediate-fit galaxies show spectra with prominent sky line residuals in the proximity of the \ion{H$\alpha$}{} or \ion{H$\beta$}{} absorption lines that may drive the fitting algorithm to a wrong solution. In addition, the distribution of the S/N of the perturbed spectra is broader and the final estimate of \seff\ might be slightly overestimated since the spectra have S/N\,<\,10. The unconstrained-fit galaxies are faint and some present a disturbed morphology. Since the S/N of the extracted spectrum and the vast majority of the perturbed spectra have S/N\,<\,10, only the value \vlos\ is actually reliable. 

Figure~\ref{fig:distribution} (left panel) shows the \seff\ distribution of the galaxies of the LEWIS sample compared to that of the UDGs studied in literature through spectroscopy \citep[see references in][]{Gannon2024}. The light blue histogram corresponds to the galaxies with constrained fit (C in Table~\ref{tab1_stellar_kinematics}). The peak of the LEWIS sample is at \seff$\sim20-30$\,km s$^{-1}$, similar to that of the distribution of the literature galaxies. The majority of the values of \seff\, are below the actual MUSE resolution. However, all the obtained values are measured from sources with high-quality spectra (S/N\,>\,15) with constrained fit, thus the values are reliable \citep[see also][]{Iodice2023}.

\begin{table*}
\setlength{\tabcolsep}{5pt}
\renewcommand{\arraystretch}{1.22}
\caption{1D and 2D stellar kinematics and dynamical masses of UDGs and LSBs in the Hydra I cluster.} 
\label{tab1_stellar_kinematics}
\vspace{15pt}
\begin{center}
\begin{tabular}{lccccccccccc}
\hline\hline
Object & $\lambda$ range  &   S/N   &   Fit Type  &  \Vsys  & \seff    & Type  &  $V_{\rm rms}$   &\Mdyn  &   $\log_{10}{(M_{\rm dyn} / L_{V, {\rm eff}})}$ &$\Delta V$ & \\ 
       &   [\AA]       &         &             & [km s$^{-1}$]  &  [km s$^{-1}$]    &  & [km s$^{-1}$] & [$10^{9}$~M$_\odot]$ & [M$_\odot$/L$_\odot$]  &   [km s$^{-1}$]    \\
 (1)   &  (2)             &   (3)   &   (4)       &   (5)   &  (6)    &   (7)                &   (8)    &     (9)   &   (10)   & (11)      \\
\hline \vspace{-7pt}\\   
UDG1  & 4800-7000 & 32 & C & 4187$\pm$4   & 23$\pm$6           & R  &    40$\pm$18    &      2.28$\pm$2.06        &  1.62$\pm$0.39            &  44$\pm$8  \\
UDG3  & 4800-7000 & 15 & I & 3611$\pm$14  &      -             & U  &     -           &             -             &         -                 &       -     \\
UDG4  & 4800-8800 & 27 & C & 2306$\pm$3   & 13$\pm$6           & NR &     -           &      0.36$\pm$0.33        &  0.70$\pm$0.40            &  18$\pm$6   \\
UDG7  & 4800-7000 & 17 & C & 4126$\pm$5   & 61$\pm$9$^{(\ast)}$& R  &     -           &  4.91$\pm$1.51$^{(\ast)}$ &  2.69$\pm$0.13$^{(\ast)}$ &  41$\pm$21   \\
UDG8  & 4800-8800 & 37 & C & 4727$\pm$2   & 24$\pm$4           & IR &     24$\pm$3    &  0.51$\pm$0.14            &  1.20$\pm$0.12ì           &  20$\pm$2   \\
UDG9  & 4800-7000 & 19 & C & 4269$\pm$4   & 21$\pm$7           & NR &      -          &  1.04$\pm$0.70            &  1.44$\pm$0.29            &   8$\pm$5   \\
UDG10 & 4800-7000 & 15 & I & 3660$\pm$7   &      -             & NR &      -          &             -             &         -                 &  18$\pm$6   \\
UDG11 & 4800-9000 & 16 & C & 3507$\pm$3   & 20$\pm$8           & NR &      -          &  0.59$\pm$0.48            &  1.33$\pm$0.35            &   7$\pm$13  \\
UDG12 & 4800-8900 & 24 & C & 4585$\pm$3   & 38$\pm$9$^{(\ast)}$& IR &      -          &  1.86$\pm$0.89$^{(\ast)}$ &  2.08$\pm$0.21$^{(\ast)}$ &  41$\pm$21   \\
UDG13 & 4800-7000 &  9 & U & 3425$\pm$11  &      -             & -  &      -          &             -             &        -                  &       -     \\
UDG15 & 4800-7000 &  8 & U & 3625$\pm$12  &      -             & -  &      -          &             -             &        -                  &       -     \\
UDG20 & 4800-7000 & 11 & I & 4697$\pm$6   &      -             & U  &      -          &             -             &        -                  &       -     \\
UDG21 & 4800-7000 & 10 & I & 3460$\pm$112 &      -             & U  &      -          &             -             &        -                  &       -     \\
UDG22 & 4800-7000 &  - & - &    5198      &      -             & -  &      -          &             -             &        -                  &       -     \\
UDG23 & 4800-7000 & 14 & I & 3477$\pm$8   &      -             & U  &      -          &             -             &        -                  &       -     \\
\hline  \hline        
LSB1  & 4800-7000 & 11 & I &  4754$\pm$16 &      -             & U  &      -          &             -             &        -                  &       -     \\
LSB4  & 4800-7000 & 11 & I &  3296$\pm$22 &      -             & U  &      -          &             -             &        -                  &       -     \\
LSB5  & 4800-8900 & 26 & C &  3386$\pm$2  & 25$\pm$7           & NR &      -          &   0.58$\pm$0.33           &   1.75$\pm$0.25           &  3 $\pm$4   \\
LSB6  & 4800-8900 & 35 & C &  5150$\pm$2  & 28$\pm$5           & R  &     30$\pm$13   &   2.41$\pm$2.25           &   1.68$\pm$0.41           &  32$\pm$5   \\
LSB7  & 4800-8900 & 47 & C &  3541$\pm$1  & 14$\pm$5           & IR &     21$\pm$8    &   0.69$\pm$0.52           &   1.06$\pm$0.33           &  25$\pm$2   \\
LSB8  & 4800-8900 & 38 & C &  2715$\pm$2  & 12$\pm$6           & IR &     19$\pm$7    &   0.46$\pm$0.35           &   1.03$\pm$0.33           &  20$\pm$5   \\
\hline
\end{tabular}
\end{center}
\tablefoot{ Column 1 reports the target name of the galaxy in the LEWIS sample. Column 2 reports the wavelength range used to derive the stellar kinematics. Column 3 reports the average S/N of the spectrum. Column 4 reports the class of the fit: constrained (C), intermediate (I), and unconstrained (U). Columns 5 and 6 report the systemic velocity (\Vsys) and velocity dispersion derived inside 1\Reff\ (\seff), respectively. Column 7 reports the class of the kinematic feature: rotation along the major photometric axis (R), rotation along an intermediate axis (IR), no rotation (NR), and unconstrained rotation (U). Column 8 reports the luminosity-weighted second velocity moment ($V_{\rm rms}$) derived from the 2D stellar kinematics and used to compute the \Mdyn\ in R and IR type galaxies. Columns 9 and 10 report the dynamical mass (\Mdyn) and logarithm of dynamical mass-to-light ratio (\Mdyn/$L_{V, {\rm eff}}$), respectively. For non-rotating galaxies, the value for \Mdyn\, is calculated from \seff, whereas for rotating galaxies, the value of \Mdyn\, is calculated from $V_{\rm rms}$\,. Column 11 reports the semi-amplitude of the rotation curve $\Delta V$. $^{(\ast)}$: values of \seff, \Mdyn, and \Mdyn/$L_{V, {\rm eff}}$ could be overestimated (see text for details).}
\end{table*}
\subsection{Line-of-sight velocity field}
\label{sec:2d_Stellar_kin}

We measured the spatially-resolved stellar kinematics of each galaxy in the LEWIS sample. We masked all the foreground/background sources in the FOV and we spatially binned datacube spaxels using the adaptive algorithm of \cite{Cappellari2003} based on Voronoi tessellation to obtain a specific S/N per bin. Since we aim to recover only the stellar velocity field, we can relax the constraint on the threshold for the target S/N. Therefore, for each sample galaxy, we tested and adopted a different target S/N. We avoided a threshold S/N\,<\,5, since the absorption lines were not easily visible in the stacked spectrum. In the LEWIS sample, the S/N of the stacked spectra in the Voronoi bin lies in the range S/N\,=\,6-11 on average.

For each bin, we extracted the stacked spectrum and we followed the same recipe adopted for the 1D case. We fitted the spectra with {\sc pPXF} algorithm, choosing the optical spectral range (4800-7000\,\AA) or the whole spectral range (4800-8900\,\AA) according to the visibility of the CaT absorption line triplet. We used the E-MILES stellar library, the LSF derived from our MUSE data, and a combination of multiplicative and additive Legendre polynomials with the same grades used to derive the stacked stellar kinematics, as described in the previous Section. We finally reconstructed the 2D maps of the S/N and \vlos\ for each galaxy in the LEWIS sample (Figure~\ref{fig:UDG1_stellar_kinematics}, bottom left and centre). We adopted as systemic velocity the value obtained from the best fit of the stacked spectrum within 1\Reff. After subtracting \Vsys\ from the stellar velocity field, we derived velocity profiles extracting the values of \vlos\ in bins along apertures parallel to the photometric major and minor axes of the galaxy (Figure~\ref{fig:UDG1_stellar_kinematics}, bottom right).

Unfortunately, we were not able to derive the spatially-resolved stellar kinematics for all the targets. Two galaxies were extremely faint and the application of the Voronoi binning returned a single bin (i.e. UDG13 and UDG15). In some other cases, the application of the Voronoi algorithm returns only two bins which split the galaxy into two nearly symmetric halves (i.e. UDG3, UDG20, UDG21, UDG23, LSB1 and LSB4). 
For these galaxies, we do not have enough data points to properly extract a velocity profile, so we flagged these targets with `U' (unconstrained) in Table~\ref{tab1_stellar_kinematics}. 
The 2D velocity maps derived for all the LEWIS targets, including the unconstrained cases, are shown in Appendix~\ref{app:velocity_maps}. As already pointed out in Section~\ref{sec:struct_morph}, UDG6 and UDG32 are two special cases that will be analysed and presented in detail in dedicated papers (Rossi et al., in prep., Hartke et al., subm.)

\section{Results}
\label{sec:results}

In this section, based on the stellar kinematics derived from the MUSE data, we present the stellar kinematics
properties and constraints on the dynamical mass and DM content of the LEWIS sample and we compare the results with data published in the literature.

\subsection{Stellar velocity profiles}\label{sec:V_gradient}

\begin{figure*}[!h]
    \centering
    \includegraphics[scale=0.44]{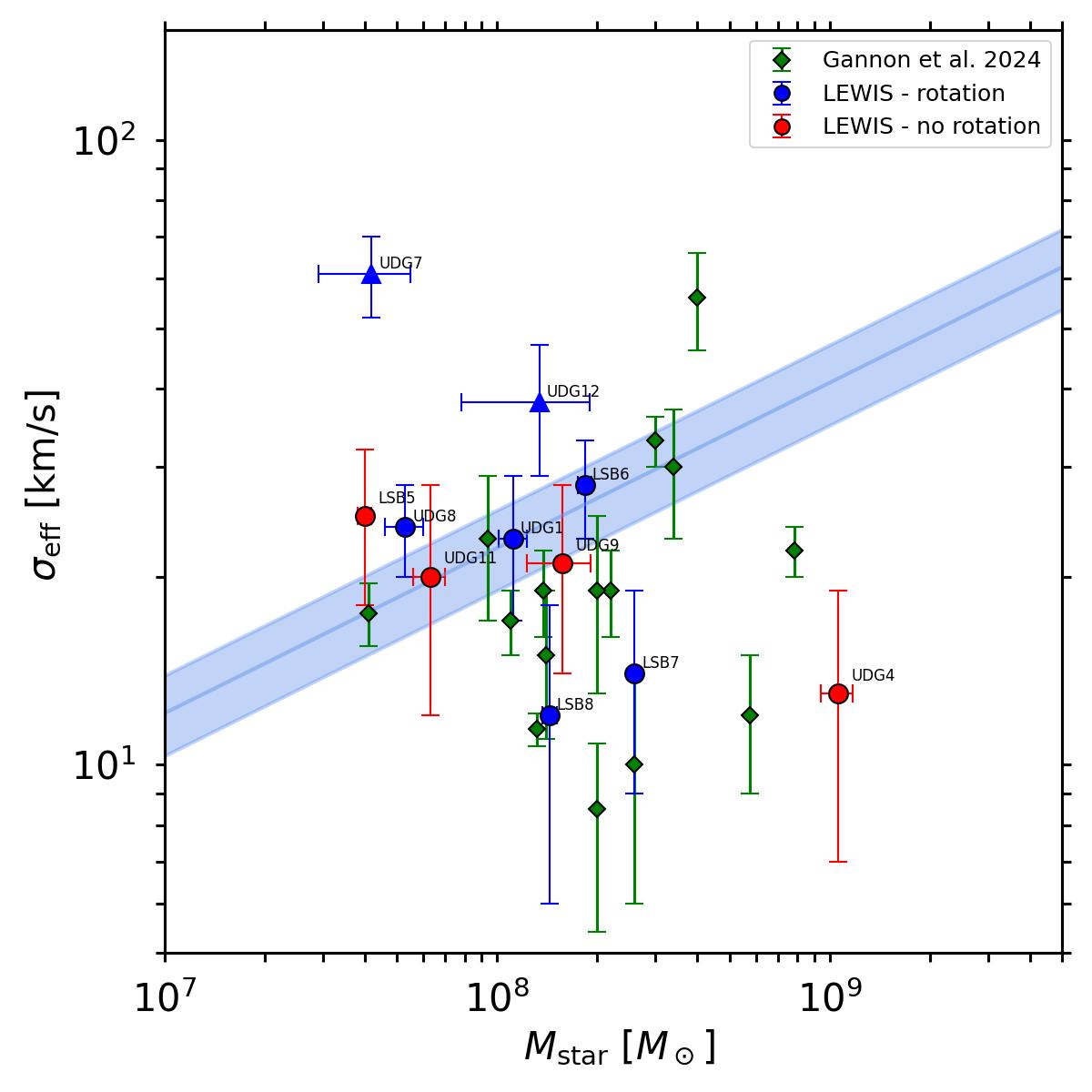}
    \includegraphics[scale=0.44]{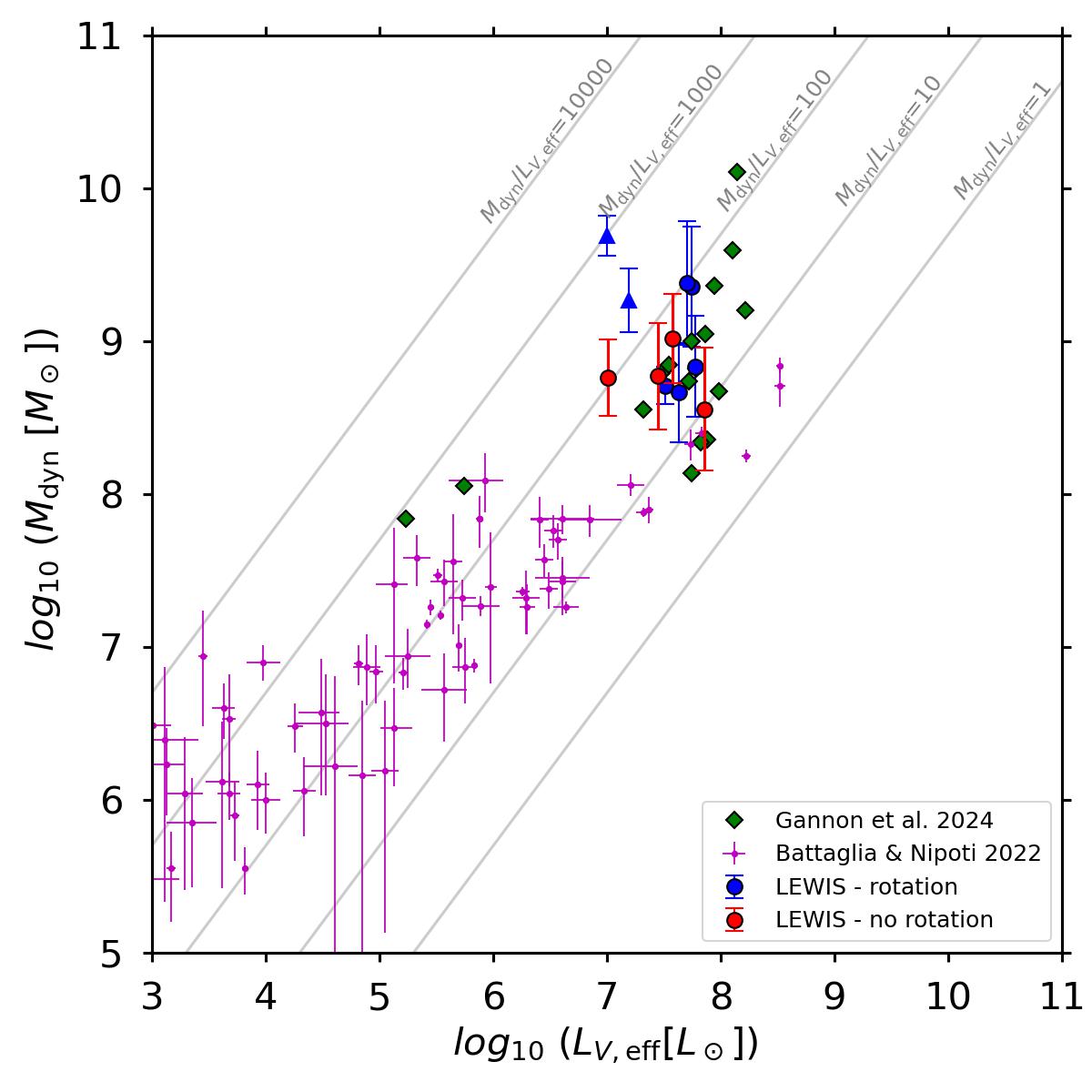}
    \caption{Left panel: Velocity dispersion \seff \ as a function of stellar masses of UDGs from literature (green diamonds) and LEWIS sample. The blue and red circles represent rotating and non-rotating LEWIS galaxies, respectively. The shaded light blue region represents the luminosity Faber-Jackson relation, extrapolated to the low-mass regime \citep[$L\sim\sigma^{2.2}$, ][]{Kourkchi2012}. The luminosities are converted into stellar masses using the \cite{Into2013} colour\,-\,$M/L$ relation. UDG7 and UDG12 are marked with triangles since their values of \seff\, could be overestimated due to their non-negligible rotation. Right panel: Dynamical mass ($M_{\rm dyn}$) as a function of $V$-band total luminosity ($L_{V, {\rm eff}}$) computed within 1\Reff\ of UDGs from literature (green diamonds) and LEWIS sample. The points are colour-coded as in the left panel. The solid grey lines mark the loci in the plane where \Mdyn/$L_{V, {\rm eff}}$ is constant and equal to 1, 10, 100, 1000, and 10000, respectively. The magenta points represent the dwarf galaxies in the Local Group analysed by \cite{Battaglia2022}.}
    \label{fig:FJ_BN}
\end{figure*}

We evaluated the stellar kinematics of the galaxy sample. We computed the semi-amplitude of the rotation curve ($\Delta V$) as the semi-difference of the first and last point of the velocity profile along the kinematic major axis. We found that galaxies in the LEWIS sample span a wide range of kinematic features. Thus we decided to classify them into three classes: 

\begin{itemize}

\item Rotation along the major photometric axis (R): some galaxies show a mild rotation along the major photometric axis (UDG1, UDG7, and LSB6) with values $\Delta V \sim 25 - 40$ km s$^{-1}$. We found that, on average, the central regions of the velocity profile appear to have mild variations in terms of velocity, values can also be consistent with \Vsys. In the outskirts, instead, the velocity reaches opposite values on the two sides, and the values are not compatible with \Vsys\, even considering their uncertainties. These galaxies are flagged with `R' in Table~\ref{tab1_stellar_kinematics}. 

\item Rotation along an intermediate axis (IR): a few UDGs in the LEWIS sample show a mild rotation along a different axis than the major photometric one (UDG8, UDG12, LSB7, and LSB8). In these cases, we plotted in an additional panel the velocity profile along this direction (e.g. green line in Figure~\ref{lsb7_2D}, for example), and we reported the corresponding value of $\Delta V$. Some of these objects have a roundish shape ($\epsilon<0.3$), thus it might be possible that this simple morphology hides a more complex internal dynamical structure. These galaxies are flagged with `IR' in Table~\ref{tab1_stellar_kinematics}. 

\item No rotation (NR): other UDGs do not show clear evidence of rotation along any direction (UDG4, UDG9, UDG11, and LSB5). The velocity profiles along the major, minor, or intermediate axis do not show a clear velocity difference between the two sides. In addition, some velocity values have large uncertainties which are consistent with \Vsys . These galaxies are flagged with `NR' in Table~\ref{tab1_stellar_kinematics}.

\end{itemize}

The values for $\Delta V$ are measured on the sky-plane and are reported in Table~\ref{tab1_stellar_kinematics}.

We stress that the classification of the kinematic feature is based on the presence of a significant 
velocity difference along a certain axis. Due to the extremely faint nature of UDGs and the effectiveness of the MUSE 
data, for the majority of the targets, the derived \vlos\, map does not extend beyond the \Reff\, of the galaxy. Furthermore, 
in some cases, the detection of a velocity difference could be induced by an external tidal distortion rather than 
the internal stellar kinematics. These events could affect the morphology (e.g. LSB4 and LSB6) or produce 
kinematic signals along different axes \citep[see also][]{Lokas2024}.

Figure~\ref{fig:distribution} (right panel) shows the distribution of $\Delta V$ for UDGs in LEWIS. The pink 
and blue histograms correspond to the UDGs with mild rotation along any axis (R and IR type) and those with 
no clear rotation (NR type), respectively. The majority of the galaxies in the sample (7 out of 18) show a rotation with values of 
$\Delta V \sim 30-40$\,km s$^{-1}$.

\subsection{The Faber-Jackson relation}\label{sec:scaling_rel}

In Figure~\ref{fig:FJ_BN} (left panel) we show the effective velocity dispersion \seff\, as a function 
of stellar masses ($M_\ast$) of galaxies from the LEWIS sample and literature 
\citep[see references in][]{Gannon2024}. 
As noted in Section~\ref{sec:1D_stellar_kin}, for galaxies with
intermediate fit and unconstrained fit (I and U fit types, respectively), the values of \seff\, and thus all the derived
quantities are overestimated. In the forthcoming analysis, we excluded
these cases before drawing our final considerations. Blue and red circles represent rotating and non-rotating galaxies, 
respectively. The stellar masses are estimated by \cite{Iodice2020b} and \cite{LaMarca2022a} using galaxy luminosities 
and the \cite{Into2013} colour\,-\,$M/L$ relation. 

We identified three groups of objects in the $M_\ast$\,-\,\seff\, plane. A group of UDGs (6 out of 11) are consistent
within the errorbars with the prediction of the Faber-Jackson relation extrapolated to the low-mass regime
\citep[$L\sim\sigma^{2.2}$, ][]{Kourkchi2012}. Two objects have \seff\ higher than the predicted
one (UDG7 and UDG12). These UDGs have a non-negligible rotation velocity ($\Delta V\sim~30-40$\,km s$^{-1}$), 
thus their values of \seff\ could be contaminated by the contribution of rotation (see also Section~\ref{sec:DM}). 
A few galaxies are instead below the prediction of the Faber-Jackson
relation (LSB7 and LSB8). They have a non-negligible
rotation velocity (i.e. $\Delta V~\sim~20-25~$km~s$^{-1}$). UDG4 is another outlier of the Faber-Jackson relation. 
It is a genuine UDG which shows no hint of rotation along any axis. Despite its position in the 
$M_\ast$\,-\,\seff\, plane and the large scatter, it is consistent with other UDGs from the literature.

\begin{figure*}[!h]
    \centering
    \includegraphics[scale=0.44]{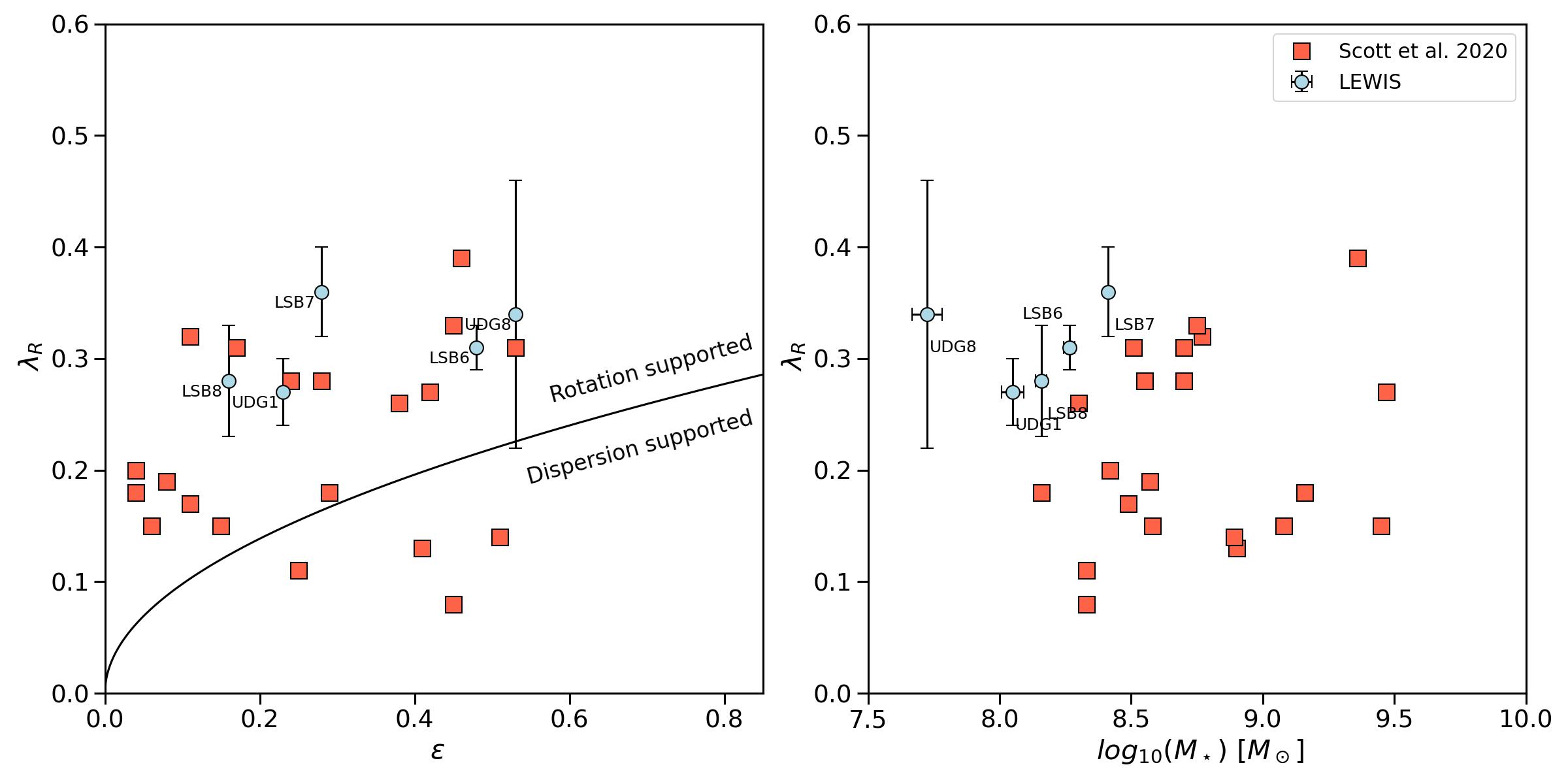}
    \caption{Projected specific angular momentum ($\lambda_{\rm R}$) as a function of galaxy ellipticity (left panel) and stellar mass (right panel). The light blue circles mark the LEWIS galaxies, while the pink squares represent the dwarf galaxies studied by \cite{Scott2020}. The black curve in the left panel ($\lambda_{\rm R}=0.31\sqrt{\epsilon})$ split the galaxies in rotation-supported (above the line) and dispersion-supported (below the line) systems \citep{Emsellem2011}.}
    \label{fig:lambda}
\end{figure*}

\subsection{Dark matter content}\label{sec:DM}

To constrain the DM content, we computed the dynamical mass (\Mdyn)  by applying the formula proposed in
\cite{Wolf2010}: $\displaystyle M_{\rm dyn} = 4R_{\rm eff,\,c}\,\sigma_{\rm eff}^2 / G$, where \Reff$_{\rm,\,c}$=\Reff$\sqrt{q}$ 
is the circularised half-light radius calculated through the galaxy axial ratio $q$, and $G$ is the 
gravitational constant. Thus, \Mdyn\ is the mass enclosed in a sphere with a radius of a circularised deprojected half-light radius. 
The \cite{Wolf2010} formula is widely applied in many literature works and is valid for dispersion-supported systems.

Nevertheless, we found that a significant fraction of galaxies in LEWIS show a rotation amplitude that is comparable to the value
for \seff \, ($\Delta V\gtrsim$\,\seff) thus they cannot be considered dispersion-supported systems. With some variation, 
the formula proposed in \cite{Wolf2010} can be used to derive a good estimate for the enclosed mass in any case. As described 
in \cite{Courteau2014}, \seff\ in real galaxies does not correspond to the stellar velocity dispersion, but can
be considered as an approximation of the luminosity-weighted second velocity moment 
($\sigma^2_{\rm eff} \approx \langle V^2_{\rm rms} \rangle = \langle V^2 + \sigma^2\rangle$), with $V$ and $\sigma$ the observed
mean stellar velocity and the corresponding dispersion, respectively. Thus, it already
includes the rotation and velocity dispersion of the stars and it is sometimes erroneously termed as $\sigma$. 

We tested if the approximation provided in \cite{Courteau2014} is still valid for the class of rotating UDGs, to investigate the effect 
on the dynamical masses. To this aim, we decided to repeat the extraction of the stellar kinematics to measure both \vlos\ and \slos .
However, to obtain an unbiased estimate of \slos, we required that all the spectra of the 
Voronoi bins have a sufficiently high S/N$\gtrsim$15. This is the case for five out of the seven rotating galaxies in LEWIS. For these galaxies, we extracted a map of the \slos\, (see Appendix~\ref{app:vlos_slos_maps}) and we were able to compute
an independent measure of the effective velocity dispersion averaging the \slos\ values in the various bins within an 
elliptical aperture with semi-major axis $a=R_{\rm eff}$. The values of \seff\, are systematically smaller than 
the estimate provided by the 1D stacked spectrum. Finally, we computed the luminosity-weighted second velocity moment
$V_{\rm rms}$ and calculated the dynamical mass using $V_{\rm rms}$. For UDG7, UDG12, and galaxies
with no clear rotation (NR-type), we computed the dynamical mass from \seff.

We calculated the $V$-band total luminosity in the ($L_{V, {\rm eff}}$) after converting the 
$r$-band magnitude using the formula proposed in \cite{Kostov2018}. In this way, we can compute 
the dynamical mass-to-light ratio (\Mdyn/$L_{V, {\rm eff}}$) and evaluate the amount of baryonic and DM
content bound to the system. Values for \Mdyn\, and \Mdyn/$L_{V, {\rm eff}}$ are reported in Table~\ref{tab1_stellar_kinematics}.

In Figure~\ref{fig:FJ_BN} (right panel) we show \Mdyn\, as a function of the
total luminosity in $V$-band of galaxies from the LEWIS sample and literature \citep{Gannon2024}. 
In addition, we plot the catalogue of dwarf galaxies in the Local Group studied by 
\cite{Battaglia2022}. The grey lines mark the loci where \Mdyn/$L_{V, {\rm eff}}$ is constant and equal to 1, 
10, 100, 1000, and 10000, respectively. 
The majority of galaxies in the LEWIS sample have DM content larger than the Local
Group dwarf galaxies with similar total luminosity (\Mdyn/$L_{V, {\rm eff}}\sim$10-100\,M$_\odot$/L$_\odot$), 
as found in previous works \citep{Gannon2021}. It is worth noticing that galaxies taken from \cite{Battaglia2022} are dwarf galaxies in the Local Group, i.e. located at a distance $D<1$\,Mpc from the Milky Way whose internal dynamics were investigated by using spatially-resolved kinematics of individual stars. The values of \seff\, inferred for these objects rely on precise measurements of the velocities of the stars, whereas the values of \seff\, for the galaxies in the LEWIS sample were estimated by measuring the global stellar motions along the line-of-sight. With LEWIS, an additional constraint on the DM content can be provided by analysing the GC dynamics (Mirabile et al., in prep.).

Two objects, UDG7 and UDG12, have an exceptionally large value of the effective velocity dispersion, 
being \seff~=~61~$\pm$~9~km~s$^{-1}$ and \seff~=~38~$\pm$~9~km~s$^{-1}$, respectively.
These UDGs are characterised by a mild rotation along the photometric major axis ($\Delta V \sim 30-40$\,km s$^{-1}$), 
thus the values of \seff\, can be contaminated by the contribution of rotation. These are the 2 UDGs for which we were not able 
to derive the \slos\, field and thus $V_{\rm rms}$. We estimated in any case the derived quantities using \seff . The values 
of \Mdyn\ and \Mdyn/$L_{V, {\rm eff}}$ can be overestimated by one order of magnitude. We flagged UDG7 and UDG12 in Table~\ref{tab1_stellar_kinematics}.

\subsection{Kinematic support of LEWIS galaxies}\label{sec:lambda}
We extracted the \vlos\, and \slos\, maps for the five galaxies of the LEWIS sample with sufficiently high S/N 
(UDG1, UDG8, LSB6, LSB7, and LSB8). For these galaxies, we can provide information on the kinematic
support by calculating the projected specific angular momentum ($\lambda_{\rm R}$) defined as:

\begin{equation*}
    \nonumber
    \displaystyle \lambda_{\rm R} = \frac{\langle r_\perp |V_{\rm LOS}|\rangle}{\Bigl \langle r_\perp \sqrt{V^2_{\rm LOS} + \sigma^2_{\rm LOS}} \Bigr \rangle}
\end{equation*}

\noindent where $r_\perp$ is the projected distance of the bin to the centre of the galaxy and $\langle$ $\rangle$ represents 
the flux-weighted average over the galaxy image \citep{Emsellem2011}. This quantity requires a 2D stellar kinematics 
map and accurate measurements of \slos\, and it is a diagnostic of the kinematic support. According to this parameter,
large values of $\lambda_{\rm R}$ correspond to rotation-supported galaxies, i.e. the kinematic structure is 
characterised by ordered rotational motion. On the other hand, small values for $\lambda_{\rm R}$ correspond 
to the dispersion-supported system, i.e. the kinematic structure is characterised by random motions. 

In Figure~\ref{fig:lambda} we show $\lambda_{\rm R}$ as a function of galaxy ellipticity $\epsilon$ (left panel) and stellar
mass (right panel). The black curve in the left panel ($\lambda_{\rm R}=0.31\sqrt{\epsilon})$ split the galaxies in
rotation-supported (above the line) and dispersion-supported (below the line) systems \citep{Emsellem2011}. Light blue
circles represent the LEWIS galaxies, while pink squares represent the dwarf ellipticals from \cite{Scott2020}. 
This sample is composed of dwarf galaxies which belong to the Virgo cluster of galaxies ($D\sim20$\,Mpc) morphologically 
classified as ellipticals (dE) or lenticulars (dS0), with an effective $r$-band surface-brightness between 
$18~<~\mu_r~<~24 $\,mag~arcsec$^{-2}$, and stellar masses between $10^{8} < M_\ast < 10^{9.5}$~M$_\odot$. With 
a similar approach, the authors apply on their datacubes the Voronoi binning algorithm with a S/N\,=\,10 to derive the 
mean \vlos\, and \slos\, and thus retrieve the kinematic support of the galaxies 
by computing the relative angular momentum, i. e. the $\lambda_R$ parameter. 
They found that, on average, values of $\lambda_{\rm R}$ in dwarf galaxies span a wide range ($\sim0.1-0.4$) and reported a mild
correlation between the kinematic support and stellar mass. Combining their catalogue of dwarf galaxies with several catalogues
of bright and massive galaxies, \cite{Scott2020} claimed that $\lambda_{\rm R}$ reaches a maximum around $M_\ast\sim10^{10}$~M$_\odot$,
and it decreases toward the low-mass and very high-mass regimes. The LEWIS galaxies span a wide range of $\epsilon$
and are characterised by values $0.25 \lesssim \lambda_{\rm R} \lesssim 0.35$.
We do not see a distinction between the two UDGs and the three LSB galaxies.

\begin{figure*}
    \centering
    \includegraphics[scale=0.315]{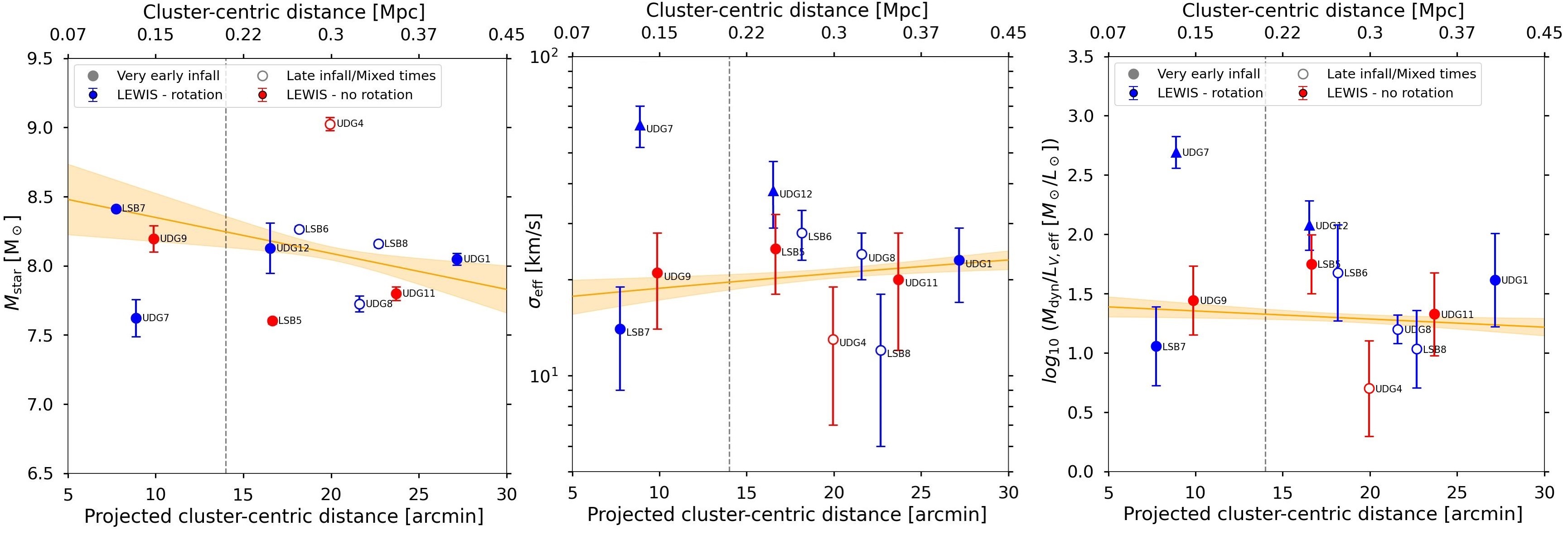}
    \caption{Left panel: Stellar mass as a function of the cluster-centric distance. Middle panel: \seff\, as a function of the cluster-centric distance. Right panel: dynamical mass-to-light ratio \Mdyn/$L_{V, {\rm eff}}$ as a function of the cluster-centric distance. The vertical grey dashed line marks the radial distance within which the X-ray emission of the cD galaxy dominates \citep[14 arcmin,][]{Spavone2024}. The labels identify the LEWIS galaxies. The filled circles represent galaxies classified as `very early infall', while the empty circles represent galaxies classified as `late infall/mixed times'. The blue and red circles represent rotating and non-rotating galaxies, respectively. The yellow line represents a linear fit obtained by excluding the values for UDG7 and UDG12. Uncertainties on the fitting lines are computed using a bootstrap technique. In the central and right panels, UDG7 and UDG12 are marked with triangles since their values of \seff\, and \Mdyn/$L_{V, {\rm eff}}$ could be overestimated due to their non-negligible rotation.}
    \label{fig:cluster_centric_dist}
\end{figure*}

\begin{figure*}
    \centering
    \includegraphics[scale=0.312]{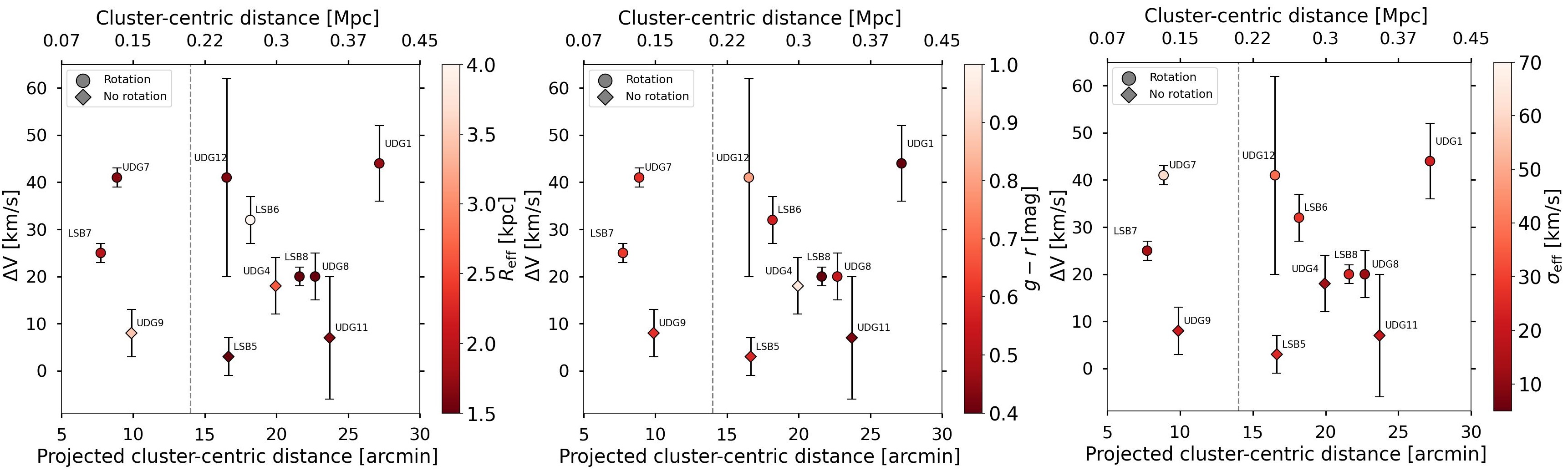}
    \caption{Semi-amplitude of the rotation curve $\Delta V$ as a function of the cluster-centric distance. The data-points are colour-coded according to the values of the \Reff\ (left panel), colour $g-r$ (central panel), and \seff\ (right panel). As in Figure~\ref{fig:cluster_centric_dist}, the vertical dashed line marks the extension of the X-ray emission. The circles represent galaxies that show a rotation along the photometric major axis or intermediate axis (R and IR-type), triangles are instead the galaxies with no rotation (NR-type).}
    \label{fig:cluster_centric_dist2}
\end{figure*}

\section{Discussion: two kinematical classes of UDGs in the Hydra I cluster}
\label{sec:discussion}

Based on the 2D stellar kinematics, we can claim that the LSB galaxies and UDGs in the Hydra I
cluster can be grouped, on average, into two classes: 
galaxies showing mild rotation and those not showing evidence of rotation along any axis.

We find a narrow distribution for \seff\, in the LEWIS galaxies. The majority of UDGs and LSB galaxies have 
similar DM content (\Mdyn/$L_{V, {\rm eff}}\sim$10-100\,M$_\odot$/L$_\odot$), 
which is larger than that observed in the dwarf galaxies of comparable luminosity. 

The main result of the paper is the existence of kinematically different types of galaxies. This suggests 
different formation pathways for UDGs and LSB galaxies in the Hydra I cluster. Unfortunately, due to their
extremely faint nature, the literature lacks systematic studies on the kinematic support of UDGs. Only few theoretical works provide predictions on stellar kinematics. 

\cite{Cardona-Barrero2020} studied the kinematic support of isolated UDGs
in NIHAO simulations \citep{Wang2015}. 
The authors found that UDGs span a continuous distribution of $\lambda_{\rm R}$ values for different
inclination angles from 0$^\circ$ up to 90$^\circ$. In particular, for galaxies with an inclination  
between 30$^\circ$ to 60$^\circ$, $\lambda_{\rm R}$ varies from 0.2 to 0.7. 
These values are consistent with those derived for the LEWIS rotating galaxies, which have similar inclinations, 
and $0.25 \leq \lambda_{\rm R} \leq 0.35$ (Figure~\ref{fig:lambda}). 
It is worth noting that LEWIS galaxies are located in a cluster environment, whereas \cite{Cardona-Barrero2020} explored the kinematical properties of field UDGs.

In addition, \cite{Cardona-Barrero2020} did not find any correlation of 
$\lambda_{\rm R}$ with the stellar mass, but they found correlations with the morphological properties, HI gas content, and 
size. Larger UDGs are more rotation-supported, have large HI gas content, and present a disk-like morphology. On the contrary, 
dispersion-supported UDGs have small HI gas content and resemble triaxial spheroids. The kinematic prediction suggests that isolated UDGs in NIHAO simulations formed due to intense gas outflow.

\cite{Benavides2023} adopted a similar approach to studying UDGs in broader environmental conditions in TNG50 simulations. 
They did not measure $\lambda_{\rm R}$, but
they took into account the ratio between the rotational support and total kinematic energy of
the galaxy. This quantity cannot be measured in real observations since it relies on quantities which are related 
to star particles. However, it allows us to distinguish rotation and dispersion-supported systems similarly to the $\lambda_{\rm R}$
parameter. \cite{Benavides2023} did not find differences in terms of kinematic support between field UDGs 
and UDGs in most dense regions, suggesting that the environmental effects act more quickly on stellar 
distribution than on the internal dynamics. This result might reconcile with comparable values of $\lambda_{\rm R}$ for the field UDGs
by \cite{Cardona-Barrero2020} and for our LEWIS galaxies in the Hydra I cluster.

In addition, \cite{Benavides2023} found a correlation between the kinematic support and stellar mass, pointing out that more massive UDGs
($M_\ast\gtrsim10^{8.5}$\, M$_\odot$) are more rotation-supported and have a disk-like morphology. Since this trend was also found in populations of  non-UDGs, they wonder if this is due to a specific parameterisation of baryonic modelling in TNG50 
simulations, or if it is a peculiarity of the class of UDGs. As in \cite{Benavides2023}, we find a mild trend of the rotational support
with increasing mass (Figure~\ref{fig:lambda}, right panel), but more systematic studies are required
to address and confirm the kinematic support in UDGs.

The distinction in the two classes of UDGs proposed by \cite{Sales2020} is 
based on the distributions of effective velocity dispersion and 
metallicity, whereas no indication of the stellar rotation is provided. Therefore, no comparison between the observed stellar 
kinematics of LEWIS galaxies and this set of simulations can be made.

\subsection{Correlations within the cluster environment}\label{sec:env}

In Figure~\ref{fig:cluster_centric_dist} we show the stellar mass (left panel), \seff\, (middle panel) 
and \Mdyn$/L_{V, {\rm eff}}$ (right panel) as a function of the cluster-centric distance.\footnote{Consistently with \cite{Forbes2023}, 
we have assumed RA = 159.17842, Dec. = –27.528339 as the centre of the Hydra I cluster, which corresponds to the centre of the
bright cluster member NGC\,3311.} In these plots, we marked the innermost regions of the cluster dominated by the X-ray 
emission, which corresponds to $R\sim14$~arcmin \citep[][]{Spavone2024}. 

We performed a linear fit to data and their uncertainties to  
highlight the presence of possible correlations between the derived properties and
position in the cluster. We considered both early and late infallers. 
They are cluster members according to their \Vsys\ and despite their location in the phase-space diagram.
Since the \Mdyn\ values for UDG7 and UDG12 are overestimated due to a non-negligible contribution of rotation velocity,
we discarded them from the fit. We found a weak trend of the stellar mass 
with the cluster-centric distance, where more massive UDGs are close to the centre. The values of the stellar
masses will be derived independently from the analysis of the stellar population (Doll et al., in prep.), thus an 
updated version of this figure will be shown in a forthcoming paper. We found a mild trend of \seff\  with the 
cluster-centric distance, where galaxies with smaller values of \seff\ are located in the inner regions. 
An opposite, but also weak, correlation appears in the $\log_{10}$(\Mdyn\,/$L_{V, {\rm eff}}$) - distance plane, 
suggesting that DM-dominated galaxies are located at smaller distances from the cluster core.

According to the predictions by \citet{Sales2020}, T-UDGs populate the centre of the clusters 
and, at a given stellar mass, have lower \seff, higher metallicity, and lower DM fraction with respect to the B-UDG. 
Since we found a significant fraction of galaxies with non-negligible rotation, we cannot directly compare our findings 
with the results provided in \cite{Sales2020}. Indeed, they assumed that UDGs are dispersion-supported systems
to estimate \seff . Our rotating UDGs cannot be considered dispersion-supported systems, thus the estimate 
of \seff\ is affected by the rotation and is not representative of the true stellar velocity dispersion. 
In addition, taking into account that the LEWIS sample is located within 0.4$R_{\rm 200}$, 
our analysis is mainly focused on the innermost regions of the Hydra I cluster.

In Figure~\ref{fig:cluster_centric_dist2} we show the semi-amplitude of the 
velocity profile $\Delta V$ as a function of the cluster-centric distance. 
In this plot, we colour-coded the data-points 
according to the values of the \Reff\ (left panel), colour $g-r$ (central panel), and \seff\ (right panel). 
In this way, we can test for a correlation of the dynamical and structural properties of the UDGs 
and LSB galaxies in the Hydra~I cluster. 

If galaxies have a disk-like memory of their past evolutionary pathway, we might expect larger values of $\Delta V$ and
\Reff , smaller values of \seff\ and bluer $g-r$ colours. This is because galaxies with disk morphology/dynamics are expected
to have younger stellar populations with more ordered motions. We do not find any clear trend considering the different
explored quantities, nor concerning the projected cluster-centric distance. In particular, UDGs and LSB galaxies with rotation are found 
at any distance from the core of the cluster.

\section{Summary and concluding remarks}
\label{sec:conclusion}

In this paper, we performed a kinematic analysis for a homogeneous and almost 
a complete sample of UDGs and LSB galaxies in the Hydra I cluster of galaxies with MUSE data from the LEWIS project \citep{Iodice2023}.
With the observing campaign $\sim92$\% completed, we were able to provide values of \Vsys\, for
more objects, with respect to the first release (Paper I). We found
that nearly all LEWIS targets have systemic velocities consistent with the overall velocity distribution of the 
giant and dwarf galaxy populations in the cluster (Figure~\ref{fig:histogram_vsys_phase_space}). 
In particular, 15 objects out of 23, have \Vsys\, within 1$\sigma_{\rm Hydra}$. 
The remaining eight galaxies have larger relative \Vsys, but within 2$\sigma_{\rm Hydra}$.

This paper focused on the stellar kinematics of the LEWIS targets to derive the rotation velocity and velocity dispersion.
Thanks to the integral-field nature of the LEWIS data, we were able to derive 2D velocity 
maps for most of the sample galaxies. This represents the first large sample of 2D stellar velocity maps for UDGs and LSBs
galaxies. To date, a stellar velocity field was extracted only for UDG NGC~1052~-~DF2 and DF44. NGC~1052~-~DF2 shows mild rotation with $\Delta V \sim 6$\,km s$^{-1}$, whereas DF 44 shows no evidence of rotation \citep{Emsellem2019, vanDokkum2019b}.

The main results are summarised below.
\begin{itemize}
    \item The peak of the distribution of \seff\, ranges between 20 and 30 km s$^{-1}$, consistent with the distribution of values from literature data (Figure~\ref{fig:distribution}, left panel). There are two LSB galaxies in the sample with extremely small values of \seff\, (\seff~$\leq 15$~km~s$^{-1}$) and two UDGs with very large values of \seff\,(\seff~$\geq 40$~km~s$^{-1}$). These sources have high-quality spectra (S/N\,>\,15), thus the estimate of their \seff\, is reliable.
    \item Based on the 2D map of the stellar velocity, 7 out of 18 LEWIS galaxies show a mild rotation ($\Delta V \sim$~25~-~40~km s$^{-1}$). The rotation of three of them is along the photometric major axis, whereas for the other four is along an intermediate axis. Five LEWIS galaxies do not show evidence of rotation, and for the remaining six we got unconstrained results (Figure~\ref{fig:distribution}, right panel) 
    \item Compared to the Faber-Jackson relation, we found three groups of UDGs. A group of UDGs in Hydra I is consistent with the relation extrapolated to the low-mass regime, i.e. $L \sim \sigma^{2.2}$ \citep{Kourkchi2012}. A few galaxies are outliers of the Faber-Jackson relation. UDG7 and UDG12 have larger values of \seff\ than expected for their stellar mass. These values are overestimated due to a non-negligible contribution of the rotation. UDG4, LSB7, and LSB8 have instead values for \seff\, below the relation (Figure~\ref{fig:FJ_BN}, left panel).
    
    \item The majority of the galaxies of the LEWIS sample (i.e. both the UDGs and LSB galaxies) have a DM content larger than dwarf galaxies with similar total luminosity \Mdyn/$L_{V, {\rm eff}}\sim$10-100\,M$_\odot$/L$_\odot$ (Figure~\ref{fig:FJ_BN}, right panel). 

    \item For 5 of the 7 rotating galaxies in LEWIS, we derived the $\lambda_{\rm R}$ parameter, which is a diagnostic of the kinematic support. They are characterised by values $0.25 \lesssim \lambda_{\rm R} \lesssim 0.35$, suggesting that these galaxies are rotation supported.
    We do not see a distinction between UDGs and LSBs (see Figure~\ref{fig:lambda}).
    
    \end{itemize}

Considering the enormous technical and observational challenge of obtaining spectra for LSB galaxies, 
which have a surface brightness that is only a small fraction of the sky level, the results from LEWIS 
have demonstrated that combining the large collecting area of the Very Large Telescopes and high efficiency of the MUSE 
integral-field spectrograph truly paved the way for the spectroscopic characterization
of this class of faint objects.

We have provided an extended census of the 
stellar kinematics for UDGs in a cluster environment. 
In particular, we have found, for the first time, a considerable
number of UDGs and LSB galaxies that are rotation-supported. 
These results 
point towards the existence of two kinematical classes of UDGs in the 
Hydra I cluster, which might have a different origin. 

However, the stellar kinematic properties are not sufficient to discriminate between
the formation channels, proposed by theoretical works on UDGs.
Stringent constraints on any possible relation with the environment where UDGs
reside, can be obtained by combining the stellar kinematics with other UDG properties.
The stellar population analysis and GCs content will be investigated 
with the LEWIS data and presented in forthcoming papers of the LEWIS series
(Doll et al., in prep., Mirabile et al., in prep.). \cite{Iodice2023} have already demonstrated that with LEWIS, 
it will be possible to confirm GCs pre-selected through photometry, spectroscopically identify new GC systems 
and study their stellar kinematics and population.

Results presented in this paper have also opened a new topic to be investigated for
UDGs. Upcoming works will focus on the analysis of the available
cosmological simulations to extract the missing information on the stellar kinematics
to be compared with results derived from MUSE data for the UDGs in the Hydra I cluster.
This kind of analysis would also trigger additional interest in integral field spectroscopic
data for this class of galaxies.

%-------------------------------------------------------------------
\begin{acknowledgements}
We wish to thank the anonymous Referee whose comments helped us to improve the clarity of the manuscript. Based on observations collected at the European Southern Observatory under ESO programmes 108.222P.001, 108.222P.002, 108.222P.003. 
The authors wish to thank L. Buzzo, L. Coccato, V. Debattista, E. Emsellem, A. Ferre-Mateu, J. Gannon, L. Greggio, F. Marleau, O. Muller, T. Puzia, R. Rampazzo for the useful comments and discussions on the work presented in this paper. 
E.I. acknowledges support by the INAF GO funding grant 2022-2023. 
E.I., E.M.C. and M.P. acknowledge the support by the Italian Ministry for Education University and Research (MIUR) grant PRIN 2022 2022383WFT “SUNRISE”, CUP C53D23000850006.
J.H. and E.I. acknowledge the financial support from the visitor and mobility programme of the Finnish Centre for Astronomy with ESO (FINCA), funded by the Academy of Finland grant nr 306531. J.H. wishes to acknowledge CSC-IT Center for Science, Finland, for computational resources.
E.M.C. acknowledges the support from MIUR grant PRIN 2017 20173ML3WW-001 and Padua University grants DOR 2021-2023.
G.D. acknowledges support by UKRI-STFC grants: ST/T003081/1 and ST/X001857/1. 
D.F. thanks the ARC for support via DP220101863 and DP200102574. 
J.F-B. acknowledges support from the PID2022-140869NB-I00 grant from the Spanish Ministry of Science and Innovation. 
This work is based on the funding from the INAF through the GO large grant in 2022, to support the LEWIS data reduction and analysis (PI E. Iodice). 
The authors thank \citet{Gannon2024} for the compilation of their catalogue of UDG spectroscopic properties. The catalogue includes data from: \citet{mcconnachie2012, vanDokkum2015, Beasley2016, Martin2016, Yagi2016, MartinezDelgado2016, vanDokkum2016, vanDokkum2017, Karachentsev2017, vanDokkum2018, Toloba2018, Gu2018, Lim2018, RuizLara2018, Alabi2018, FerreMateu2018, Forbes2018, Martin2019, Chilingarian2019, Fensch2019, Danieli2019, vanDokkum2019, torrealba2019, Iodice2020, Collins2020, Muller2020, Gannon2020, Lim2020, Muller2021, Forbes2021, Shen2021, Ji2021, Huang2021, Gannon2021, Gannon2022, Mihos2022, Danieli2022, Villaume2022, Webb2022, Saifollahi2022, Janssens2022, Gannon2023, FerreMateu2023, Toloba2023, Shen2023}. 
The authors acknowledge the use of the following Python scripts: 
{\sc ASTROPY} \citep{astropy:2013, astropy:2018}, {\sc MATPLOTLIB} \citep{matplotlib}, {\sc MPDAF} \citep{Bacon2016, Piqueras2017}, {\sc NUMPY} \citep{numpy}, {\sc PHOTUTILS} \citep{photutils}, {\sc SCIPY} \citep{SciPy}, and {\sc ZAP} \citep{Soto2016}.

\end{acknowledgements}

%--------------------------------
\bibliographystyle{aa.bst}
\bibliography{LEWIS_paperII.bib}

%--------------------------------

%
\begin{appendix}
\section{Estimation of the MUSE LSF}
\label{app:LSF}

To test the robustness of the measure of \slos\, it is crucial to have a precise measurement of the 
spectral resolution of the MUSE spectrograph, which is parameterised via the line-spread function (LSF).
Our observing strategy consists of obtaining first shallow observations to address the candidate cluster
membership, and then collecting and completing the observations in the subsequent periods 
\citep[see][for details]{Iodice2023}. Thus, the targets in the LEWIS sample have a different number 
of OBs that depends on the surface brightness of the galaxy and on the depth level we want to reach. 
This means that for each galaxy, the different exposures have been taken in different sky quality 
conditions and on different nights. We aim to investigate possible systematic variations in the MUSE 
LSF among the various exposures and the final combined cube. 

We chose UDG1 as a starting test case to measure the instrumental LSF. We performed the data reduction 
on all the 9 exposures of UDG1 with the {\sc ESOREFLEX} pipeline routine \citep{Weilbacher2020, Freudling2013} 
without including the sky background subtraction step. For each exposure, we extracted a sky spectrum in a 
circular aperture of a radius of 15 pixels in different regions in the whole FOV. In this way, we explored 
possible spatial dependences of the LSF. We selected a set of 14 not-blended sky emission lines sampling a 
broad wavelength range, and we measured the FWHM of the line after interpolating the line profile
with a second order polynomial. We finally obtained an unique dataset by combining all the 
exposures and adopting the mean and standard deviation values of the distributions of FWHMs
measured for each sky emission line. We measured a constant LSF by calculating 
the weighted mean average of the FWHMs across the wavelength range. In addition, we compared the resulting 
LSF with the one measured from the all-exposures combined datacube, adopting the same 
fitting strategy. The two measured LSFs are consistent. 
 
Since previous results confirmed the agreement between the LSF measured from the single exposures and the combined cube, we derived the LSF directly from the combined cubes still containing the sky background for all the UDGs in the LEWIS sample, and we repeated the same analysis. We found similar trends for datasets belonging to different UDGs. This means that the instrumental LSF is not affected by the chosen observing strategy or by the nature of the target itself. We refined the estimation of the MUSE LSF by fitting data with a polynomial function of grade equal to 2, as done in \cite{Bacon2017}. From the best-fitting polynomial function, we found:

\begin{equation}
    {\rm FWHM} (\lambda) = 1.185\cdot10^{-8}\lambda^2 - 1.916\cdot10^{-4} \lambda + 3.397
    \label{eq:1}
\end{equation}

\noindent Our best-fit polynomial function appears to be shallower than the one obtained by \cite{Bacon2017}. The difference might be due to the fact that the two sky emission lines at $\lambda=4861.3$\AA\ and $\lambda=5577.3$\AA\ drove the fit to shallower values. Therefore we repeated the fit considering data points in the spectral region $\lambda>5700$\AA, and we obtained:

\begin{equation}
    {\rm FWHM} (\lambda) =3.740\cdot10^{-8}\lambda^2 - 5.879\cdot10^{-4} \lambda + 4.917
    \label{eq:2}
\end{equation}

\begin{figure}[!h]
    \centering
    \includegraphics[scale=0.28]{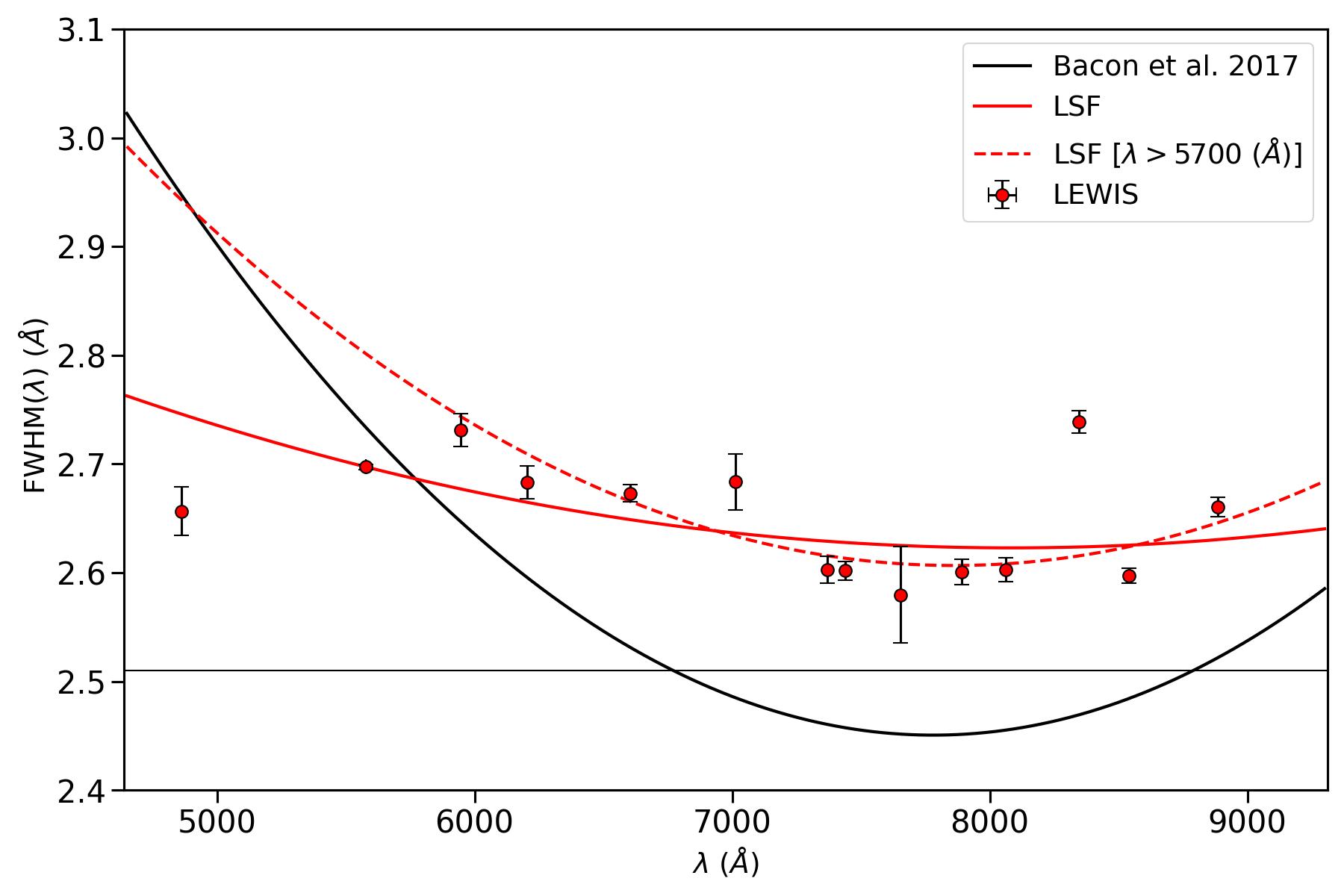}
    \includegraphics[scale=0.28]{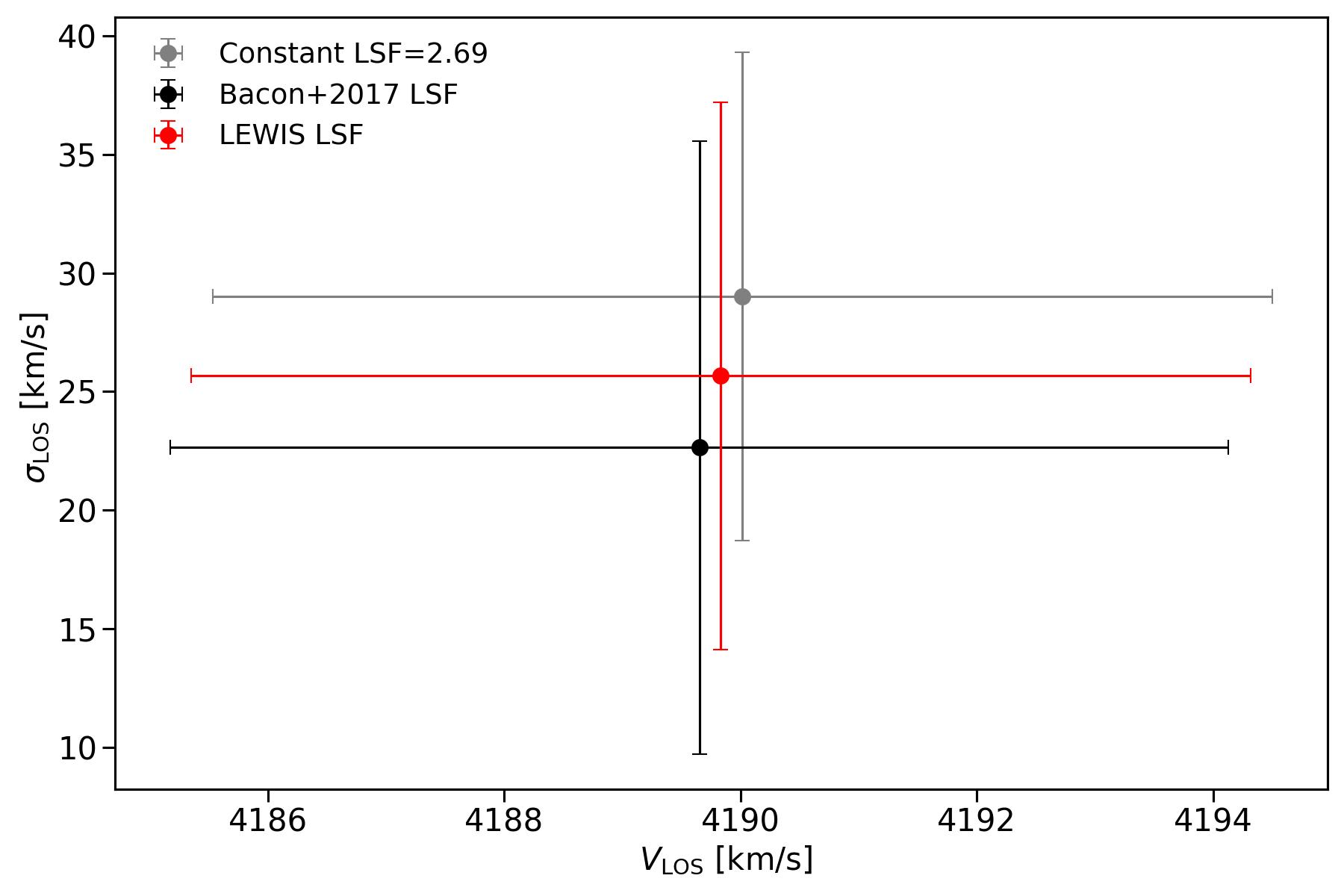}
    \caption{Top panel: MUSE LSF from \cite{Bacon2017} (thick black line) and in LEWIS data (thick red line, Eq.~\ref{eq:1}).  The red dashed line represents the MUSE LSF derived restricting the polynomial fit to $\lambda>5700$~\AA\,(Eq.~\ref{eq:2}). The thin horizontal black line marks the spectral resolution of the E-MILES stellar library (FWHM=2.51\AA). Bottom panel: comparison of {\sc pPXF} results for \cite{Bacon2017} LSF (black), LEWIS LSF (red, Eq.~\ref{eq:2}) and constant LSF (grey, FWHM[4800-7000] = 2.69\AA).}
    \label{fig:LSF}
\end{figure}

In the top panel of Figure~\ref{fig:LSF} we present the LSF measured in \cite{Bacon2017} (thick black line) and the MUSE LSF 
in the LEWIS data (thick red line, Eq.~\ref{eq:1}). The red dashed line represents the MUSE LSF derived restricting the polynomial fit 
to $\lambda>5700$~\AA~(Eq.~\ref{eq:2}). The thin horizontal black line marks the spectral resolution of the E-MILES stellar 
library (FWHM=2.51\AA), Our instrumental resolution is slightly better than the one measured by \cite{Bacon2017} in the bluest 
spectral region ($\Delta \sigma \sim 4 $ km s$^{-1}$ at $\lambda=5000$\AA), but slightly worse at longer wavelengths ($\Delta \sigma \sim 3 $
km s$^{-1}$ at $\lambda=7000$\AA). In most of our cases, since we extracted the stellar kinematics limiting the wavelength range to the optical
one, it is reasonable to adopt a constant value for the LSF (FWHM[4800-7000] = 2.69\AA). 

We finally tested the effect of the LSF on our results by fitting with the {\sc pPXF} algorithm the 1\Reff\ stacked spectrum
of UDG1 and deriving the first (\vlos) and second (\slos) order velocity moment (Figure~\ref{fig:LSF}, bottom panel). Results are
obtained by using the LSF in \cite{Bacon2017} and the LEWIS LSF, both the constant value and the second-order polynomial fit. 
The final fitted parameters are consistent within the errorbars. Through the whole analysis, we will adopt the LSF measured 
from LEWIS data (Eq.~\ref{eq:2}). 

\section{Stellar velocity maps}
\label{app:velocity_maps}

In this section, we present the stacked and spatially-resolved stellar kinematics map for all the galaxies in the LEWIS sample.

\begin{figure*}
    \centering
    \includegraphics[scale=0.28]{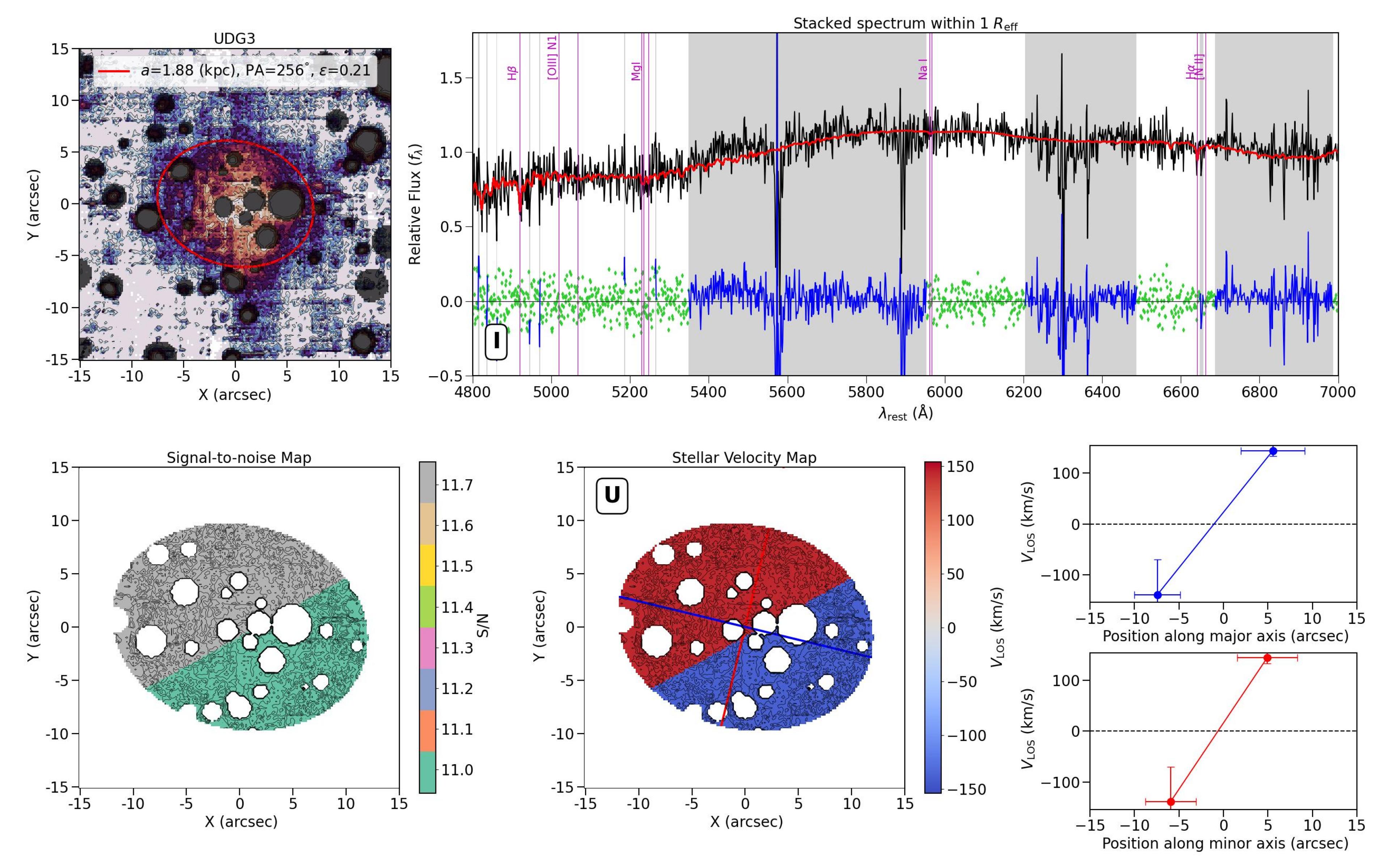}
    \caption{Same as Figure~\ref{fig:UDG1_stellar_kinematics}, but for UDG3.}
    \label{udg3_2D}
    \includegraphics[scale=0.28]{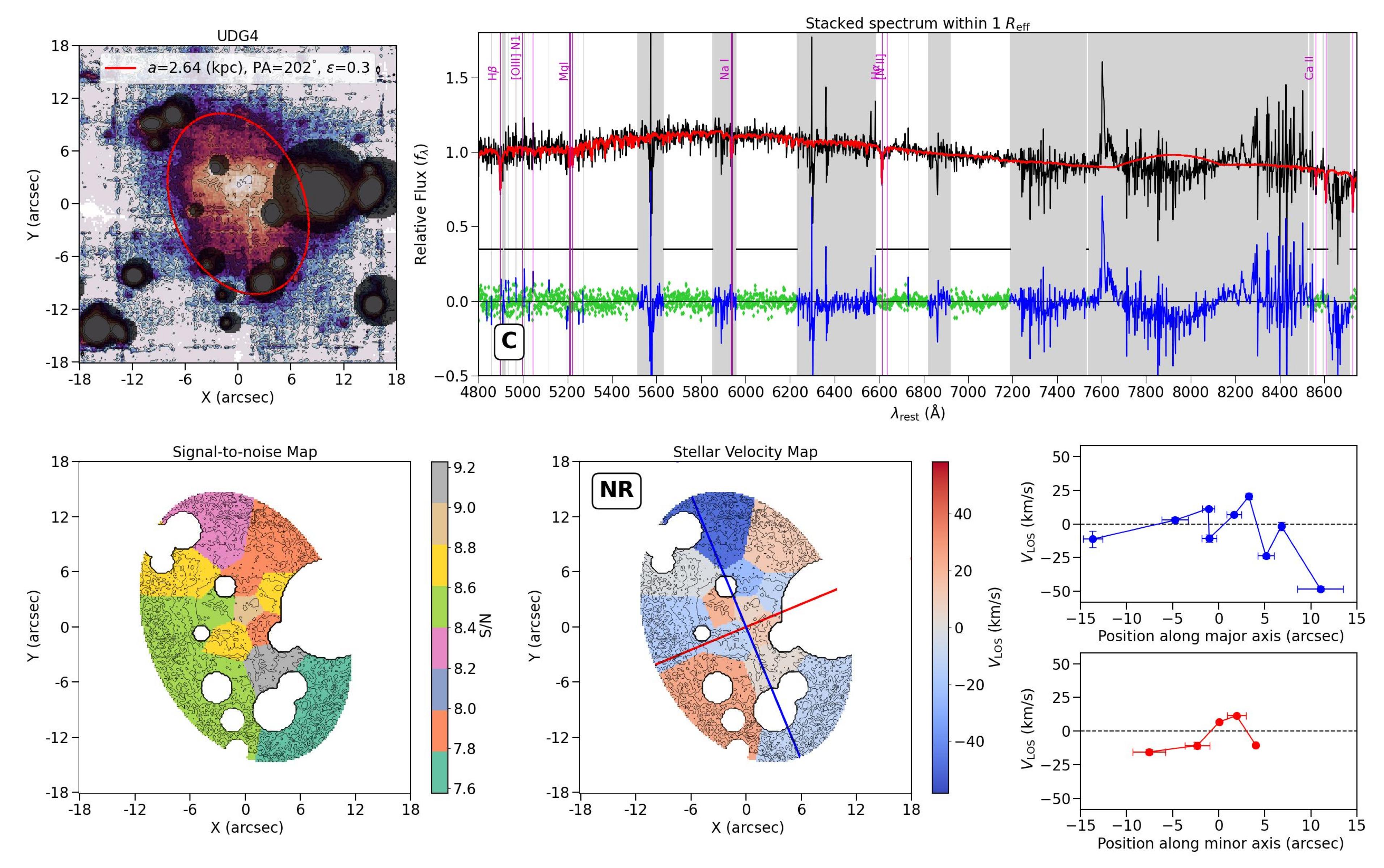}
    \caption{Same as Figure~\ref{fig:UDG1_stellar_kinematics}, but for UDG4.}
    \label{udg4_2D}
\end{figure*}

\begin{figure*}
    \centering
    \includegraphics[scale=0.28]{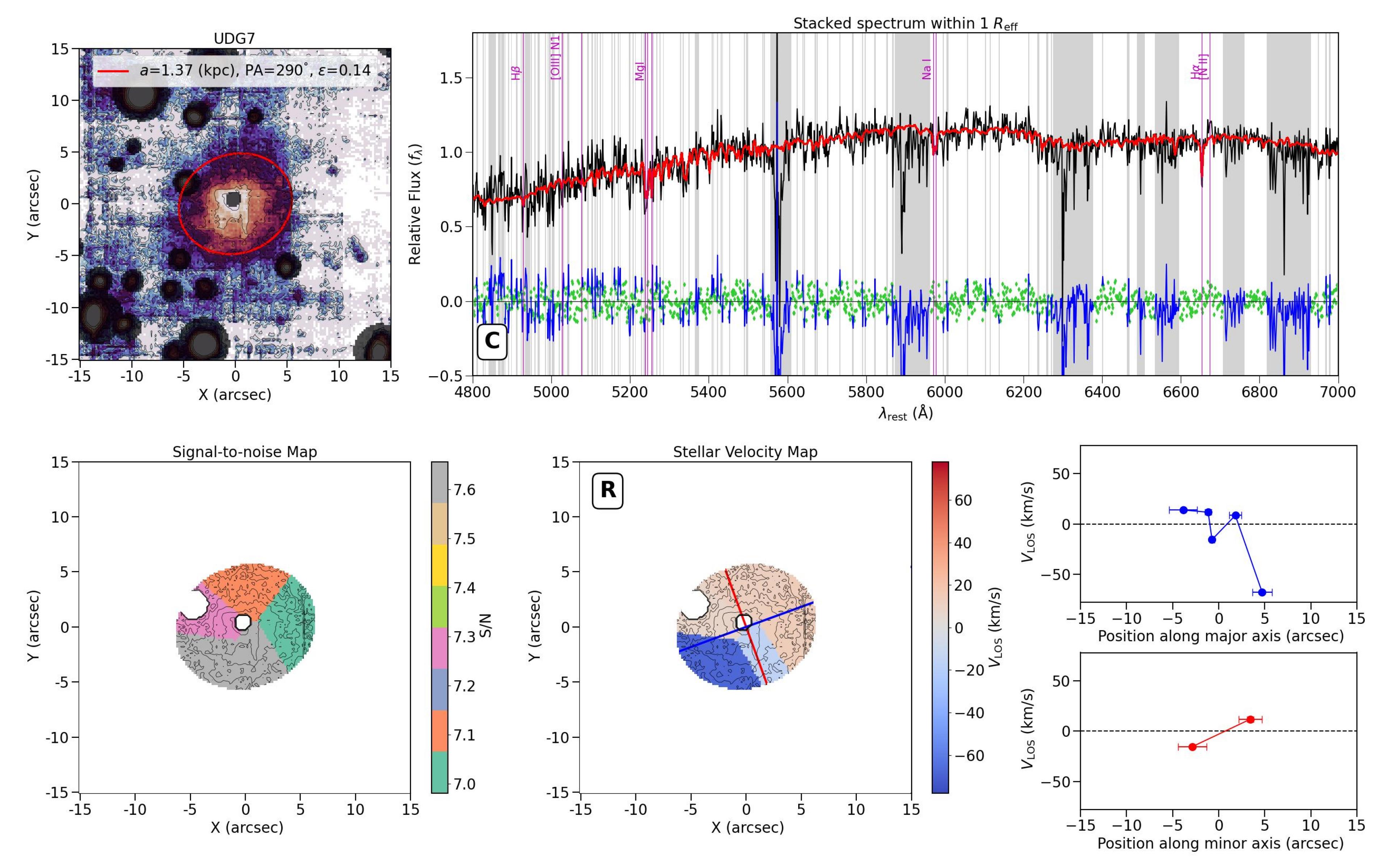}
    \caption{Same as Figure~\ref{fig:UDG1_stellar_kinematics}, but for UDG7.}
    \label{udg7_2D}
    \includegraphics[scale=0.28]{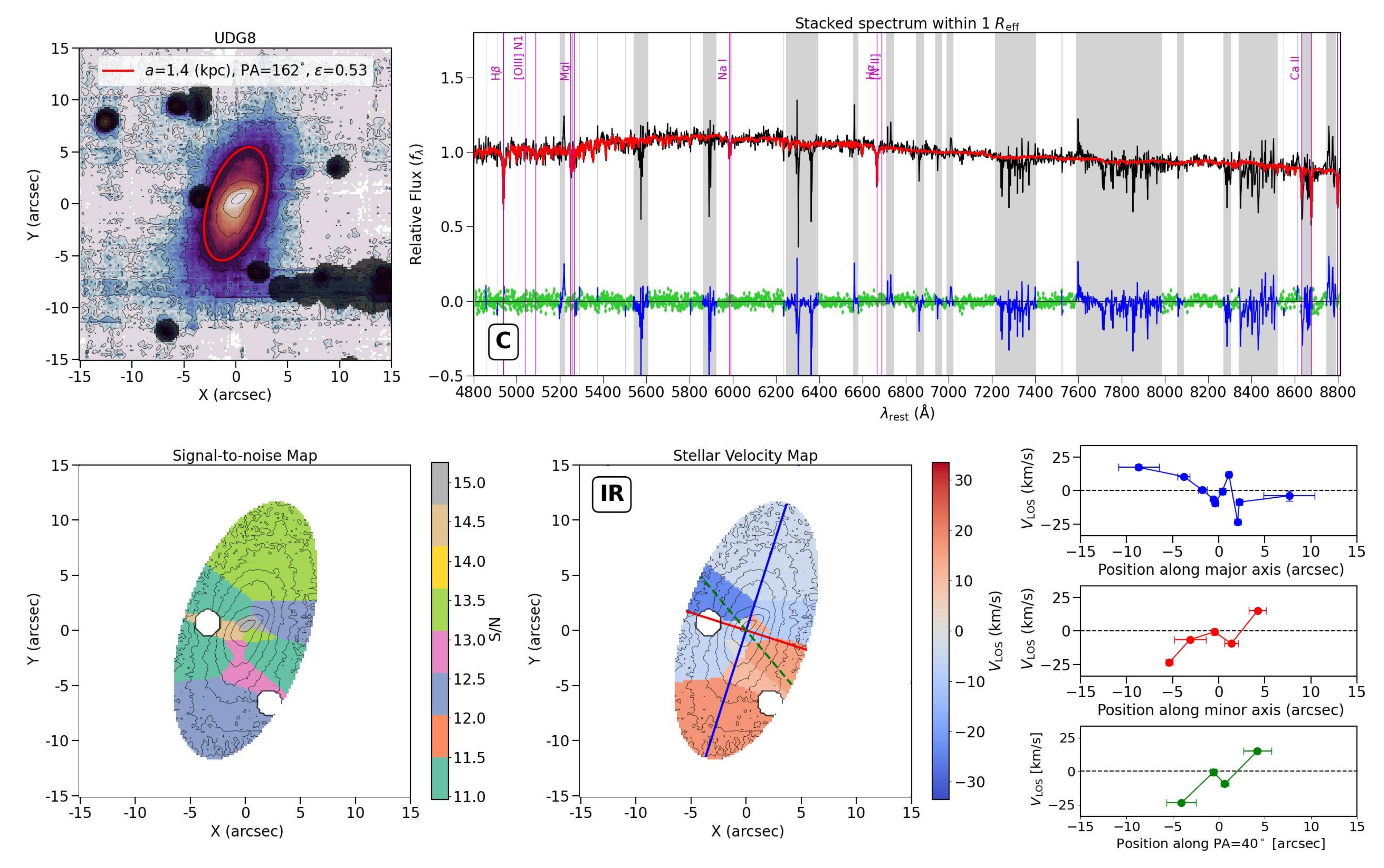}
    \caption{Same as Figure~\ref{fig:UDG1_stellar_kinematics}, but for UDG8.}
    \label{udg8_2D}
\end{figure*}

\begin{figure*}
    \centering
    \includegraphics[scale=0.28]{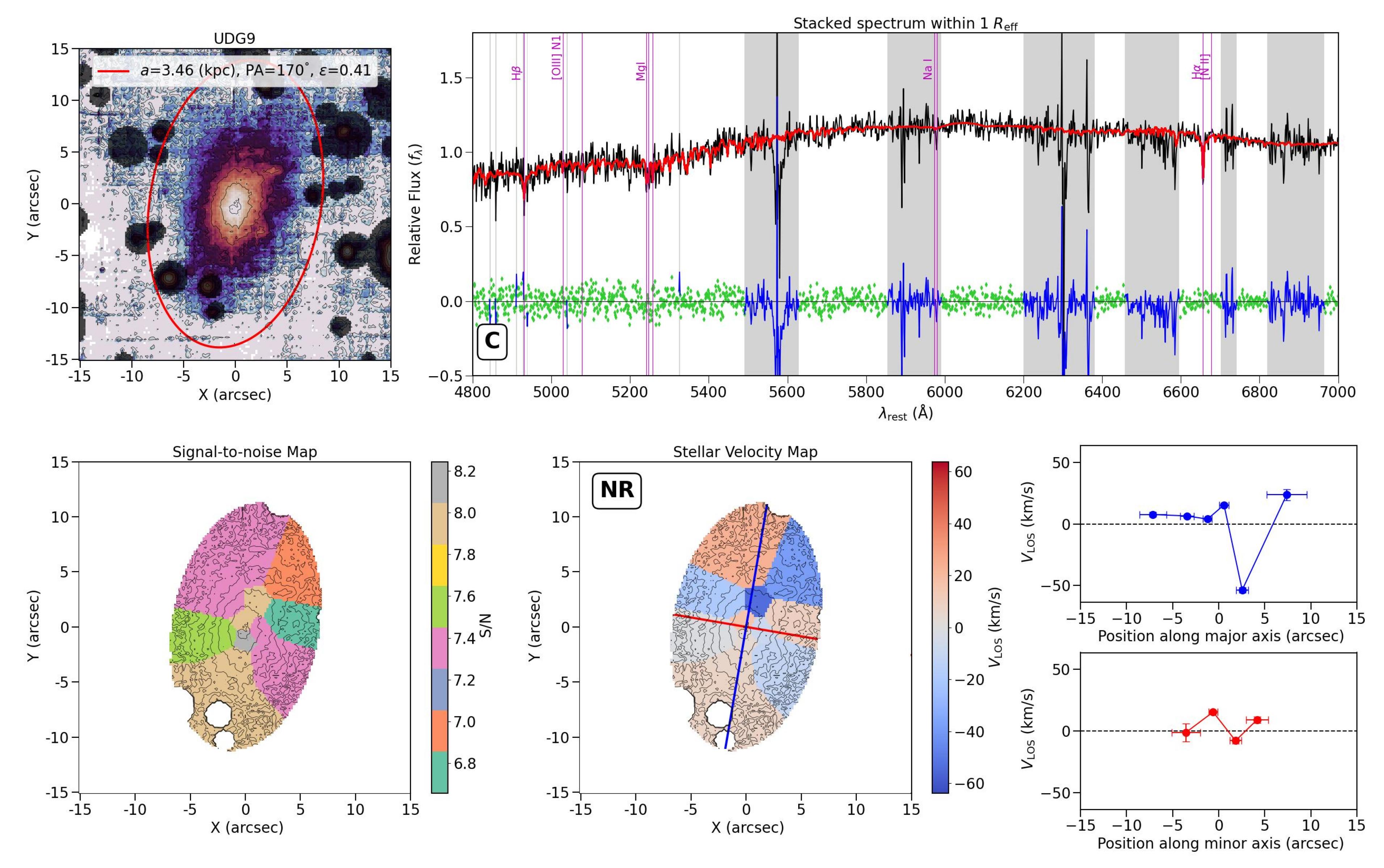}
    \caption{Same as Figure~\ref{fig:UDG1_stellar_kinematics}, but for UDG9.}
    \label{udg9_2D}
    \includegraphics[scale=0.28]{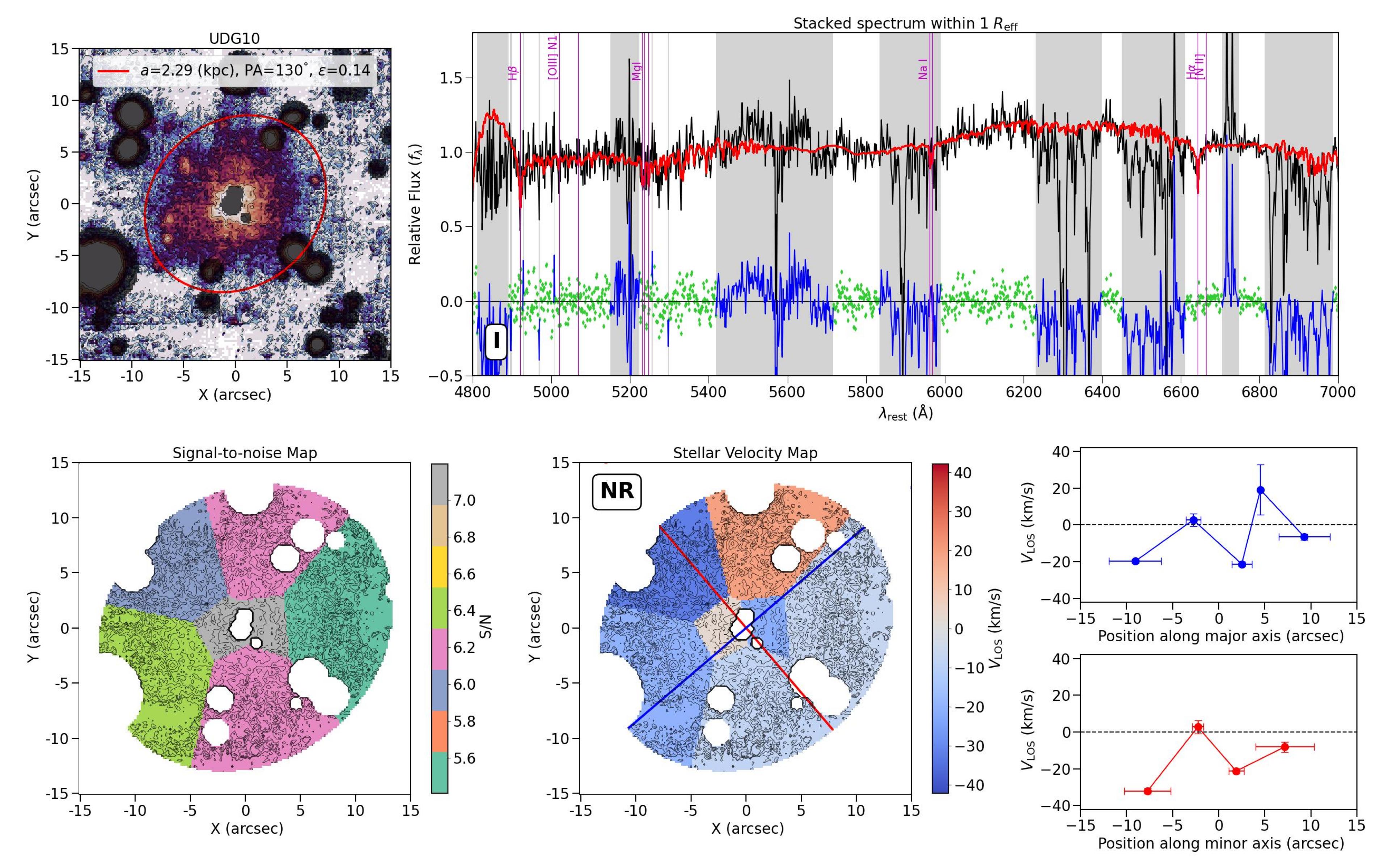}
    \caption{Same as Figure~\ref{fig:UDG1_stellar_kinematics}, but for UDG10.}
    \label{udg10_2D}
\end{figure*}

\begin{figure*}
    \centering
    \includegraphics[scale=0.28]{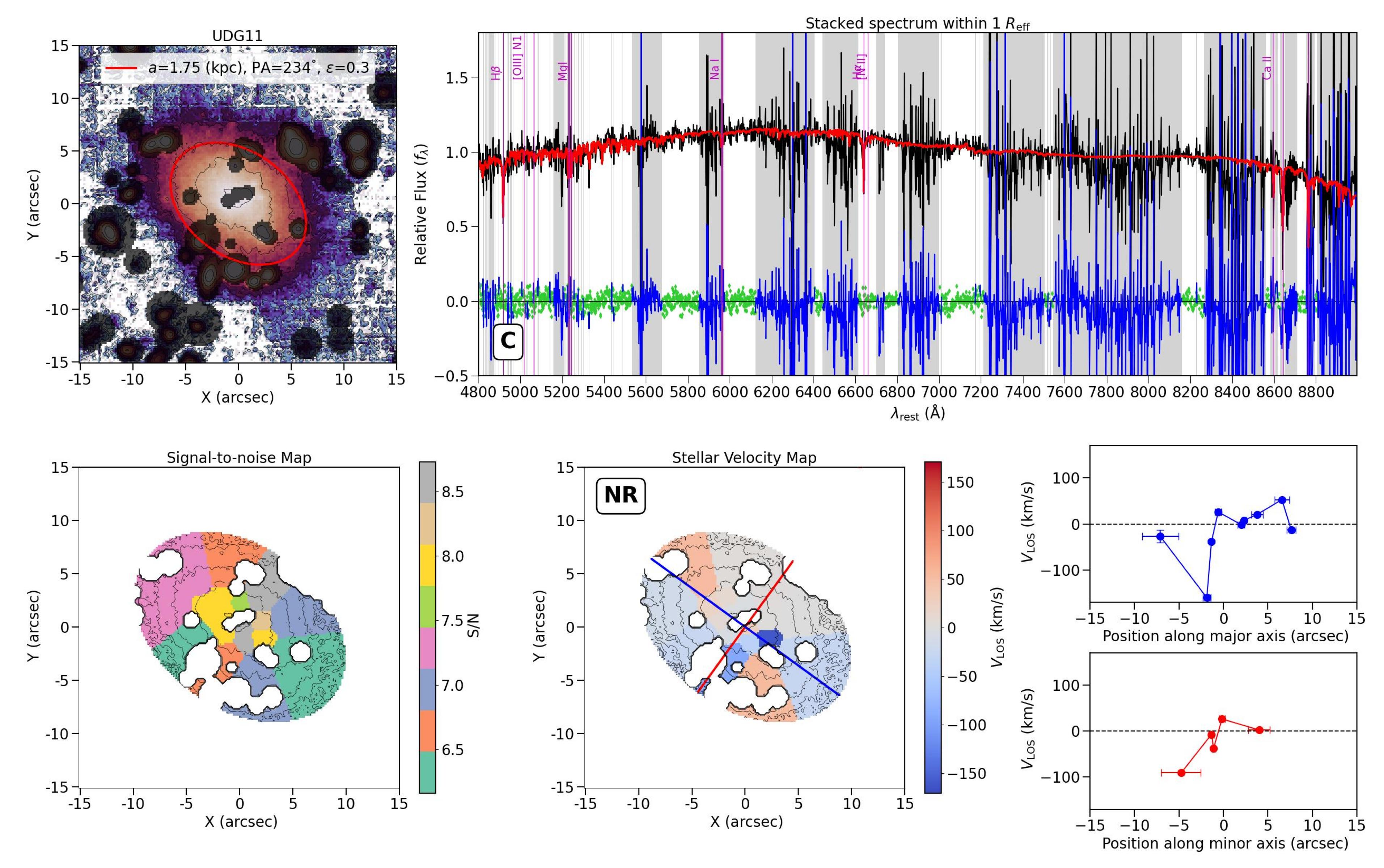}
    \caption{Same as Figure~\ref{fig:UDG1_stellar_kinematics}, but for UDG11.}
    \label{udg11_2D}
    \includegraphics[scale=0.28]{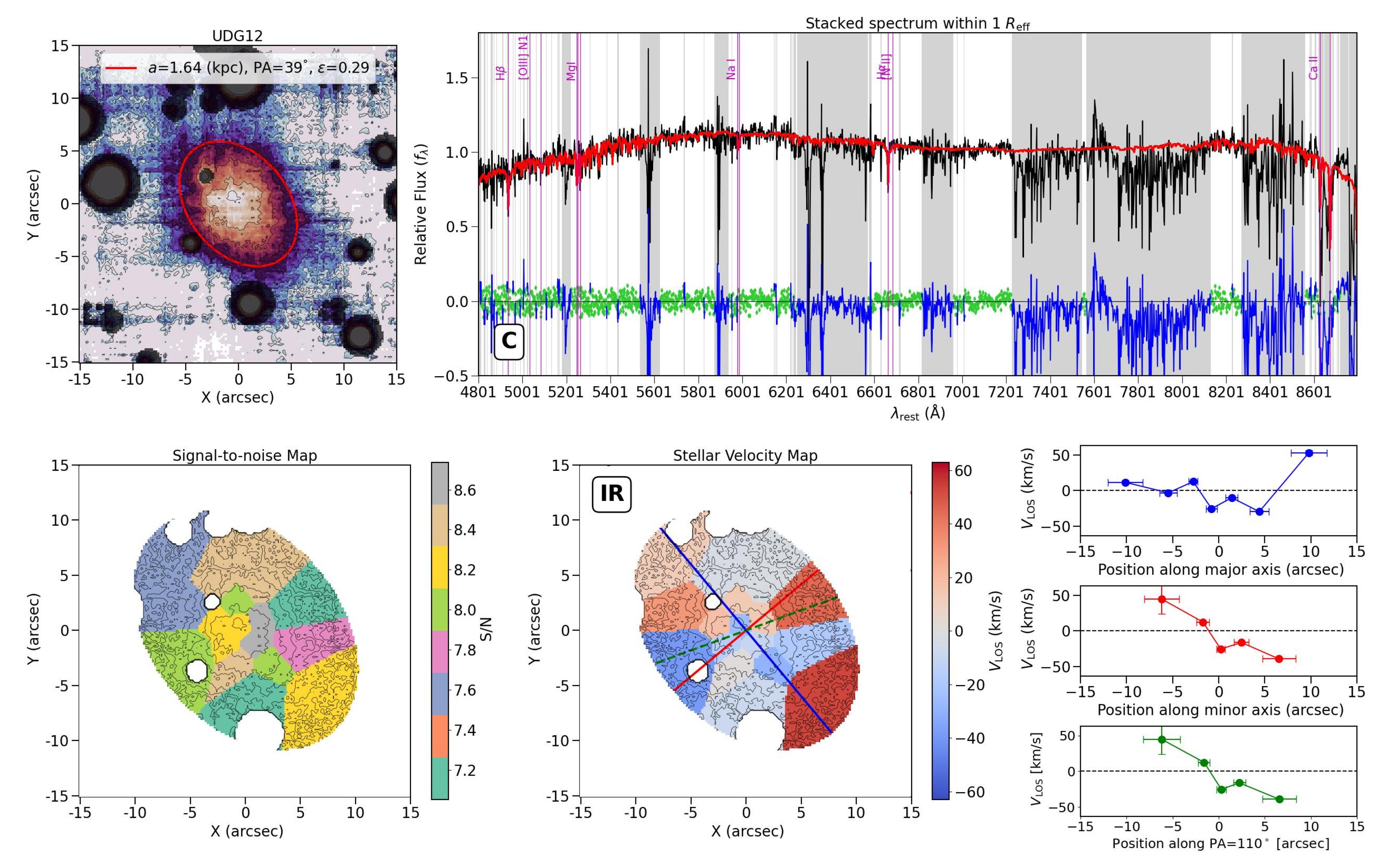}
    \caption{Same as Figure~\ref{fig:UDG1_stellar_kinematics}, but for UDG12.}
    \label{udg12_2D}
\end{figure*}

\begin{figure*}
    \centering
    \includegraphics[scale=0.28]{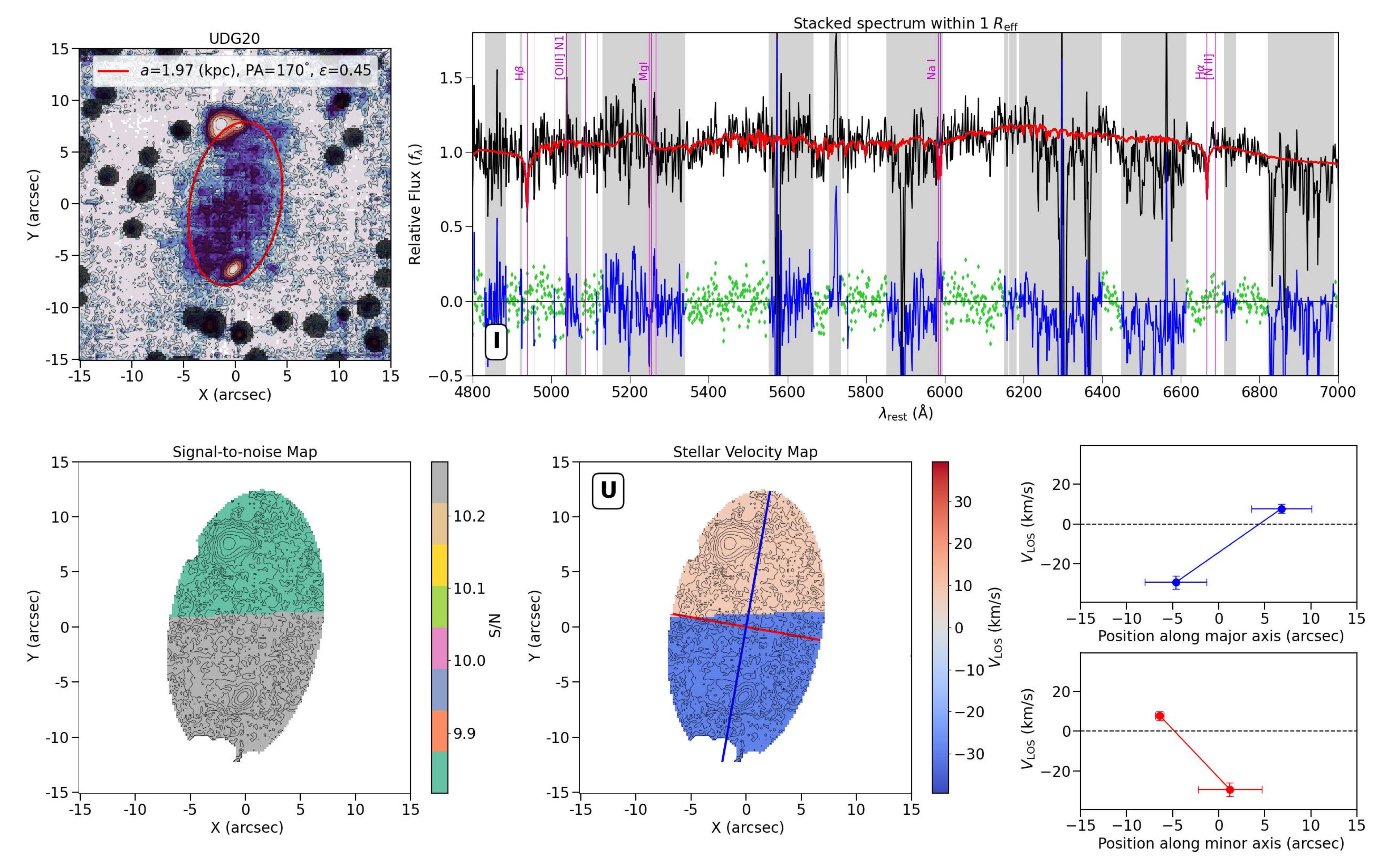}
    \caption{Same as Figure~\ref{fig:UDG1_stellar_kinematics}, but for UDG20.}
    \label{udg20_2D}
    \includegraphics[scale=0.28]{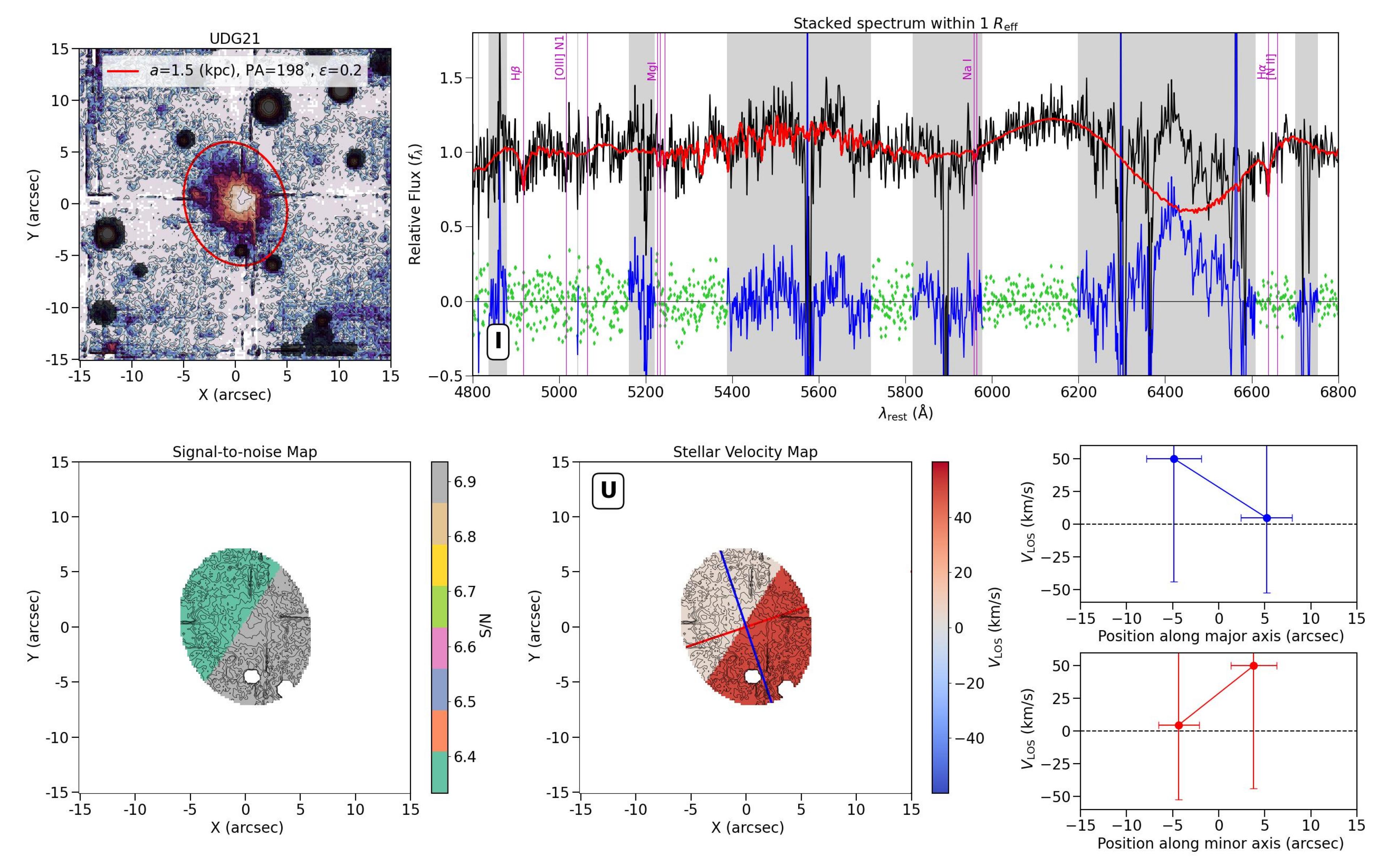}
    \caption{Same as Figure~\ref{fig:UDG1_stellar_kinematics}, but for UDG21.}
    \label{udg21_2D}
\end{figure*}

\begin{figure*}
    \centering
    \includegraphics[scale=0.28]{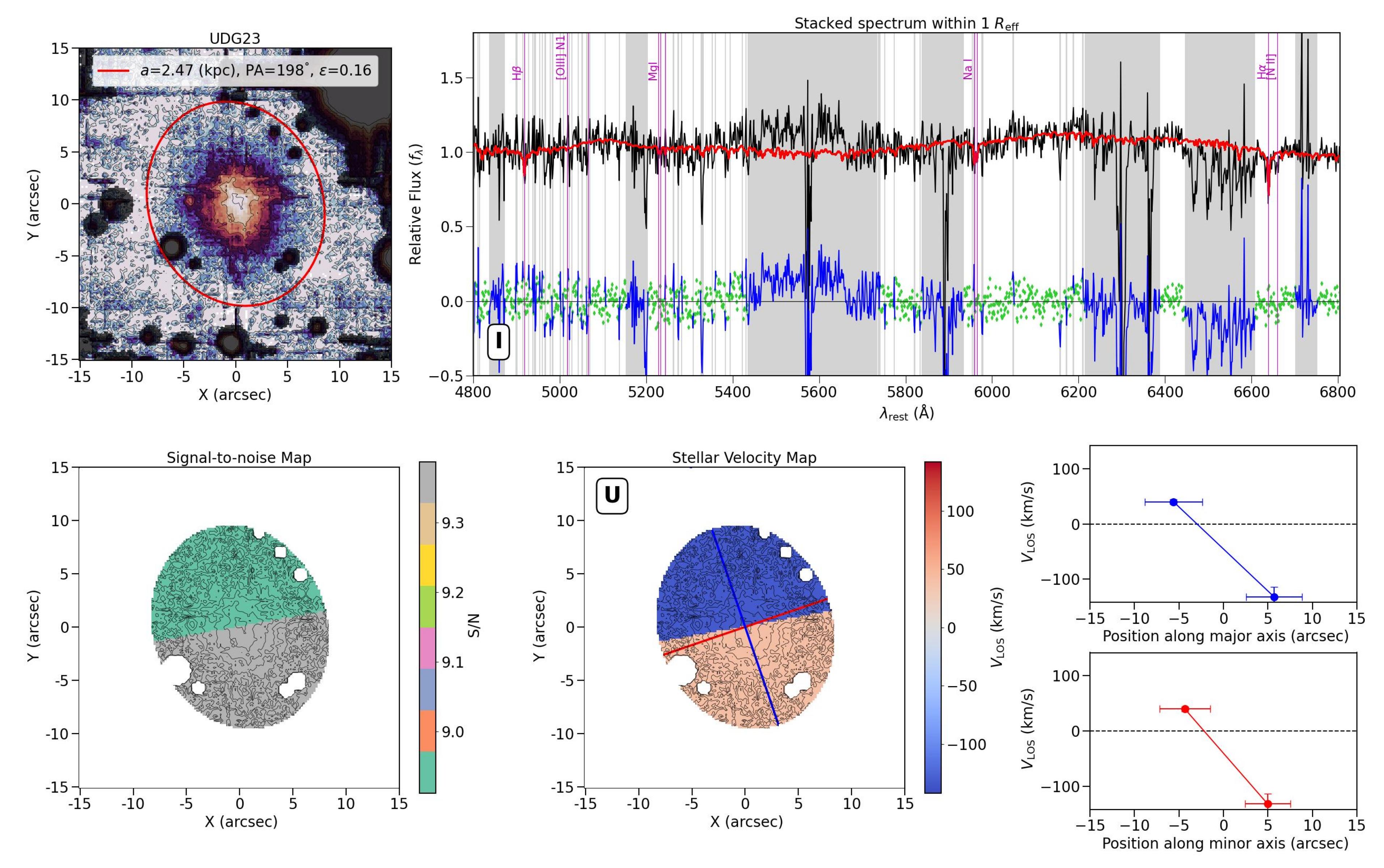}
    \caption{Same as Figure~\ref{fig:UDG1_stellar_kinematics}, but for UDG23.}
    \label{udg23_2D}
    \includegraphics[scale=0.28]{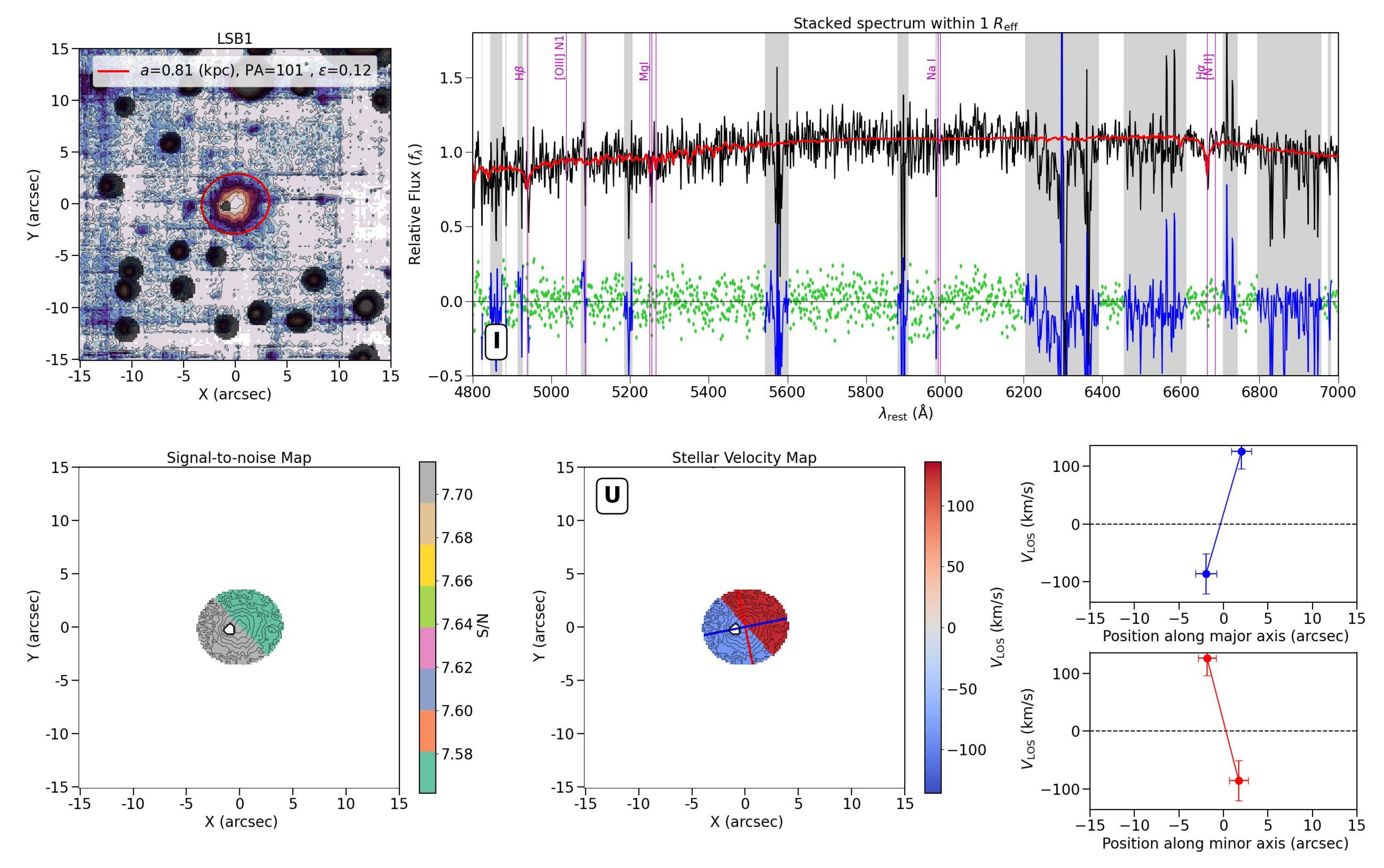}
    \caption{Same as Figure~\ref{fig:UDG1_stellar_kinematics}, but for LSB1.}
    \label{lsb1_2D}
\end{figure*}

\begin{figure*}
    \centering
    \includegraphics[scale=0.28]{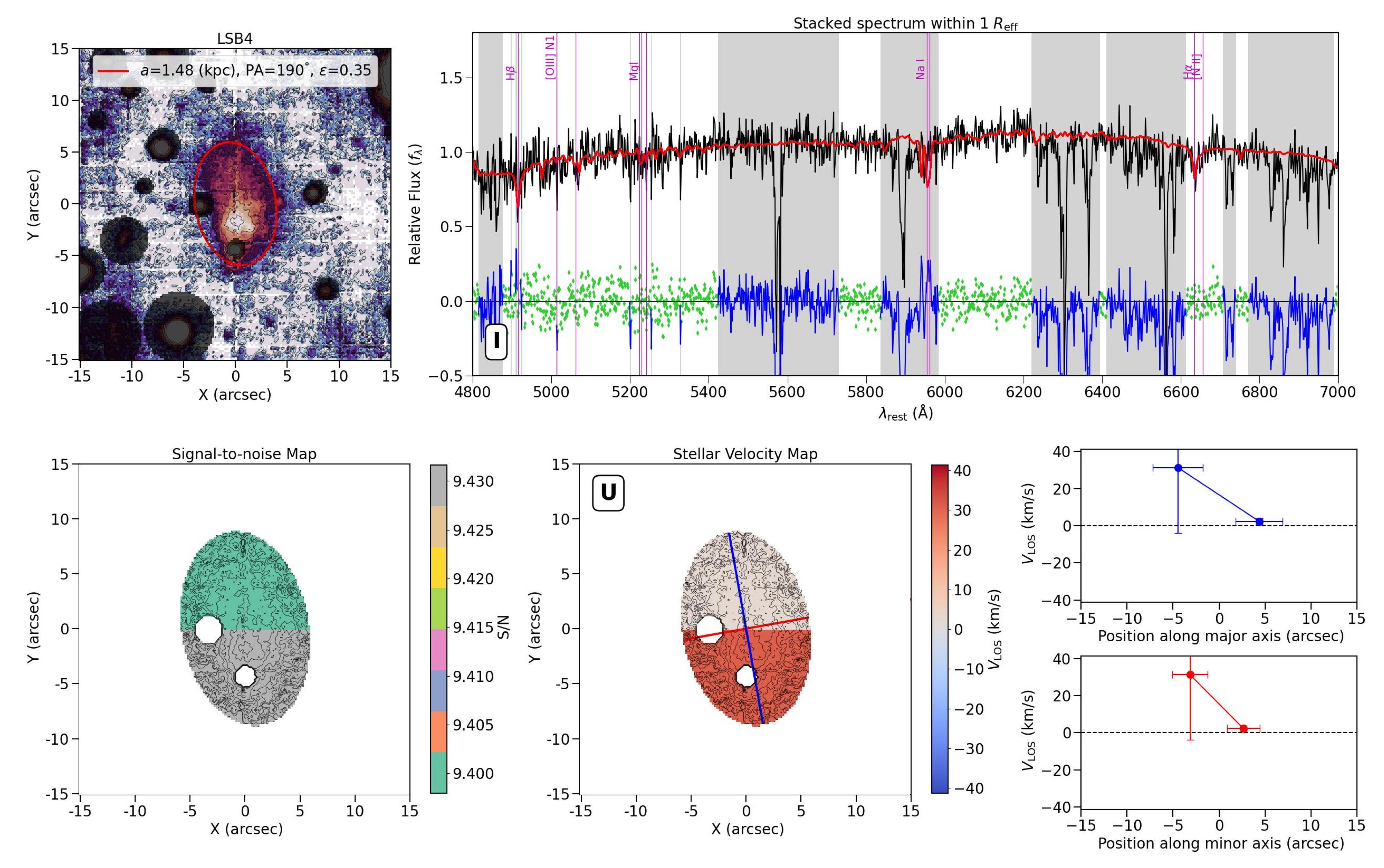}
    \caption{Same as Figure~\ref{fig:UDG1_stellar_kinematics}, but for LSB4.}
    \label{lsb4_2D}
    \includegraphics[scale=0.28]{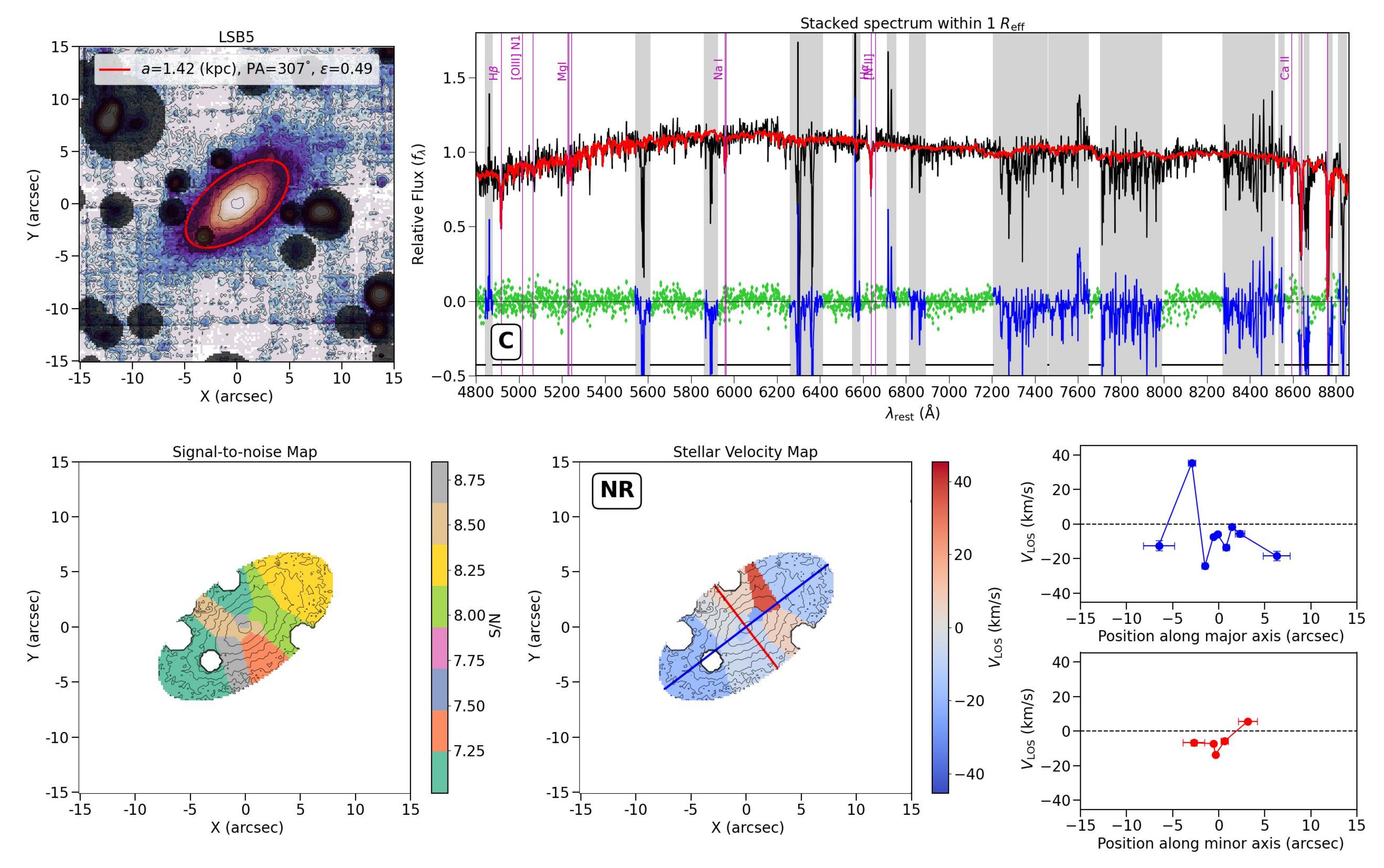}
    \caption{Same as Figure~\ref{fig:UDG1_stellar_kinematics}, but for LSB5.}
    \label{lsb5_2D}
\end{figure*}

\begin{figure*}
    \centering
    \includegraphics[scale=0.28]{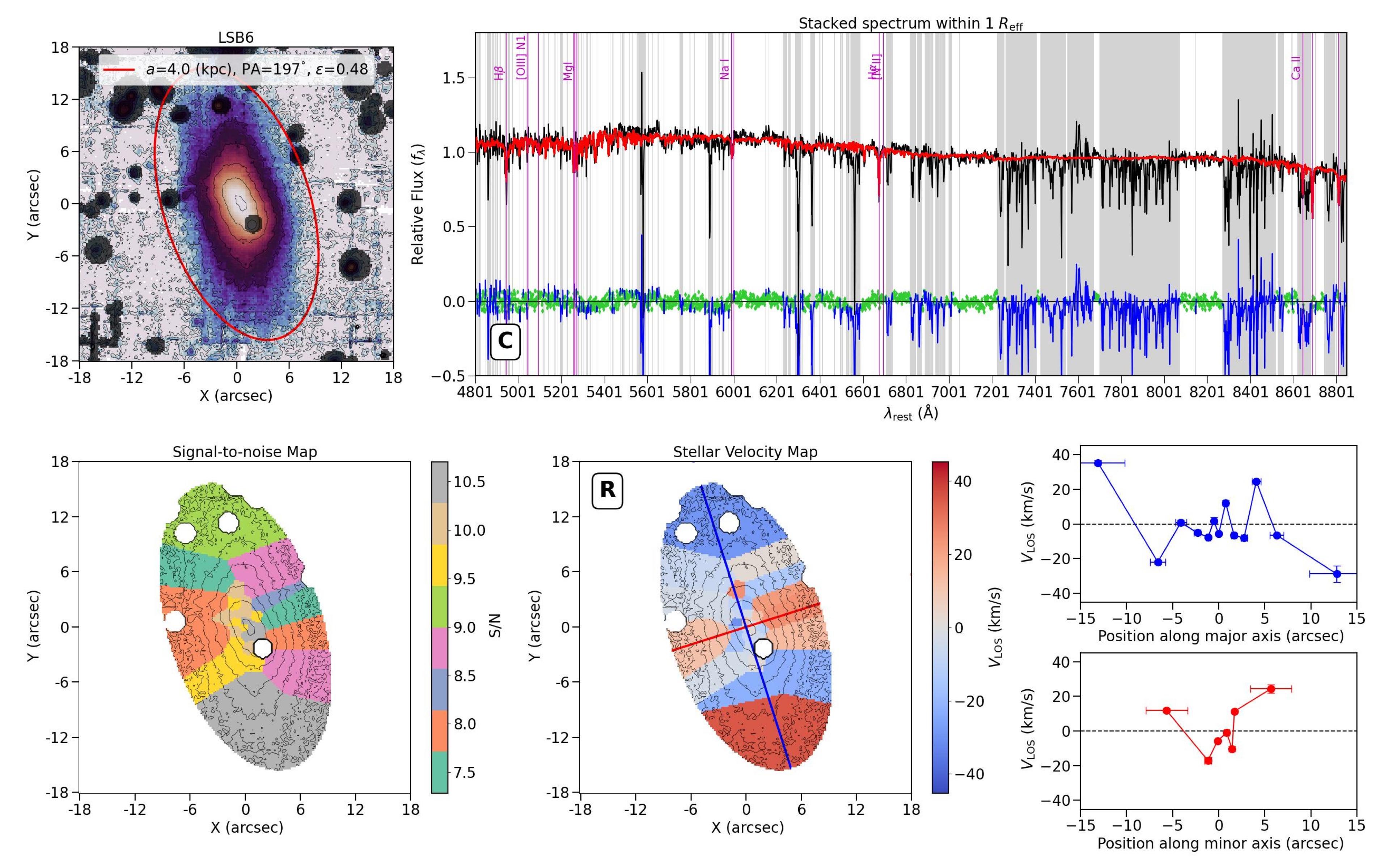}
    \caption{Same as Figure~\ref{fig:UDG1_stellar_kinematics}, but for LSB6.}
    \label{lsb6_2D}
    \includegraphics[scale=0.28]{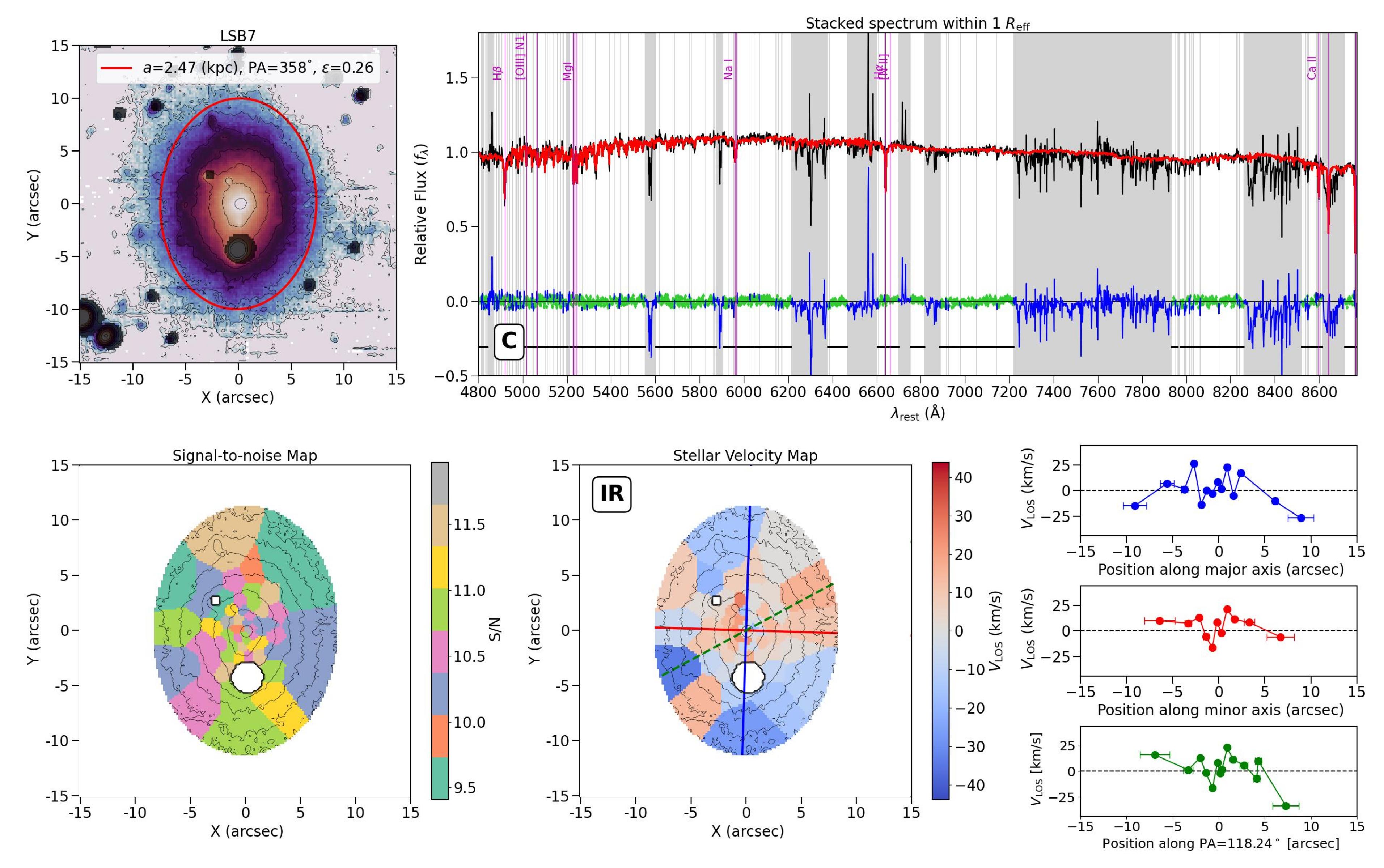}
    \caption{Same as Figure~\ref{fig:UDG1_stellar_kinematics}, but for LSB7.}
    \label{lsb7_2D}
\end{figure*}

\begin{figure*}
    \centering
    \includegraphics[scale=0.28]{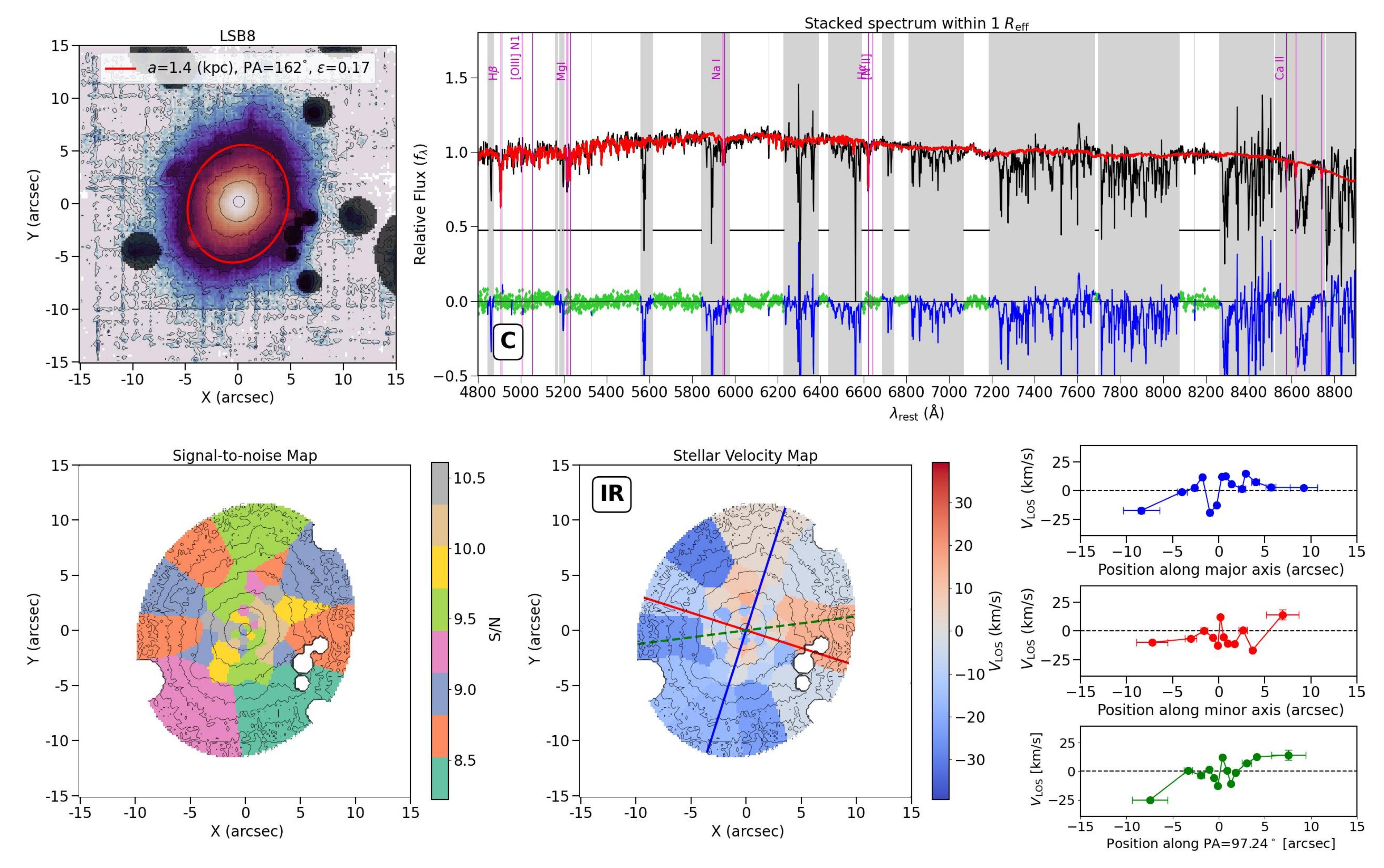}
    \caption{Same as Figure~\ref{fig:UDG1_stellar_kinematics}, but for LSB8.}
    \label{lsb8_2D}
\end{figure*}

\section{Stellar velocity and velocity dispersion maps}
\label{app:vlos_slos_maps}

In this section, we present the \vlos\, and \slos\, maps for a sub-set of galaxies in the LEWIS sample. To ensure an unbiased measure for \slos, the Voronoi binning threshold was set to S/N$\geq$15.

\begin{figure*}
    \centering
    \includegraphics[scale=0.28]{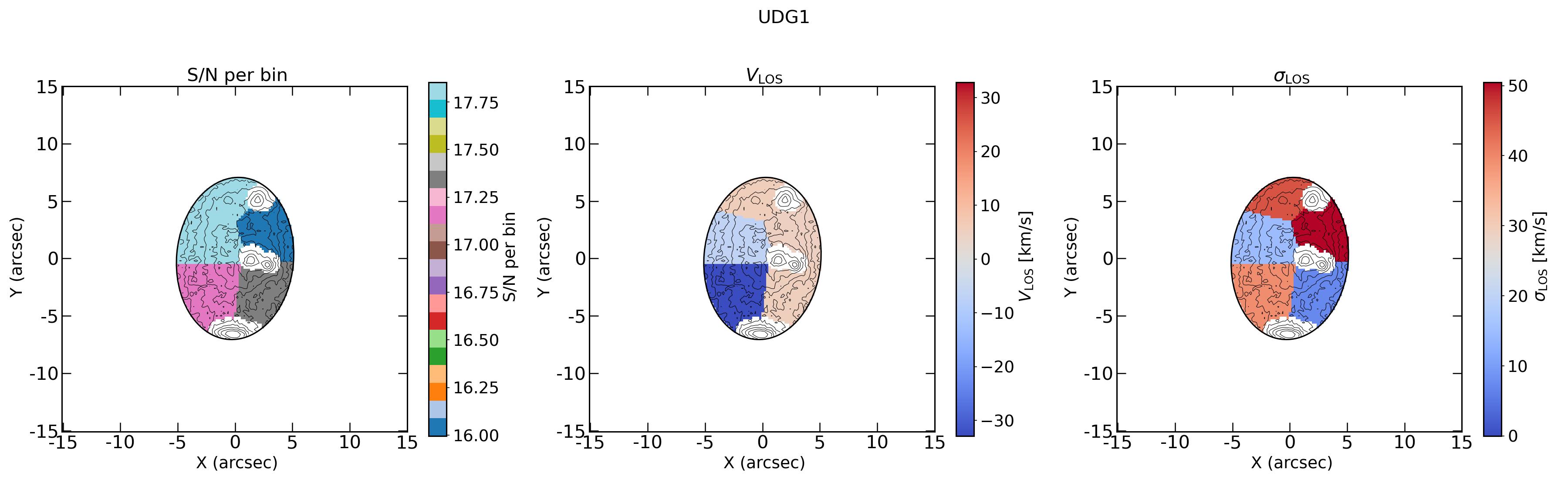}
    \caption{Stellar \vlos\, and \slos\, maps for UDG1. Left panel: Voronoi binned map of the S/N with threshold S/N\,=\,15. Central panel: stellar \vlos\, map subtracted from systemic velocity \Vsys . Right panel: stellar \slos\, map. The black ellipse represents the elliptical region used to extract the $V_{\rm rms}$ with semi-major axis $a$\,=\,\Reff.}
    \label{fig:UDG1_vlos_slos}
\end{figure*}

\begin{figure*}
    \centering
    \includegraphics[scale=0.28]{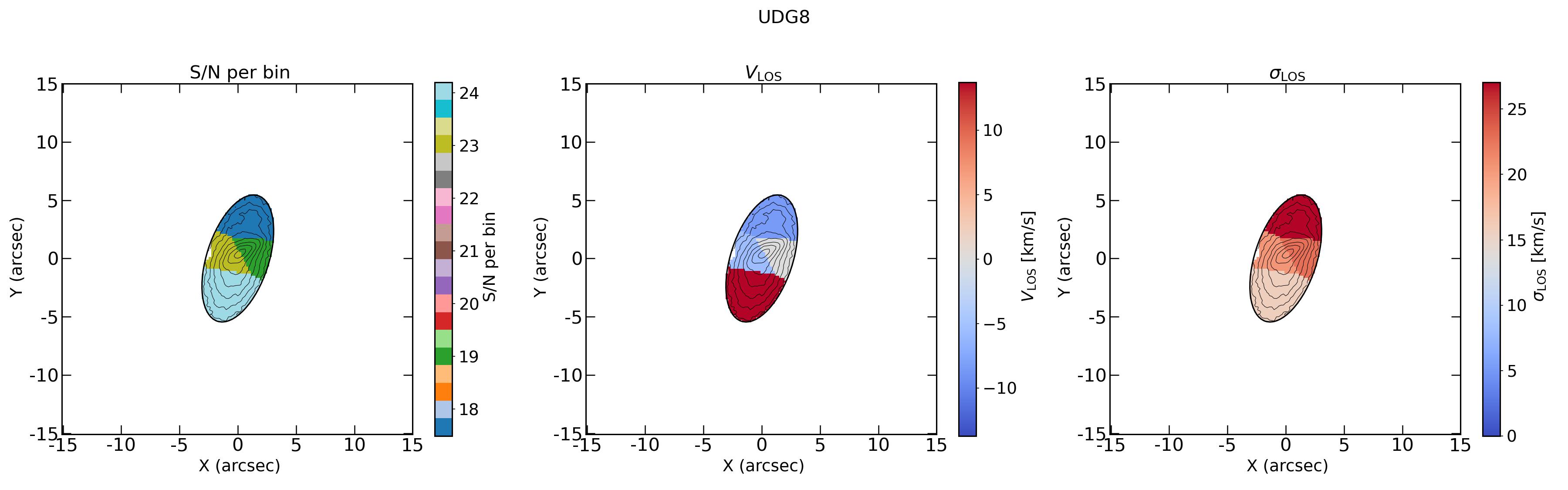}
    \caption{Same as Figure~\ref{fig:UDG1_vlos_slos}, but for UDG8.}
    \label{fig:UDG8_vlos_slos}
\end{figure*}

\begin{figure*}
    \centering
    \includegraphics[scale=0.28]{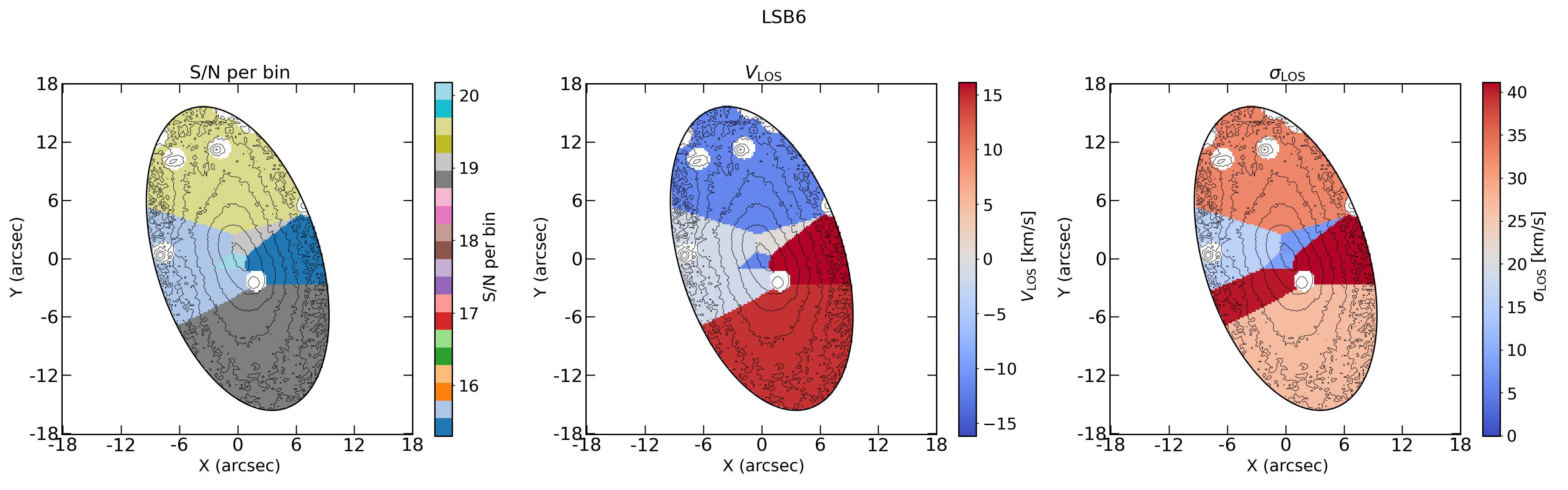}
    \caption{Same as Figure~\ref{fig:UDG1_vlos_slos}, but for LSB6.}
    \label{fig:LSB6_vlos_slos}
\end{figure*}

\begin{figure*}
    \centering
    \includegraphics[scale=0.28]{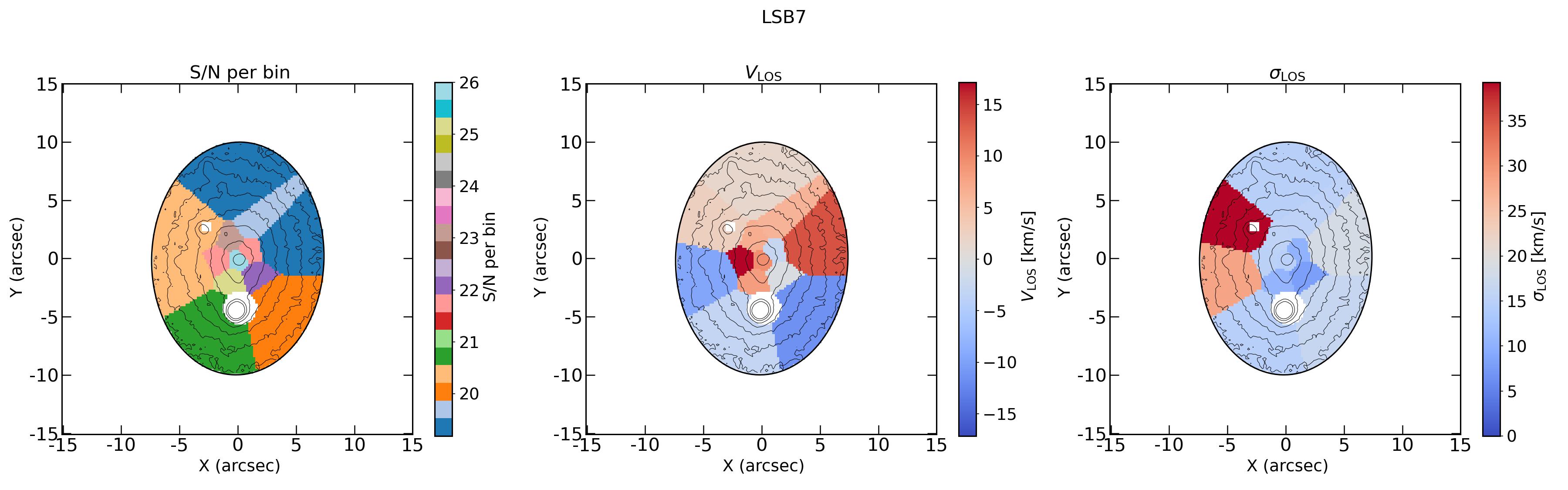}
    \caption{Same as Figure~\ref{fig:UDG1_vlos_slos}, but for LSB7.}
    \label{fig:LSB7_vlos_slos}
\end{figure*}

\begin{figure*}
    \centering
    \includegraphics[scale=0.28]{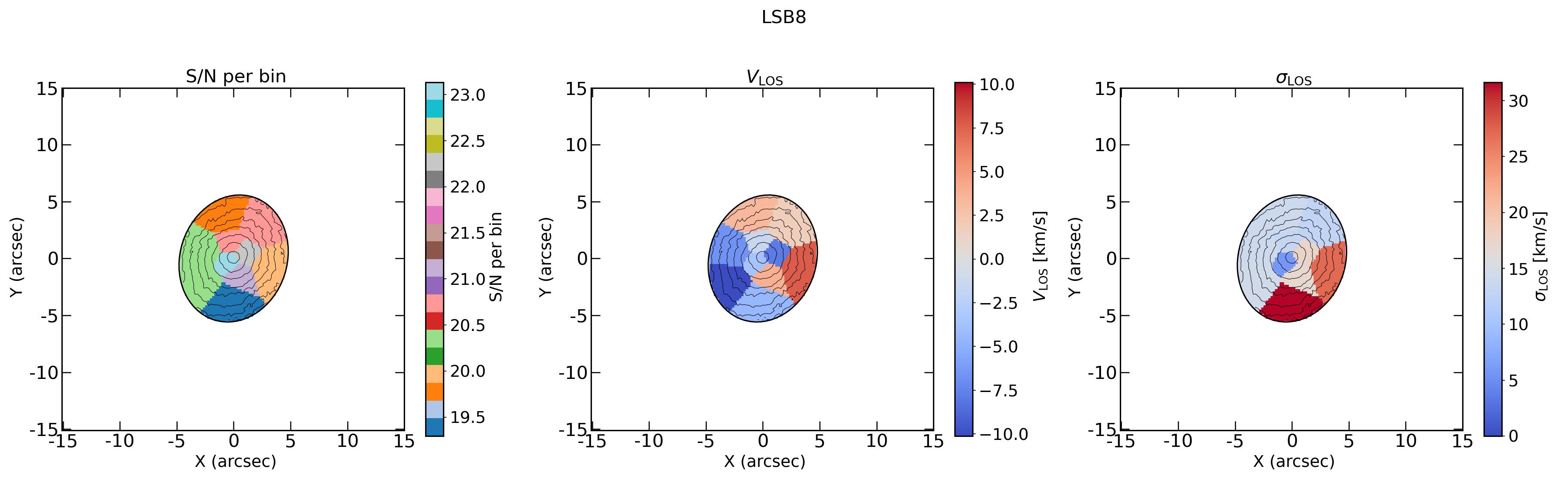}
    \caption{Same as Figure~\ref{fig:UDG1_vlos_slos}, but for LSB8.}
    \label{fig:LSB8_vlos_slos}
\end{figure*}

\end{appendix}

\end{document}